\documentclass[
12pt, % The default document font size, options: 10pt, 11pt, 12pt
oneside, % Two side (alternating margins) for binding by default, uncomment to switch to one side
english, % ngerman for German
singlespacing, % Single line spacing, alternatives: onehalfspacing or doublespacing
headsepline, % Uncomment to get a line under the header
]{MastersDoctoralThesis} % The class file specifying the document structure

\usepackage{fourier} % Use the Palatino font by default
\usepackage[T1]{fontenc} % Output font encoding for international characters
\usepackage[utf8]{inputenc} % Required for inputting international characters
\usepackage{siunitx}
\usepackage{mathtools,amsthm}
\usepackage{bm}
\usepackage{algorithm}
\usepackage{algpseudocode}
\usepackage{subfig}
\usepackage{caption}
\usepackage{float}
\usepackage[inline, shortlabels]{enumitem}
\usepackage[final]{microtype}
\usepackage[backend=bibtex,style=alphabetic,natbib=true]{biblatex} % Use the bibtex backend with the authoryear citation style (which resembles APA)

\usepackage[autostyle=true]{csquotes} % Required to generate language-dependent quotes in the bibliography

\newcommand{\vecr}{\bm{r}} %% r vector
\newcommand{\vecp}{\bm{p}} %% p vector (momento)
\newcommand{\veck}{\bm{k}}
\newcommand{\nnet}{N_{\theta}(\bm{r})}
\newcommand{\rdf}{$g(r) \ $}
\newtheorem{theorem}{Theorem}
\thesistitle{Using Computational Intelligence for solving the Ornstein-Zernike equation} % Your thesis title, this is used in the title and abstract, print it elsewhere with \ttitle
\supervisor{Dr. Luis Carlos \textsc{Padierna García}\\
	Dr. Ramón \textsc{Castañeda Priego}
} % Your supervisor's name, this is used in the title page, print it elsewhere with \supname
\examiner{} % Your examiner's name, this is not currently used anywhere in the template, print it elsewhere with \examname
\degree{Master in Applied Science} % Your degree name, this is used in the title page and abstract, print it elsewhere with \degreename
\author{Edwin Armando Bedolla Montiel} % Your name, this is used in the title page and abstract, print it elsewhere with \authorname
\addresses{} % Your address, this is not currently used anywhere in the template, print it elsewhere with \addressname

\subject{Physical Sciences} % Your subject area, this is not currently used anywhere in the template, print it elsewhere with \subjectname
 % Keywords for your thesis, this is not currently used anywhere in the template, print it elsewhere with \keywordnames
\university{\href{https://ugto.mx}{Universidad de Guanajuato}} % Your university's name and URL, this is used in the title page and abstract, print it elsewhere with \univname
\department{\href{}{University of Guanajuato}} % Your department's name and URL, this is used in the title page and abstract, print it elsewhere with \deptname
\group{\href{}{Science and Engineering Division}} % Your research group's name and URL, this is used in the title page, print it elsewhere with \groupname
\faculty{\href{}{Science and Engineering Division}} % Your faculty's name and URL, this is used in the title page and abstract, print it elsewhere with \facname

\AtBeginDocument{
\hypersetup{pdftitle=\ttitle} % Set the PDF's title to your title
\hypersetup{pdfauthor=\authorname} % Set the PDF's author to your name
\hypersetup{pdfkeywords=\keywordnames} % Set the PDF's keywords to your keywords
}

\begin{document}

\frontmatter % Use roman page numbering style (i, ii, iii, iv...) for the pre-content pages

\pagestyle{plain} % Default to the plain heading style until the thesis style is called for the body content

%----------------------------------------------------------------------------------------
%	TITLE PAGE
%----------------------------------------------------------------------------------------

\begin{titlepage}
\begin{center}

\vspace*{.035\textheight}
{\scshape\LARGE \univname\par}\vspace{1.5cm} % University name
\textsc{\Large Master's Thesis}\\[0.5cm] % Thesis type

\HRule \\[0.4cm] % Horizontal line
{\huge \bfseries \ttitle\par}\vspace{0.4cm} % Thesis title
\HRule \\[1.5cm] % Horizontal line
 
\begin{minipage}[t]{0.4\textwidth}
\begin{flushleft} \large
\emph{Author:}\\
\href{}{\authorname} % Author name - remove the \href bracket to remove the link
\end{flushleft}
\end{minipage}
\begin{minipage}[t]{0.4\textwidth}
\begin{flushright} \large
\emph{Supervisor:} \\
\href{}{\supname} % Supervisor name - remove the \href bracket to remove the link  
\end{flushright}
\end{minipage}\\[3cm]
 
\vfill

\large \textit{A thesis submitted in fulfillment of the requirements\\ for the degree of \degreename}\\[0.3cm] % University requirement text
\textit{in the}\\[0.4cm]
\groupname\\\deptname\\[2cm] % Research group name and department name
 
\vfill

{\large \today}\\[4cm] % Date

\vfill
\end{center}
\end{titlepage}

%----------------------------------------------------------------------------------------
%	ABSTRACT PAGE
%----------------------------------------------------------------------------------------

\begin{abstract}
\addchaptertocentry{\abstractname} % Add the abstract to the table of contents
The main goal of this thesis is to provide an exploration of the use of 
\emph{computational intelligence} techniques to study the numerical solution of the 
Ornstein-Zernike equation for simple liquids. In particular, a continuous model of the hard 
sphere fluid is studied. There are two main proposals in this contribution. First, the use 
of \emph{neural networks} as a way to parametrize closure relation when solving the 
Ornstein-Zernike equation. It is explicitly shown that in the case of the hard sphere 
fluid, the neural network approach seems to reduce to the so-called Hypernetted Chain 
closure. For the second proposal, we explore the fact that if more physical information is 
incorporated into the theoretical formalism, a better estimate can be obtained with the use 
of \emph{evolutionary optimization} techniques. When choosing the modified Verlet closure 
relation, and leaving a couple of free parameters to be adjusted, the results are as good 
as those obtained from molecular simulations. The thesis is then closed with a brief 
summary of the main findings and outlooks on different ways to improve the proposals 
presented here.
\end{abstract}

%----------------------------------------------------------------------------------------
%	ACKNOWLEDGEMENTS
%----------------------------------------------------------------------------------------

% \begin{acknowledgements}
% \addchaptertocentry{\acknowledgementname} % Add the acknowledgements to the table of contents
% \textcolor{red}{Pendiente por redactar \dots}
% \end{acknowledgements}

%----------------------------------------------------------------------------------------
%	LIST OF CONTENTS/FIGURES/TABLES PAGES
%----------------------------------------------------------------------------------------

\tableofcontents % Prints the main table of contents

% \listoffigures % Prints the list of figures

% \listoftables % Prints the list of tables

%----------------------------------------------------------------------------------------
%	THESIS CONTENT - CHAPTERS
%----------------------------------------------------------------------------------------

\mainmatter % Begin numeric (1,2,3...) page numbering

\pagestyle{thesis} % Return the page headers back to the "thesis" style

% Include the chapters of the thesis as separate files from the Chapters folder
% Uncomment the lines as you write the chapters

\chapter{Introduction}
\label{Cap1}

Since the mainstream adoption of \emph{Machine Learning} (ML) methods
on common tasks such as object recognition, computer vision,
and human-computer interactions~\cite{lecunDeepLearning2015},
scientists have tried to adopt most of these techniques to further research
in their respective fields. From drug development~\cite{redaMachineLearningApplications2020}
to genetics and biotechnology~\cite{libbrechtMachineLearningApplications2015},
multiple applications of ML to current research problems have seen
widespread interest for their generalization and automatic discovery attributes.
It is with the inspiration from these applications that physicists have attempted to use 
such methods in diverse Physics fields~\cite{carleoMachineLearningPhysical2019a,dunjkoMachineLearningArtificial2018,carrasquillaMachineLearningPhases2017a}.

Most of the attempts and successes of using ML methods in Physical sciences come from
the direct application of common ML pipelines and uses, such as \emph{classification},
\emph{regression}, and \emph{unsupervised learning}~\cite{hastieElementsStatisticalLearning2009}, just to name some.
Such is the case of the determination of the
critical point of the Ising model as means of a classification task~\cite{carrasquillaMachineLearningPhases2017a}.
Similar is the case of the use of \emph{computer vision} and \emph{deep learning} 
techniques in particle physics, which have seen great applications when dealing with 
experimental data~\cite{radovicMachineLearningEnergy2018}.
In each of the previous examples, scientists have taken the most common applications
of ML methods and have adjusted them for their respective research problems.
This has the advantage that such ML techniques have been extensively researched
and developed, so physicists know that these methods are robust and useful for
the problems they have been developed for.
However, it turns out that not all ML techniques can be readily applied to the problem
at hand, and physicists should instead try to capitalize on the Physics of the problem and
use it along with the ML method to boost its usefulness, flexibility and accuracy.

It is with this perspective that physicists have preferred to incorporate most of the 
Physics into the ML method, and thus create a new form of
\emph{physics-inspired machine learning}~\cite{karniadakisPhysicsinformedMachineLearning2021a}.
One such example is the Behler-Parrinello neural network approach for energy surfaces
in the Density Functional Theory framework~\cite{behlerGeneralizedNeuralNetworkRepresentation2007a}.
Within such proposal, Behler and Parrinello chose to use some functions whose 
definition and composition are based on the properties of the system studied, which were 
atoms and their components,
and use such information as input to a \emph{regression} scheme to approximate the
energy surface of the studied system. At the moment of publication, this approach
defined a new paradigm of ML application within the physical sciences. It was no longer
the fact that simple learning tasks were used, but by including physical descriptors
in the ML methods, new ways of obtaining the same results were found.
Not only do ML methods provide mostly the same results as the physical framework they
are modeling, they also provide solutions much faster and more efficiently~\cite{zhuPhysicsconstrainedDeepLearning2019}.

Of all the research fields within Physics, in this thesis we focus
our attention to the field of \emph{Condensed Matter Physics}, and in particular, to the
field of Liquid State Theory and Soft Matter. Before we do that, however, we should
mention briefly some of the precursors to the applications of ML to those fields.
A review by J\"{o}rg Behler~\cite{behlerPerspectiveMachineLearning2016a}
describes in great detail some of these precursors. Most applications have dealt with
materials science, computational chemistry and chemistry, and the computational aspect
of condensed matter physics. It is within these fields that new ways of using ML methods
have been developed from various needs in research. Another prime example is the coupling
of computer simulations and ML methods, such as the work by Li \emph{et al}~\cite{liMolecularDynamicsOntheFly2015}.
In that work, whenever the quantum-mechanical information is needed it is computed with
first-principles calculations. These calculations are then added to a dataset to be used
by ML methods, which in turn are used as approximators for the computer simulation.
This approach not only efficiently uses Physics in its most pure form, but it also
adopts the ML best attributes and uses them to its advantage.
For a more complete overview of some of the most impactful applications, the
review by Bedolla \emph{et al}~\cite{bedollaMachineLearningCondensed2020}
covers these and some other important aspects of ML methods applied to
condensed matter physics.

Although these applications form a tiny subset of all the modified applications of ML 
techniques in physical sciences research, we should note that these applications show an 
important aspect in common between them. If we wish to assimilate the physics of the 
problem at hand, variations and modifications to the common ML methods and techniques are 
needed. Exploration and testing, trial and error, are an important part of the search and 
application of ML methods to Physics research. In the case of Liquid State Theory and Soft 
Matter, we can consider that most research is still in the exploration and testing stage,
although some applications have seen great success in specific scenarios.
One such successful application is the use of \emph{Support Vector Machines}\textemdash
an explicit method useful for classification and regression based on kernels and
quadratic optimization~\cite{steinwartSupportVectorMachines2008}\textemdash
in the description of the properties of glassy dynamics~\cite{schoenholzStructuralApproachRelaxation2016}.
The reason for research in Soft Matter and ML still being part of the exploration step is 
that in Soft Matter and Liquid Theory it is hard to find suitable descriptors that 
actually tell us useful information from the system or phenomena~\cite{dijkstraPredictiveModellingMachine2021a}.
As such, most of the time a descriptor-based approach might not be feasible for every 
possible system. Instead, we need to explore a diverse range of possibilities when using ML 
methods within the context of Soft Matter and Liquid Theory.

\section{Recent research in Soft Matter and Machine Learning}
Even though most research is still exploration and testing, there have been interesting
amalgamations and developments in the fields of Soft Matter, with Liquid Theory dragging
behind. In a sense, this is expected, due to the fact that Liquid Theory might be thought
of as \emph{solved}, although there is still research done within the field.
Let us focus first on the developments of Soft Matter. Instead of referring to specific
uses of ML within Soft Matter, it is more fruitful to mention some of the research groups
that have delved deep into using ML methods in Soft Matter. The group of Marjolein
Dijkstra at Utrecht University is an excellent example. Having done impactful research
in the field of Colloidal Soft Matter~\cite{dijkstraPhaseDiagramHighly1999,leunissenIonicColloidalCrystals2005},
the group is now focusing on using all the research and knowledge built and trying to
understand the best way to enforce the physics of the systems into ML techniques.
The group has explored with \emph{evolutionary algorithms} and their uses in patchy
colloids~\cite{bianchiPredictingPatchyParticle2012}. We refer to evolutionary algorithms as
derivative-free optimization algorithms that are useful for nonlinear optimization
problems~\cite{yuIntroductionEvolutionaryAlgorithms2010}.
Unsupervised methods, such as \emph{principal component analysis}~\cite{hastieElementsStatisticalLearning2009},
have been used for choosing the best descriptors in supercooled liquids~\cite{boattiniAveragingLocalStructure2021},
as well as the detection of local structure in colloidal systems~\cite{boattiniUnsupervisedLearningLocal2019a}.
All in all, these methods simplify the process of dealing with these research problems.
ML methods make it simpler and easier to identify structure and attributes from a
system. However, it is important to note that not only do these methods make it simpler,
they can also contribute to finding new things that were previously not as obvious or easy
to see.
Another work from the group is the use of classification methods 
to identify different types of crystal phases using
a mix of supervised and unsupervised methods~\cite{hastieElementsStatisticalLearning2009},
such as in the work by van Damme \emph{et al}~\cite{vandammeClassifyingCrystalsRounded2020}.
This work is quite interesting because it is a great example of using physical descriptors,
such as bond order parameters~\cite{steinhardtBondorientationalOrderLiquids1983,lechnerAccurateDeterminationCrystal2008},
along with ML methods that actively select and distinguish between the best descriptors
for the system.

Another important group that has done several advances in the use of ML within Soft Matter
is the one lead by Thomas Truskett from the University of Texas.
Their work on the use of ML methods for inverse design of soft materials~\cite{shermanInverseMethodsDesign2020a}
is in similar ways helping out the work by the group of Dijkstra in the same research
problem~\cite{APSAPSMarcha}.
However, the work by the Truskett group is in fact more focused on the definition and
foundations of better descriptors for soft materials and off-lattice systems~\cite{jadrichUnsupervisedMachineLearning2018}.
An important topic that the group explores is the inverse design of self-assembly
systems. \emph{Self-assembly} is the property of soft materials to order their
components, such as particles, atoms or cells, without any external interactions,
into functional structures~\cite{grzybowskiSelfassemblyCrystalsCells2009}.
In this direction, their interest is particularly focused on self-assembly and how these 
phenomena can be dealt with using ML methods.
In a particular work by Lindquist \emph{et al} with the Truskett group~\cite{lindquistCommunicationInverseDesign2016},
optimization-based methods were used along with standard molecular dynamics computer
simulation methods to understand how certain materials can self-assemble.
Instead of using optimization-based methods, another work from the Truskett group
deals with probabilistic ML methods for some of the same phenomena~\cite{jadrichProbabilisticInverseDesign2017}.
The difference between both these uses is that probabilistic methods are better at dealing
with probability distributions that stem from the Statistical Mechanics description of the
problem. With the use of specialized frameworks, handling these descriptors are easier
and more flexible.

\section{Liquid Theory as a basis for Soft Matter}
Up until this point, we have talked about Soft Matter and its interaction with machine
learning methods, and most of all the way it has been disrupting this research field
with novel ways to do things. Liquid State Theory was briefly mentioned before,
but it is time to state the relation between both research fields.
Why talk about Liquid State Theory and Soft Matter as if these were related somehow?
In a fascinating note by Evans, Frenkel and Dijkstra~\cite{evansSimpleLiquidsColloids2019}
they talk about the precise relation between liquids, their theory, and how by studying
such simple, but also complicated systems, soft matter research arises quite naturally.
Liquids are non-trivial systems, that are also dense, disordered and spatially correlated
but exhibit peculiar characteristics that make them different from crystals, solids,
and gases. The first attempt to understand and apply liquid theory,
along with computer simulations of liquids, was to study the
classic Lennard-Jones system to model Argon~\cite{mcdonaldCalculationThermodynamicProperties1967,verletComputerExperimentsClassical1967a}. 
Computer simulation methods, theoretical frameworks, and experimental setups helped 
understanding the true nature of liquids, giving them a special place in Physics research. 
Physicists were attempting to construct a rigorous framework which were used to 
further create new research fields.
Such fields were then created by the outstanding work of Nobel Laureate Pierre Gilles-de 
Gennes. In his Nobel Lecture~\cite{degennesSoftMatter1992} he mentioned that by studying 
simple systems it was possible to extend what was known of such systems to new, different 
systems that exhibit similar properties. His work on liquid crystals and polymer physics 
opened up a whole new research field, which we now call Soft Matter.
Then again, behind all the bewildering phenomena found in Soft Matter, and all the 
innovative research done to understand it, there exists a solid theory of the most simple 
systems known as liquids. And yet, these systems are not \emph{simple} at all, they
have evolved into active research fields such as ionic liquids~\cite{rogersIonicLiquidsSolvents2003}, complex fluids~\cite{gelbartNewScienceComplex1996},
and many others.

Despite all the research done in Liquid State Theory, most research has been focused on
understanding it through computational and theoretical frameworks. There is not much
interaction between Liquid State Theory and ML research at this point.
Yet, there seems to be an area of opportunity where Liquid State Theory can greatly
benefit from all the current research in ML methods. Arguably, the first work to delve deep 
into the intersection between Liquid State Theory and ML theory was the work done by
Goodall and Lee.
In their work, a dataset is built from Molecular Dynamics computer simulations using several
interaction potentials. Then, a ML method was used to predict properties from a new,
unknown system, using most of the information available from the dataset.
The method worked great, but the precision was not the best in some cases. Granted, this
is just the first attempt to build a solution to the problem of closure relations
for the theoretical framework of the Ornstein-Zernike equation~\cite{hansenTheorySimpleLiquids2013}.
It was then extended, in the same work, to solve an inverse design problem of liquids, 
which indeed solved the problem but was lacking in accuracy, and was not tested in
more challenging problems, which would be interesting to see the generalization of
such methods.
After all, the attempt is a novel way to look at things, and most of all, a new way to
look at Liquid State Theory and its intersection with ML theory.
Exploring this area is the sole purpose of this thesis, and in particular, in the same
line of understanding how can ML methods be used in theoretical frameworks that model
simple liquids. It seems like a worthwhile task to test various approaches and look for a
way to streamline the methods used so far in Liquid State Theory. This is the basis for
this work, to explore the use of ML methods in the theoretical formalism of the
Ornstein-Zernike equation, a rigorous framework that provides most of the structural and thermodynamical information from
a simple liquid system. We wish to understand if some of the most common ML methods
can be used, how they can be used, if they can even provide physically-relevant
solutions or not.

\section{Thesis outline}
To achieve this goal, this thesis is organized as follows. In chapter 2, 
the theory of simple liquids will be outlined, focusing primarily on the Ornstein-Zernike
formalism, as well as the theory of integral equations.
A description of the reference system\textemdash the hard sphere fluid\textemdash 
will be discussed. The thermodynamical properties of simple liquids will also be described 
in full. Computer simulation methods will also be discussed as they are the standard tool 
for current research when a baseline result is needed. Computer simulation methods can 
provide similar results to those obtained from experimental data, which is why they are 
used as a reference.
Then, in chapter 3, an overview of ML theory is presented.
In particular, those methods used in the current thesis, namely neural networks and
optimization methods. These methods have their mathematical frameworks, which will be
outlined as well, without going into much mathematical rigor.
The most common learning tasks such as supervised and unsupervised learning will be
touched only slightly, in a manner that is consistent with the current presentation
of the topics.
After the theoretical background, a discussion is carried out on the first proposal of using 
ML methods in Liquid State Theory in chapter 4.
The proposal is to use a neural network as a 
parametrization for the closure relation within the Ornstein-Zernike formalism.
It is found that the proposal works, but if the neural
network is not given too much information or description about the system, the neural
network training dynamics will reduce to a common closure relation, which is not
thermodynamically self-consistent.
In chapter 5, an exploration of including physical information to the problem gives
a more direct solution. Instead of using a neural network to parametrize the closure
relation, a known approximation is used, and let an evolutionary algorithm find the
best parameters for it. When using an approximation that is already shown to give
accurate results, the ML method is better suited to work correctly, but not without
its drawbacks, which are discussed in detail.
The thesis closes in chapter 6, with important conclusions and future ideas
to improve the proposals presented here. A few other proposal are outlined, although
not with results, or rigorous theoretical frameworks. These can be thought as
suggestions that might somehow become new ways to use ML methods within Liquid
State Theory. %% Introduction
\chapter{Liquid State Theory}
\label{Cap2}

In this chapter, a brief description of Liquid State Theory is carried out. In particular, 
the focus of the chapter is to state what a liquid is, its thermodynamical properties and
how equilibrium statistical mechanics is used to understand them. Then, a description of 
the hard-sphere fluid is mentioned, which is the fundamental system studied in this work.
After this, a brief description of computer simulation methods is outlined, both
Molecular Dynamics and Monte Carlo methods are mentioned, giving more importance to
Monte Carlo methods which are extensively used in this work.
The chapter is closed with a presentation of the important formalism that is the 
Ornstein-Zernike integral equation. Its important aspects are mentioned, along with
a discussion of several closure relations and the important role of the
\emph{bridge function.}

\section{Equilibrium Statistical Mechanics}
\label{sec:eq-statmech}

Consider a system of $N$ spherical particles in three dimensions
where each particle is characterized by its position $\vecr$ and momentum $\vecp$.
The \emph{Hamiltonian} of the system is given by
\begin{equation}
    \mathcal{H} \left( \vecr^{N}, \vecp^{N} \right) = 
    K \left( \vecp \right) + U \left( \vecr \right)
    \label{eq:hamiltonian}
\end{equation}
with $K$ being the kinetic energy and $U$ the potential energy of the system.
All together, the 6$N$ variables define a \emph{phase point} in a 6$N$-dimensional
\emph{phase space}. The state point is then described by a
\emph{phase space vector} $\Gamma \left( \vecr, \vecp \right)$, however, considering that a conservative system where all
$N$ particles move according to Newton's equations of motion, $\Gamma$ is
a function of time, or $\Gamma(t)$. Using this phase space vector, \emph{time averages}
can be obtained for a given observable $A$ by means of the following expression
\begin{equation}
    \left< A \right> = \lim_{t \to \infty} \frac{1}{t} 
    \int_{0}^{t} A \: \Gamma(t') \: dt' \, .
    \label{eq:time-average}
\end{equation}
If instead the complete set of state points, also known as the \emph{ensemble} of state
points, is considered, then the average $\left< A \right>$ can be rewritten in
terms of this ensemble.
This ensemble of state points is distributed in phase space according to a probability 
distribution that is specified by the \emph{thermodynamic ensemble.} Then, if the time 
evolution of $\Gamma(t)$ is such that all states are visited eventually, irrespective of 
its initial conditions, the system satisfies the weak ergodic
theorem~\cite{kittelElementaryStatisticalPhysics2004},
and the time average in \autoref{eq:time-average} can then be rewritten with an
\emph{ensemble average} that reads
\begin{equation}
    \left< A \right> = \sum_{\Gamma} A(\Gamma) \: \rho_{ens} (\Gamma)
    \: ,
    \label{eq:ensemble-average}
\end{equation}
where the sum is for all state point vectors $\Gamma$ and $\rho_{ens} (\Gamma)$
is the probability density function for the ensemble. This probability function is
a weight function for the averaging procedure and should be normalized,
\begin{equation}
    \sum_{\Gamma} \rho_{ens} (\Gamma) = 1 \, .
    \label{eq:normalized}
\end{equation}
With this information, it is now time to introduce the \emph{canonical ensemble}, which
is the main ensemble used throughout this work.

\subsection{Canonical ensemble}
The canonical ensemble is established as a system of $N$ particles in a fixed volume $V$
at temperature $T$ that can exchange energy with a heat bath.~\cite{mcquarrieStatisticalMechanics2000} This ensemble has a
probability density function associated with it,
\begin{equation}
    \rho_{NVT} = \frac{e^{-\beta \mathcal{H}(\Gamma)}}{\sum_{\Gamma} e^{-\beta \mathcal{H}(\Gamma)}}
    \: ,
    \label{eq:canonical-density}
\end{equation}
where $\beta=1/k_{B} T$, and with $k_{B}$ being the Boltzmann's constant. 
In the classical limit of continuous distribution functions, the denominator from 
\autoref{eq:canonical-density} transforms into
\begin{equation}
    Q(N,V,T) = \frac{1}{N! \, h^{3N}} \int d \vecp^{N} \, d \vecr^{N} \,
    e^{- \beta \mathcal{H} \left( \vecr^N, \vecp^N \right)}
    \: ,
    \label{eq:canonical-partition}
\end{equation}
which is known as the \emph{canonical partition function}~\cite{huangStatisticalMechanics1987}.
Here, $h$ is an arbitrary but predetermined constant with the units of energy $\times$
time. As a side note, in the original formulation by Gibbs, the value of $h$ was set
to $h=1$~\cite{gibbsElementaryPrinciplesStatistical2014}, however, since the advent
of quantum mechanics, it is now taken to be Planck's constant~\cite{tolmanPrinciplesStatisticalMechanics1979}
in order to show a correspondence between the classical and quantum formulations.
Indeed, the integrals from \autoref{eq:canonical-partition} can be separated, and the
integral over the momentum coordinates can be carried out analytically giving,
\begin{equation}
    Q(N,V,T) = \frac{1}{N! \, \Lambda^{3N}} \int d \vecr^{N} \,
    e^{- \beta U \left( \vecr^N \right)}
    \: ,
    \label{eq:canonical-partition-position}
\end{equation}
with $\Lambda=\sqrt{h^2 / 2 \pi m k_{B} T}$ the thermal wavelength, also known as the
\emph{de Broglie} wavelength.~\cite{mcquarrieStatisticalMechanics2000} This again shows a correspondence between classical
and quantum formulations. The remaining integral over the positions is called the
\emph{configuration integral},
\begin{equation}
    Z(N,V,T) = \int d \vecr^{N} \, e^{- \beta U \left( \vecr^N \right)} \, .
    \label{eq:configuration-int}
\end{equation}
Finally, we arrive at the canonical ensemble density function in the continuum limit
which is
\begin{equation}
    \rho_{NVT} = \frac{e^{- \beta U(\vecr^{N})}}{Z(N,V,T)} \, .
    \label{eq:canonical-limit}
\end{equation}
Using this probability density function, the ensemble average for an observable $A$ is
computed as
\begin{equation}
    \left< A \right> = \frac{\int e^{- \beta U(\vecr^{N})} \, A(\vecr^{N}) \, d \vecr^{N}}{Z(N,V,T)} \, .
    \label{eq:average-canonical}
\end{equation}

\section{Distribution functions}

Still, the probability density function in \autoref{eq:canonical-limit} provides far more
information from the system than necessary for the calculation of thermodynamical functions
and structure properties. Instead, a focus on a small set of particles $n \ll N$ is 
preferred, in which case a \emph{reduced probability density} is defined as
\begin{equation}
    \rho^{(n)}_{N} (\vecr^{n}) \coloneqq \frac{N!}{(N-n)!} \frac{1}{Z(N,V,T)}
    \int e^{- \beta U(\vecr^{N})} \, d \vecr^{(N-n)} \; ,
    \label{eq:reduced-canonical}
\end{equation}
where $\rho^{(n)}_{N} (\vecr^{n})$ is also known as the equilibrium 
$n$\emph{-particle density}.
The quantity $\rho^{(n)}_{N} (\vecr^{n}) \, d \vecr^{n}$ defines the probability of finding
$n$ particles of the system with coordinates in a volume element $d \vecr^{n}$ from the
phase space, regardless the positions and momenta of the remaining particles.
As a result of this contraction of the probability density function, it is now possible to
provide a complete description of the \emph{structure} of a fluid, while the knowledge
of low-order particle distribution functions is sufficient to calculate thermodynamic
quantities~\cite{mcquarrieStatisticalMechanics2000}.

From the definition of the $n$-particle density in \autoref{eq:reduced-canonical}, the
normalization condition is
\begin{equation}
    \int \rho^{(n)}_{N} (\vecr^{n}) \, d \vecr^{n} = \frac{N!}{(N-n)!} \; ,
    \label{eq:normalization-reduced}
\end{equation}
and in particular, the \emph{single-particle} density is
\begin{equation}
    \int \rho^{(1)}_{N} (\vecr) \, d \vecr = N \; ,
    \label{eq:single-particle-density}
\end{equation}
For a homogeneous fluid, \autoref{eq:single-particle-density} is then simplified
to the following expression
\begin{equation}
    \rho^{(1)}_{N} = N / V \equiv \rho
    \label{eq:homogeneous-density}
\end{equation}
where $\rho$ is the \emph{particle number density}.

The $n$-particle distribution function $g^{(n)}_{N} (\vecr^{(n)})$ is defined in terms
of the particle densities by
\begin{equation}
    g^{(n)}_{N} (\vecr^{(n)}) \coloneqq \frac{\rho^{(n)}_{N} (\vecr_{1}, \dots , \vecr_{n})}
    {\prod_{i=1}^{n} \rho^{(1)}_{N} (\vecr_{i})} \; ,
    \label{eq:gr-def}
\end{equation}
which again for a homogeneous fluid it reduces to
\begin{equation}
    \rho^{n} g^{(n)}_{N} (\vecr^{(n)}) = \rho^{(n)}_{N} (\vecr^{n}) \; .
    \label{eq:homogenous-gr}
\end{equation}
Particle distribution functions measure the extent to which the structure of a fluid 
deviates from complete randomness~\cite{hansenTheorySimpleLiquids2013}.
If the system is also isotropic, which it shall be the consideration henceforth,
the pair distribution function $g^{(2)}_{N} (\vecr_{1}, \vecr_{2})$ is a function only
of the separation $r = r_{12} = \lvert \vecr_{2} - \vecr_{1} \rvert$ between particles
in positions $\vecr_{1}$ and $\vecr_{2}$; here $\lvert \cdot \rvert$ is the Euclidean
distance or norm between two $n$-dimensional vectors.
The function $g^{(2)}_{N} (\vecr_{1}, \vecr_{2})$ is the
\emph{radial distribution function}, and it is referred to as 
\rdf for the rest of this work.
This radial distribution function is of crucial importance in Liquid State Theory for
several reasons. First, the radial distribution function can be obtained experimentally
for liquids using X-ray diffraction, digital video-microscopy, and light scattering
experiments~\cite{mcquarrieStatisticalMechanics2000}.
Second, thermodynamic properties of liquids can be determined using integrals that
contain the radial distribution function. Third, the radial distribution function can
be easily computed in computer simulations~\cite{allenComputerSimulationLiquids2017}, 
which is the standard way to obtain these quantities. Finally, the radial distribution
function can be obtained analytically using integral equations such as the Ornstein-Zernike
equation, which shall be discussed in a later section.

\section{Thermodynamic properties} \label{sec:thermodynamics}
As mentioned in the previous section, thermodynamic properties can be defined in terms of
distribution functions, and in particular, in terms of the radial distribution
function~\cite{hansenTheorySimpleLiquids2013}.
It is now time to discuss such expressions, and the relation between thermodynamical
quantities and the radial distribution function.

The \emph{total energy} $E$ of a three-dimensional system of $N$ particles can be defined 
in terms of the radial distribution function as
\begin{equation}
    E = E_{ideal} + E_{excess} = \frac{3 N k_{B} T}{2} +
    \frac{N}{2} \int_{0}^{\infty} \rho \, u(r) \, g(r) \, 4 \pi r^2 \, dr
    \, ,
    \label{eq:total-energy-rdf}
\end{equation}
where $E_{ideal}$ is the ideal gas contribution, whose result comes directly from
the energy equipartition theorem~\cite{mcquarrieStatisticalMechanics2000};
$E_{excess}$ is an interaction or excess contribution that can be understood
as the interaction energy between a central particle for all the $N$ particles,
and all the neighbors located in a spherical shell of radius $r$ and thickness
$dr$. The total number of neighbors is given by $4 \pi r^2 \rho g(r) \, dr$.
The integration from $0$ to $\infty$ gives all the interaction energy, and
the factor of $1/2$ accounts for double counting of particle pairs.
The function $u(r)$ is the \emph{pairwise interaction potential} which comes
from the fact that the total potential energy $U(\vecr^{N})$ can be expressed
in terms of the individual particle interactions as follows
\begin{equation}
    U(\vecr^{N}) = \sum_{i} u_{1} (\vecr_{i}^{N}) +
    \sum_{i} \sum_{j>i} u_{2} (\vecr_{i}^{N}, \vecr_{j}^{N}) +
    \dots
    \label{eq:pairwise-energy}
\end{equation}
These kind of interactions are the most commonly researched and studied, given that most
systems can be modeled after such interaction potentials, such as the
hard-sphere, Lennard-Jones, Yukawa~\cite{huangStatisticalMechanics1987}
and several others. The hard-sphere interaction potential will be discussed
in a later section.
In the current work, the two-particle interaction potential
$u_{2} (\vecr_{i}^{N}, \vecr_{j}^{N})$ is referred to simply as $u(r)$.
It is important to note that this is not the norm, and interaction potentials
that are not pairwise additive are also studied. These many-body potentials are
notoriously hard to study, but recent advances in Machine Learning have made it
possible to do so~\cite{boattiniModelingManybodyInteractions2020}.

The \emph{pressure equation} is a relation between the thermodynamic pressure
$P$ and the radial distribution function, defined for a three-dimensional system as:
\begin{equation}
    P=P_{ideal}+P_{excess}= \frac{\rho}{\beta} - \frac{2 \pi \beta \rho}{3}
    \int_{0}^{\infty} u'(r) \, g(r) \, r^3 \, dr \, ,
    \label{eq:pressure-equation}
\end{equation}
where $P_{ideal}$ is the kinetic pressure of an ideal gas, and $P_{excess}$
is the excess pressure that can be derived using classical mechanics through the
virial theorem~\cite{goldsteinClassicalMechanics2002}.

Another important quantity that must be mentioned is the
\emph{isothermal compressibility}. Its thermodynamic definition
is~\cite{mcquarrieStatisticalMechanics2000}
\begin{equation}
    \chi_{T} = - \, \frac{1}{V} { \left( \frac{\partial V}{\partial P} \right) }_{T} =
    \frac{1}{\rho} { \left( \frac{\partial \rho}{\partial P} \right) }_{T}
    \; ,
    \label{eq:isothermal-chi}
\end{equation}
and it is a physical measure of the relative change in volume due to a change
in pressure or stress.
It can also be computed using the radial distribution function using the following
expression~\cite{hansenTheorySimpleLiquids2013}
\begin{equation}
    \frac{\chi_{T} \, \rho}{\beta} = 1 + \rho \int d \vecr \, \left[ g(r) - 1 \right]
    \; .
    \label{eq:chi-rdf}
\end{equation}
We call this the \emph{compressibility equation}, and it is a standard way of computing
the isothermal compressibility.

\section{The hard-sphere model} \label{sec:hard-sphere}
In order to describe the behavior of materials, pairwise interaction potentials are chosen 
according to specific physical properties. These interaction potentials provide a 
particular functional form for the potential energy $U$ of the system.
With this information, all the theory described so far can be used to understand the
physical properties of the system, and compute thermodynamic quantities.

Out of all the possible interaction potentials, there is one that stands out for its
simplicity and surprising ability to generalize to complex systems. This model is
called the \emph{hard sphere model}, and shall be the topic of discussion for this
section.
The \emph{hard sphere model} is a simple pairwise interaction potential, defined as
\begin{equation}
    u_{hs} = 
    \begin{cases}
        \infty \, , &r < \sigma \\
        0 \, , &r \geq \sigma \; ,
    \end{cases}
    \label{eq:hard-sphere}
\end{equation}
where $\sigma$ is the diameter of the particles in the system, and $r$ is the distance
between the centers of two particles, as previously stated.

This interaction is a particular model; it is an approximation of the intrinsic
behavior of particles.
It describes the excluded volume interaction of particles, similar to what
happens with billiards balls in three dimensions.
These particles seem to hit each other only to be separated at the contact point, which
is the value of $\sigma$. Hence, this model is by definition a \emph{repulsive model}.
Apart from being a reference model system for the liquid
state~\cite{hansenTheorySimpleLiquids2013}, hard spheres are extremely useful in
colloid science, as mentioned in the introduction section. This is due to the fact
that it has been demonstrated that the hard sphere interaction gives rise to a
fluid-crystal phase transition around a volume fraction of 50\% of hard 
spheres~\cite{hooverMeltingTransitionCommunal1968a,gastSimpleOrderingComplex1998,roblesNoteEquationState2014a}.
This phase transition was the subject of an extensive discussion in the early
1950's and was for the first time discovered in computer
simulations~\cite{alderPhaseTransitionHard1957a}.
The relationship between this simple model and colloid science was discovered
experimentally by Pusey and van Megen~\cite{puseyPhaseBehaviourConcentrated1986},
in dense colloidal suspensions of sterically stabilized particles in a solvent.

In this work, a similar potential is used, though defined differently. Looking closely
at \autoref{eq:hard-sphere}, it is straightforward to see that the potential is a
discontinuous function. In mathematical proofs and statistical mechanics frameworks,
this poses no problem. It might be hard to manipulate, mathematically speaking,
but it is possible to do so. Yet, the problem lies with the use of computer simulations.
The technical issues arise mostly in Molecular and Brownian Dynamics computer
simulations~\cite{allenComputerSimulationLiquids2017}, so a different formulation
must be employed. Computer simulations of liquids are presented in the next section.
In 2018, Báez \emph{et al}~\cite{baezUsingSecondVirial2018} used the Extended Law
of Corresponding States~\cite{valadez-perezReversibleAggregationColloidal2018}
to map the second virial coefficient of the hard sphere model to a continuous
model. By doing so, most of the dynamic and thermodynamic properties of the hard sphere
model are "passed along" to the continuous function, which in turn is suited for all
kinds of computer simulation algorithms with a finite time step. This new continuous function is used
throughout this work, and it is defined as follows
\begin{equation}
    u_{CP} = 
    \begin{cases}
        A \, \epsilon \left[ {\left(\frac{\sigma}{r}\right)}^{\lambda} -
        {\left(\frac{\sigma}{r}\right)}^{\lambda - 1} \right] + \epsilon \, , 
        &r < \sigma \, B \, , \\
        \hspace{1.5cm} 0 \, , &r \geq \sigma \, B \, ,
    \end{cases}
    \label{eq:cont-hs}
\end{equation}
with
\begin{equation}
    A = \lambda {\left(\frac{\lambda}{\lambda -1}\right)}^{\lambda - 1} \, ,
    \quad
    B = \left(\frac{\lambda}{\lambda -1}\right) \, .
    \label{eq:ab-params}
\end{equation}
The value of $\lambda$ is fixed to $\lambda=50$, following the findings in the work of
Báez \emph{et al}~\cite{baezUsingSecondVirial2018}; and the value of $\epsilon$ is
not fixed explicitly, instead the reduced potential is employed, 
i.e., $u^{*}=u_{CP} / \epsilon$. It is also important to define a particular quantity,
the reduced temperature \(T^{*}=k_{B} T \, / \epsilon\), which used in the original work
by Báez \emph{et al} to ensure that the second virial coefficient is accurately reproduced.
This quantity is defined as,
\begin{equation}
    T^{*} = a + \frac{b}{\lambda^c}
    \label{eq:t-star}
\end{equation}
and for a three-dimensional system the parameters used are \(a=1.4543, b=1.199, c=1.0545 .\)

\section{Computer simulations of liquids}

Liquid State Theory is usually supplemented with results from computer simulations
because these methods come from first-principles Physics and yield accurate estimations
of the behavior of the physical systems. Also, given the availability of computational
resources nowadays, these methods have become a standard tool for research.
There has been an extensive amount of research in these computer simulation methods,
following the pioneering work of Alder and Wainwright~\cite{alderPhaseTransitionHard1957a}.
There are two main computer simulations methods used in Liquid State Theory and
Soft Matter research: \emph{Molecular Dynamics} and \emph{Monte Carlo methods}~\cite{allenComputerSimulationLiquids2017,frenkelUnderstandingMolecularSimulation2001}.
These methods serve different purposes, and because of that, a brief description of each
one will be outlined here. However, the main focus of this work is the use of
the Monte Carlo simulation technique. \emph{Brownian dynamics} methods are
equally as important to Soft Matter research, but they are beyond the scope of this work.
The book by Jan Dhont~\cite{dhontIntroductionDynamicsColloids1996} provides a rigorous
explanation of the Brownian dynamics framework.

\subsection{Molecular dynamics}
Molecular Dynamics simulations use the basic laws of Physics to understand the behavior
of liquids. In particular, Newton's laws are used to evolve a system of particles through
time and space. For a system of $N$ particles, each with mass $m_i$, $i=1,2,\dots,N$,
the equations of motion are
\begin{equation}
    \bm{F}_i (t) = m_i \frac{d^2 \vecr_i}{dt^2} \; ,
    \label{eq:newton}
\end{equation}
with
\begin{equation}
    \bm{F}_i (t) = - \nabla_{\vecr_i} U(\lvert \vecr_i - \vecr_j \rvert)
    , \quad U = \sum_{i} \sum_{j} u(\lvert \vecr_i - \vecr_j \rvert)
    \label{eq:newton-potential}
\end{equation}
the force on particle $i$ with position $r_i$ exerted by particle $j$ with position
$\vecr_j$, and $U$ the potential energy, which in the context of liquids this is the
interaction potential between the particles. In most cases, this interaction potential
can be modeled as the pairwise interaction between two particles.

To integrate these equations of motion, different integration schemes have been developed.
At first, Alder and Wainwright~\cite{alderPhaseTransitionHard1957a} used the simple
Euler method to integrate these equations. It was later understood that such scheme
does not yield \emph{stable} results, i.e., the integration scheme gives different results
if used with different initial conditions each time.
Integration methods must conserve important physical quantities, such as energy,
momentum and phase space properties~\cite{razafindralandyReviewGeometricIntegrators2018}.
The Verlet method was later introduced in 1967 by Loup Verlet~\cite{verletComputerExperimentsClassical1967a},
in order to alleviate the drawbacks of the Euler method. These methods are crucial to obtain
correct results from molecular dynamics computer simulations.

The main appeal of molecular dynamics computer simulations is the fact that
\emph{dynamical quantities} can be computed effortlessly, given the temporal nature 
of the formulation. Physical observables such
as the \emph{mean square displacement} and the \emph{dynamic structure factor}~\cite{dhontIntroductionDynamicsColloids1996,hansenTheorySimpleLiquids2013}
are of great importance to the understanding of transport properties in liquids and
more complex systems in Soft Matter.

\subsection{Monte Carlo methods}
Monte Carlo methods make use of randomness to explore a possible solution to a
given problem. The basic idea is to exploit random sampling to obtain numerical
results. Monte Carlo methods have successfully been applied to
three main problems: optimization, numerical integration, and probability density
function estimation~\cite{kroeseWhyMonteCarlo2014}. It might seem odd that Statistical
Physics is not mentioned here, but the reason for that is that it actually belongs to
the wider range of probability density function estimation techniques. It is the
purpose of this section to explore Monte Carlo methods and their application to
the simulation of liquids, mainly because this is the method used in this
work, and because in this work there is no dynamical physical quantity involved,
only statical ones, i.e., the radial distribution function.

The basic idea of Monte Carlo methods in the computer simulation of liquids is to
generate a sequence of configurations for a given system. The idea is to arrive at a
probability distribution function close to the ensemble probability distribution
function chosen to model the fluid. For instance, if a system
is under the $NVT$ ensemble, the sequence of Monte Carlo steps are expected to converge
to the distribution function given by \autoref{eq:canonical-limit}. Having arrived at this
estimation, observables can the be computed, as in \autoref{eq:average-canonical}.
There is also other possibilities of ensemble distribution functions that can be
used with Monte Carlo computer simulations, such as the isobaric-isothermal ensemble
and the Gibbs ensemble~\cite{frenkelUnderstandingMolecularSimulation2001}.

In general, Monte Carlo methods are random in nature. However, liquids cannot be
subjected to complete randomness due to their intermolecular interactions. In the case
of the hard sphere fluid, the denominator of \autoref{eq:canonical-limit} would be zero
for at least 50\% of the configurations sampled, which would incur in numerical instabilities and observables
would not get measured properly. So, a \emph{bias} must be introduced in order to
\emph{guide} the sequence of configurations to a path which could avoid unnecessary
configurations. This bias is called \emph{importance sampling}, and the most common rule,
as well as the one used here, is the Metropolis rule~\cite{landauGuideMonteCarlo2021}.
Nevertheless, when a bias is introduced, there is one important rule that must
\emph{always} be obeyed: the condition of \emph{detailed balance}. Detailed balance
is a strong condition stating that if there is a transition from a state $o$ to a
state $n$, this should be balanced with the number of accepted trial moves that go
from state $n$ to state $o$. This can also be summarized in mathematical form,
\begin{equation}
    p(o) \, T(o \Longrightarrow n) = p(n) \, T(n \Longrightarrow o) \, ,
    \label{eq:detailed-balance}
\end{equation}
where $p(\alpha)$ is the probability to be at state $\alpha$, and 
$T(\alpha \Longrightarrow \beta)$ denotes the probability to go from a state 
$\alpha$ to a state $\beta$. It might look like a simple rule, but this condition
is what guarantees the ergodicity of the computer
simulation~\cite{landauGuideMonteCarlo2021}, i.e., that the phase space is explored
correctly and with physical significance. If this condition is not met, the results
do not represent the physical properties of the system, and the observables cannot
be computed properly.

In the $NVT$ ensemble, the Metropolis rule that satisfies detailed balance is
\begin{equation}
    \frac{p(n)}{p(o)} = \frac{A(o \Longrightarrow n)}{A(n \Longrightarrow o)} =
    \exp{ \left\{ -\beta \left[U(n) - U(o)\right] \right\} }
    \; ,
    \label{eq:nvt-balance}
\end{equation}
with
\begin{equation}
    A(o \Longrightarrow n) = \min{ \left(1, \exp{ \left\{ -\beta \left[U(n) - U(o)\right] \right\} } \right)} 
    \; ,
    \label{eq:nvt-acceptance}
\end{equation}
the proof that this condition satisfies detailed balance is omitted here, but it can
be found in the book by Frenkel~\cite{frenkelUnderstandingMolecularSimulation2001}.
With this acceptance rule, it is guaranteed that the configurations that have the larger
Boltzmann factor, i.e. $\exp{ \left\{ -\beta \left[U(n) - U(o)\right] \right\} }$, 
will be visited more frequently rather than the configurations with a smaller
Boltzmann factor, which will be avoided.
In practice, the core of the Monte Carlo method for the computer simulation of liquids, 
which corresponds to the Metropolis rule, is implemented as follows:
\begin{itemize}
    \item Select a particle $i$ at random.
    \item Calculate the energy cost to move particle $i$.
    \item Move particle $i$ randomly.
    \item Calculate the new energy cost for having displaced a particle.
    \item Accept or reject the move according to \autoref{eq:nvt-balance} and \autoref{eq:nvt-acceptance}.
    \item Update the total energy configuration.
\end{itemize}
Monte Carlo methods are efficient when static quantities, such as the radial distribution
function, are computed. This is because time is not part of the computer simulation method,
unlike the previously discussed Molecular Dynamics method.
Monte Carlo methods can efficiently explore a large portion of the phase
space due to their stochastic nature. Although the phase space might be large, it is often
the case that different explorations using different seed for the random number generator
are enough to obtain good estimates for an observable. In this work, Monte Carlo computer 
simulations were used for all the observables computed throughout.

\subsection{On the computation of the radial distribution function}
The procedure to obtain the radial distribution function from computer simulations is as
follows. For each Monte Carlo step a particle $i$ is chosen randomly.
Then, all the particles that are a distance $\delta r$ from the particle $i$ are counted
as a neighbor particle. After that, all the particles that are now at a distance
$2 \delta r$ from the particle $i$ are counted. This is done after all the particles
have been counted. If the interaction potential between the particles is truncated
to a cutoff radius $r_c$, then this is the last value of the distance used to count
neighboring particles. When every particle has been counted, a histogram is built,
which corresponds to a discrete distribution function of the neighboring particles. 
Now, this distribution function is not normalized, and the procedure to do so is as 
follows. A normalization constant is obtained by dividing the total number of particles at 
position $r$ with the product of the total number of particles in the simulation, 
multiplied by the total number of configurations visited throughout the simulation, 
multiplied by the total volume of the spherical shell that lies between $r$ and $r + \delta 
r$. In mathematical form, this can be summarized as follows
\begin{equation}
    g(r)=\frac{1}{\rho} \left\langle \frac{1}{N} \sum_{i=1}^{N} \sum_{i \neq j}^{N} 
    \delta (r - \vecr_{ij}) \right\rangle
    \, , \quad
    \vecr_{ij}=\vecr_{i} - \vecr_{j} \; ,
    \label{eq:rdf-simulation}
\end{equation}
where \(\left\langle \dots \right\rangle\) denotes an ensemble average.

\section{The Ornstein-Zernike Integral Equation} \label{sec:ornstein-zernike}
Up until this point, the presentation on Liquid State Theory has been driven by the
possibility of calculating a given observable for a liquid. The method to compute such
observables is to obtain a main quantity, the radial distribution function, and
make use of any of the thermodynamic equations that relate \rdf with the needed
thermodynamic observable. There has been a discussion of how the \rdf can be obtained
through computer simulations. And yet, there has not been a discussion of how the \rdf
can be obtained analytically. The Ornstein-Zernike formalism provides a way to obtain
such analytical results, and it is the focus of the present section, as well as the main 
focus of this thesis. Despite its importance, the formal derivation of the equation is too 
rigorous to be presented here, so most of the details will be omitted here but left to 
specific references on the subject~\cite{hansenTheorySimpleLiquids2013}. Nonetheless, the 
main results and physical interpretation shall be outlined here.

To understand the Ornstein-Zernike equation, a physical interpretation is defined as
follows. If there is a system of $N$ interacting particles with an interaction potential
defined by $u(r)$, then it so happens that each of them are related to each other.
It might seem like an obvious statement, but this relation has in fact two contributions.
There exists a \emph{direct} relation between particles and
their neighbors, and an \emph{indirect} relation between particles and the rest that
may be far away from them. This interaction begs the question,
\emph{is there a way to relate the indirect and direct correlations between particles?}
The answer is given recursively through the following definition. Let $h(r)$ be the
\emph{total correlation function} that "measures" the total correlations between
particles; and let $c(r)$ be the \emph{direct correlation function} that "measures"
the direct correlations between particles, then both $c(r)$ and $h(r)$ are related
with each other through the \emph{Ornstein-Zernike equation} as follows
\begin{equation}
    \begin{aligned}
        h(r) &= c(r) + \rho \int_{V} c(\vecr') \, c(\lvert \vecr - \vecr' \rvert) \, d \vecr' \\
        &+ \rho^2 \int_{V} c(\vecr'') c(\lvert \vecr - \vecr' \rvert) c(\lvert \vecr' - \vecr'' \rvert) \, d \vecr' \, d \vecr'' \\
        &+ \rho^3 \int_{V} c(\vecr''') c(\lvert \vecr - \vecr' \rvert) c(\lvert \vecr' - \vecr'' \rvert) c(\lvert \vecr'' - \vecr''' \rvert) \, d \vecr' \, d \vecr''
        \, d \vecr''' \\
        &+ \dots \; ,
    \end{aligned}
    \label{eq:hr-oz}
\end{equation}
with \(\rho \coloneqq N / V .\)
\autoref{eq:hr-oz} can be rewritten recursively by noting the following factorization
\begin{equation}
    h(r) = c(r) + \rho \int_{V} c(\vecr') \left[
    \begin{aligned}
        & \hspace{1.6cm} c(\lvert \vecr - \vecr' \rvert) \\
        &+ \rho \int_{V} c(\lvert \vecr - \vecr' \rvert) c(\lvert \vecr' - \vecr'' \rvert) \, d \vecr' \, d \vecr'' \\
        &+ \rho^2 \int_{V} c(\lvert \vecr - \vecr' \rvert) c(\lvert \vecr' - \vecr'' \rvert) c(\lvert \vecr'' - \vecr''' \rvert) \, d \vecr' \, d \vecr''
        \, d \vecr''' \\
        &+ \hspace{1.6cm} \dots
    \end{aligned}
    \right] \, d \vecr'
    \label{eq:oz-factorization}
\end{equation}
and thus arrive at the most common form of the Ornstein-Zernike equation, which for
isotropic and uniform liquids reads
\begin{equation}
    h(r) = c(r) + \rho \int_{V} c(r') \, h(\lvert \vecr - \vecr' \rvert) \, d \vecr'
    \, .
    \label{eq:ornstein-zernike}
\end{equation}

The Ornstein-Zernike integral equation provides an analytical way to obtain the \rdf 
provided an interaction potential. Yet, in \autoref{eq:ornstein-zernike} there is no 
mention or sign of any of such quantities. So how is it possible to obtain an estimate
of \rdf? It turns out that the definition of $h(r)$ is key to answering this question,
for the actual mathematical definition is $h(r) \coloneqq g(r) - 1$~\cite{hansenTheorySimpleLiquids2013}.
As for the interaction potential, the information lies within $c(r)$ and $g(r)$.
In the \emph{low density limit}, the radial distribution function has the following
asymptotic behavior
\begin{equation}
    g(r) \to e^{- \beta u(r)} \, , \quad \rho \to 0 \; .
    \label{eq:gr-asymp}
\end{equation}
Also, when $r \to \infty$ then $g(r) \to 1$. 
Now, again in the low density limit, the asymptotic behavior
for $c(r)$ that follows from \autoref{eq:ornstein-zernike} is that $c(r) \to h(r)$.
This also means that, with \autoref{eq:gr-asymp}, $c(r)$ has the following
limiting definition
\begin{equation}
    c(r) \to h(r) = g(r) - 1 = e^{- \beta u(r)} - 1 \, , \quad \rho \to 0
    \; ,
    \label{eq:cr-aysmp}
\end{equation}
where the function $f \coloneqq e^{- \beta u(r)} - 1$ is called the $f$-Mayer
function, and it is an important quantity for theoretical analysis in Liquid State
Theory. Thus, it follows from \autoref{eq:cr-aysmp} that
\begin{equation}
    c(r) = - \beta u(r) \, , \quad r \to \infty 
    \, .
    \label{eq:cr-r-asymp}
\end{equation}
An additional important fact is that the compressibility equation, \autoref{eq:chi-rdf},
can be defined in terms of the $c(r)$ functions as follows
\begin{equation}
    \frac{\beta}{\rho \chi_{T}} = 1 - \rho \int c(r) \, d \vecr
    \, ,
    \label{eq:compressibility-cr}
\end{equation}
this definition follows from the relation between $c(r)$ and the
\emph{structure factor}~\cite{hansenTheorySimpleLiquids2013}, but the details are omitted 
here.

\subsection{Closure relations}
Now that it has been stated the relation between the three quantities, 
$c(r), \, g(r), \, \text{and} \, u(r)$, the solution to the Ornstein-Zernike equation
must be discussed. Until now, the discussion of the functions $c(r)$ and $h(r)$ has been
focused on their low density limit. Still, liquids are \emph{dense} systems and most of
the time the purpose of studying them is to understand their properties for a wide range
of density values, low and high density regimes. 
Furthermore, it is expected to study the phenomena that
arises when the density varies in ways that \emph{phase transitions} are induced.
To provide a solution to this, \autoref{eq:ornstein-zernike} must be solved for all possible
values of the density. However, if looked at closely, \autoref{eq:ornstein-zernike} has
two unknown functions, $c(r)$ and $h(r)$, thus cannot be solved as it is. For this reason,
approximations must be introduced in the form of \emph{closure relations}.

Closure relations come from diagrammatic expansions and density function theory~\cite{hansenTheorySimpleLiquids2013},
and are not simple to summarize in this section. The most common closure relations will be
stated, along those that are used in this work. One of the most important closure relations
is the \emph{Percus-Yevick} approximation.
The most important fact from this closure is that it provides an analytical solution to the 
hard sphere model. The closure relation reads~\cite{percusAnalysisClassicalStatistical1958},
\begin{equation}
    c(r) = g(r) \, \left[1 - e^{\beta u(r)}\right]
    \; ,
    \label{eq:py-cr}
\end{equation}
and when used along the hard sphere potential found in \autoref{eq:hard-sphere}, it provides
an exact solution for \rdf.
Another common closure relation is the \emph{Hypernetted Chain} approximation,
which reads~\cite{hansenTheorySimpleLiquids2013},
\begin{equation}
    c(r) = h(r) - \beta u(r) - \ln{g(r)} \; .
    \label{eq:hnc-cr}
\end{equation}
The appeal of this closure relation is that it works quite well for interaction potentials 
with long-range repulsive behavior, such as the Yukawa or Lennard-Jones potentials~\cite{hansenTheorySimpleLiquids2013}.
Despite this, both closure relations have deficiencies when attempting to predict a
phase transition. Moreover, these closure relations suffer from
\emph{thermodynamic inconsistency}. This is a phenomenon that happens from the approximation
nature of closure relations. Both closure relations give different results for
a given thermodynamic quantity if computed from different routes. For instance, if the 
thermodynamic pressure $P$ is computed using the Ornstein-Zernike and the 
Percus-Yevick closure relation, then, if \autoref{eq:pressure-equation} is used, a 
particular result will be obtained.
But if \autoref{eq:chi-rdf} is used, and then \autoref{eq:isothermal-chi} to obtain the 
thermodynamic pressure, a different result will be obtained with respect to the one
computed before. Furthermore, if the energy equation is used
\textemdash \autoref{eq:total-energy-rdf} \textemdash, 
and then using thermodynamic relations to obtain the pressure, a new result will be 
obtained, different from the previous two results. This constitutes a serious drawback of
the Ornstein-Zernike formalism.

To alleviate this main drawback, Rogers and Young~\cite{rogersNewThermodynamicallyConsistent1984b}
proposed a new version of the closure relation, which in fact combines two closure 
relations, the Hypernetted Chain and the Percus-Yevick. With great success, this new
and improved closure relation can reproduce the thermodynamics of the hard sphere model
and related short-range interaction potentials. Another attempt at providing thermodynamic
consistent closure relations were provided by Zerah and Hansen~\cite{zerahSelfConsistentIntegral1986},
which combined different closure relations into one, in the same fashion as the Rogers-Young
closure relation. Both of these approaches work as follows. Two main routes to compute a 
given thermodynamic quantity are chosen, and are forced to yield the same results through 
the use of a fixed parameter, called the \emph{mixing parameter}. This value is obtained as 
an optimization problem by enforcing equality of the thermodynamic routes.

Another important closure relation is the one provided by Verlet and then modified by
Kinoshita~\cite{kinoshitaInteractionSurfacesSolvophobicity2003}. It turns out that
when hard spheres are studied, there are better closure relations than the
Percus-Yevick approximations for larger density values.
The Kinoshita modification reads,
\begin{equation}
    B(r) = - 0.5 \frac{\gamma^2 (r)}{1 + 0.8 \lvert \gamma (r) \rvert} \; ,
    \label{eq:kinoshita}
\end{equation}
where $B(r)$ is the bridge function (see next section), and $\gamma(r)=h(r)-c(r)$,
and \(\lvert \cdot \rvert\) stands for the absolute value.
This closure relation is not thermodynamically consistent, but in this work this
consistency will be explored in detail later in this work.

\subsection{The role of the bridge function}
All the possible solutions to the Ornstein-Zernike equation discussed so far 
have been just approximations.
But the Ornstein-Zernike formalism is an exact integral equation, thus there should be an
exact closure relation. Indeed, an exact relation exists which comes from diagrammatic
expansions~\cite{hansenTheorySimpleLiquids2013} and reads,
\begin{equation}
    \ln{\left[h(r) - 1\right]} = \beta u(r) + B(r) + \gamma(r) \; .
    \label{eq:exact-closure}
\end{equation}
Still, it introduces a new function and quantity, the so-called \emph{bridge function}.
What role does the bridge function play in the exact solution of the Ornstein-Zernike
equation? It turns out that if chosen correctly, the bridge function can make
\autoref{eq:exact-closure} reduce to all previous closure relations. For instance,
if $B(r)=0$, the Hypernetted Chain closure relation is recovered. If now the bridge
function takes the value of $B(r)=\ln{ \{\gamma(r) + 1\} }-\gamma(r)$, then the 
Percus-Yevick approximation is obtained. Given the true nature of the bridge function,
physicists were drawn to understand the bridge function and proposed approximations for
the bridge function, instead of a simple closure relation for the Ornstein-Zernike
equation itself. Nevertheless, by approximating the bridge function, the same problems
arise, i.e., thermodynamic consistency and deficient results for every possible
interaction potential. So the most common way to solve the Ornstein-Zernike equation is through
experience: knowing the properties of $u(r)$, a given bridge function is chosen,
then with this approximation the Ornstein-Zernike equation is solved, and a solution is
obtained. Most of the time, the solution is extremely precise. But if there is not
much understanding of the interaction potential, then several bridge function approximations
must be tested. However, most of the important interaction potentials have been
extensively studied, and it is already known how these behave and their properties.
But it is still cumbersome to attempt to use all possible bridge function approximations.
It seems like a task that could be \emph{discovered} or \emph{automated} in a way, and
fortunately this is what Machine Learning excels at. %% Liquid State Theory
\chapter{Computational Intelligence and Machine Learning}
\label{Cap3}

In this chapter, the fundamentals of Computational Intelligence and Machine Learning 
are developed. Particularly, the focus of the chapter is to present the main tools 
used in this thesis, namely \emph{neural networks} and \emph{evolutionary algorithms}. To 
reach a general understanding of these tools, a brief description of learning mechanisms
and numerical optimization is carried out.

\section{Computational Intelligence}
The first thing to address is the meaning and scope of \emph{Computational 
Intelligence} (CI). With recent advancements in fundamental research in this area
and its sub-fields, there seems to be a blurry definition of what exactly is CI, and there 
is no concrete one until now. For this reason, in this work the definition of CI is an 
umbrella term for several other applications. However, these applications are related to 
each other for the same reason that CI exists: to provide a computational solution to a 
problem using as inspiration the paradigms of nature-inspired intelligence. For instance, 
following the handbook by Kacprzyk and 
Pedrycz~\cite{kacprzykSpringerHandbookComputational2015},
the definition of CI is to be a collection of nature-inspired computational methods that
provide solutions to problems where \emph{hard computing} is inefficient or it not even
suited to provide a solution to a given problem. Here, there is an important distinction
that should be carried throughout the remaining of the work: that there are problems for
which traditional tools are insufficient for the traditional problems. In this work, this
is the philosophy used to provide solutions: that the traditional methods might seem hard 
and unfitting to provide solutions to the problems presented, and therefore new ways of 
approaching these solutions should be used.

In general, CI methods do not provide exact and accurate results, but this is expected.
It does not mean that CI is providing the ultimate, best solution to a problem. Rather, it 
is providing an approximate solution that can be used later with more robust algorithms and
methods. CI is not meant to be used as the sole method to solve a problem, but instead to
help find \emph{some} solution to a difficult problem. That is why CI is considered an 
umbrella term, because it comprises diverse fields that can be used to find such an 
approximate solution. Indeed, in this work, two main field of CI are used:
\emph{neural networks}, which is a part of ML, a sub-field of CI; and
\emph{evolutionary algorithms}, which are stochastic optimization methods inspired by the 
evolution mechanisms found in nature.

\section{Machine Learning}
In modern times, data is constantly being created and used to model the reality around us.
However, traditional methods (hard computing) have not been enough to handle the large
data sets created. For this reason, new ways of dealing with this information are needed. 
Most importantly, the need for automated discovery and \emph{pattern recognition} within the
data. Following Murphy~\cite{murphyMachineLearningProbabilistic2012}, ML 
is defined as the set of techniques that can automatically detect patterns in data, and use 
these patterns to create new predictions, or to perform other kinds of decision making.
In other words, \emph{the data creates the ML algorithm}, not the other way around.
It is with data that ML methods work, however, these methods are essentially
\emph{function approximation} methods, which shall be discussed in a later section.
For now, a brief overview of the different kinds of ML tasks and problems will be presented,
focusing primarily on \emph{supervised learning}. Nonetheless, there exist other types
of learning, such as \emph{unsupervised learning}~\cite{goodfellowDeepLearning2016,hastieElementsStatisticalLearning2009} and \emph{reinforcement learning}~\cite{suttonReinforcementLearningSecond2018,kaelblingReinforcementLearningSurvey1996},
which are out of scope of this work, but remain an important part of modern ML theory.

\section{Supervised Learning}
The most common kind of ML methods is that of \emph{predictive} learning. For a set 
\(\mathcal{D}={ \{(\bm{x}_{i}, y_i)\} }_{i=1}^{N}$ with inputs $\bm{x}\)
and outputs $y$, the goal of \emph{supervised learning} is to learn a map between
inputs and outputs. Here, $\mathcal{D}$ is the so-called \emph{training set} and $N$
is the number of \emph{training samples}.

In most common cases, the inputs $\bm{x}$ are $D$-dimensional vectors that
represent information about something. For instance, the height and weight of a
person, or the evolution of the stock market throughout the years. 
The components of these vectors are referred to as \emph{features} or \emph{attributes}.
The general form of $\bm{x}$ is not defined, it can be anything from an image, a time 
series, sentences from a text, graphs, molecules, and so on.

In a similar fashion, the \emph{response} variables $y$ can be, in general, anything. 
However, there is a clear distinction given the forms that these variables can take. For 
instance, if the variable $y$ has \emph{categorical} values, the supervised learning task 
is considered a \emph{classification} problem. Categorical values come from a finite set of 
possible values, \(y_{i} \in \{1, \dots, C\}\), and might represent any type of discrete or 
nominal value. For example, it might represent the colors in a clothing line, the gender 
between people, and so on. On the other hand, if the values of $y$ are real-valued, such 
that \(y_{i} \in \mathbb{R}\), then the learning task is dubbed a \emph{regression} problem.

\subsection{Classification}
In this section, the problem of classification is looked at with more detail. Although 
classification is not used at all in this thesis, it is helpful to discuss it as it makes
it easier to understand the importance of supervised learning. It also creates a basis on 
which the problem of regression can then be generalized from.

In the problem of classification, the computer algorithm is given a data set \(\mathcal
{D}={ \{(\bm{x}_{i}, y_i)\} }_{i=1}^{N}\), and is asked to specify to which category 
does the input belong to. Here, $\bm{x}_i$ is an $n$-dimensional feature vector. As 
mentioned before, the idea is for the algorithm to learn a map, or more precisely, a \emph
{function} such that $f \colon \mathbb{R}^n \mapsto \{1, \dots, k\}$, with $k$ the total 
number of categories, or \emph{classes}, to which the input can be assigned to. More 
specifically, when $y=f(\bm{x})$, the model assigns an input described by the 
$n$-dimensional feature vector $\bm{x}$ to a particular categorical value of $y$.

One of the most common uses of classification is object recognition. The goal of the object 
recognition problem is to decode a particular image with a specific object in it, and label 
it accordingly. An extremely popular data set for this kind of task is the MNIST 
handwritten digits data set~\cite{lecunGradientbasedLearningApplied1998a}, and its most 
common variation, the Fashion-MNIST~\cite{xiaoFashionMNISTNovelImage2017a}. 
These data sets are 
comprised of several images and categories, and the goal is to \emph{classify} each of the 
several categories. In the case of the handwritten digits, the goal is to specify which 
digit is represented in the image; in the case of the Fashion-MNIST data set, the goal is 
to specify the type of clothing. These data sets have become the standard benchmarks 
in the ML community for a long time. They are used primarily to test whether a ML 
algorithm is working properly, and if it is \emph{accurate} enough. The word 
\emph{accuracy} means something quite specific in ML theory, and will be discussed in a 
later section.

\subsection{Regression}
In \emph{regression}, the goal is for the computer algorithm to learn a map, or 
equivalently a function $f \colon \mathbb{R}^n \mapsto \mathbb{R}$, that predicts a 
numerical or real-valued output. The resemblance to the classification task is quite 
obvious: categories or classes can now be represented as a continuous value within the 
reals and there is no restriction for the number of classes. In a sense, this is the most 
general form of classification problem. So then, why make a distinction between the two 
problems? Most of the time, regression tasks are created to \emph{predict}, \emph{forecast}
, or even \emph{generate} outputs for a given data set $\mathcal{D}$.

One of the major applications of regression is \emph{time series forecast}. This can 
be found mostly in economical and financial contexts~\cite{bontempiMachineLearningStrategies2013,sezerFinancialTimeSeries2020}, but there are also 
applications in bioengineering and medical situations~\cite{mccoyAssessmentTimeSeriesMachine2018}. The goal here is to obtain a good approximation of 
the function $f$ in order to \emph{extrapolate} its 
domain to obtain new, and unseen, results. It is expected that ML methods, with their 
ability to find undiscovered 
patterns, can effectively predict and forecast outputs that are not in the data set 
$\mathcal{D}$. This is an extremely hard task, but one that has been finding a lot of 
applications and many groups have done extensive research on the matter.

Regression is an important task, and in \autoref{Cap4} it is the primary learning mechanism 
used to solve a particular problem in Liquid State Theory. The problem of regression, as 
stated before, is to learn a \emph{good} approximation of the function $f$. Again, the 
measure of \emph{good} is not clear enough, as was the case with \emph{accuracy}, and both 
these concepts shall be discussed next.

\subsection{Performance Measure} \label{sec:performance}
ML algorithms must be assessed on their abilities to perform a certain task $T$, either a 
classification or regression problem in the current discussion. 
It is common practice to extract a subset of the 
training data set $\mathcal{D}$ as the \emph{testing set}, $\mathcal{T}$, to evaluate the 
algorithm in the task $T$. The measure depends on the task $T$, as well as the type of data 
used.

In classification tasks, the \emph{accuracy} is one of the most general performance 
measures. The accuracy is simply defined as the proportion of data points in $\mathcal{T}$ 
for which the model produces the correct output. This measure is quite strict in the sense 
that if there are examples that are \emph{mislabeled}, this is carried on to the ML model.

Another common performance measure for classification tasks is the receiver operating 
characteristic, also known as the ROC~\cite{hastieElementsStatisticalLearning2009}. More 
specifically, the area under the ROC curve. This measure of performance is created by 
plotting the \emph{true positive rate} against the \emph{false positive rate}. This measure 
is used because these rates are more permissive, since they stem from the well established 
type I and II errors from statistical theory~\cite{riceMathematicalStatisticsData2006}. 
Depending on the data set, other measures can 
be used, such as the F1 score, the Jaccard index, Akaike information criterion, and 
others~\cite{murphyMachineLearningProbabilistic2012}.

Regression tasks are different when measuring performance since these methods attempt to 
approximate continuous, real-valued functions. So in a sense, one can simply use \emph
{metrics} in the mathematical analysis sense of the word. For instance, the use of the 
$L^2$ norm, also known as the so-called Euclidean norm, defined as
\begin{equation}
    { \left\lVert y - \hat{y} \right\rVert }_{2} = \sqrt{ \sum_{i=1}^{N} { \left(y_i - \hat{y}_i \right) }^2 }
    \; ,
    \label{eq:l2norm}
\end{equation}
is extremely common in regression tasks. Here, the training examples $y$ are measured 
against the output value from the ML model, $\hat{y}$. Another extremely common, and 
profoundly useful metric, is the $L^1$ norm,
\begin{equation}
    { \left\lVert y - \hat{y} \right\rVert }_{1} = \sum_{i=1}^{N} \left\lvert y_i - \hat{y}_i \right\rvert
    \; .
    \label{eq:l1norm}
\end{equation}
The $L^1$ has the amazing property that it can create \emph{sparse} representations of the 
learned function $f$. This is the basis of the LASSO method~\cite{hastieElementsStatisticalLearning2009}, a very useful and common ML method to do 
\emph{variable selection}, the discrimination of variables that are useful or not to the 
prediction of the model; and \emph{regularization}, which is adding information in order to 
solve a problem that might seem hard or impossible to solve, also known as an
\emph{ill-posed} problem~\cite{goodfellowDeepLearning2016}.
When generalizing these metrics to the full data set, they take on different names. For 
instance, the squared $L^2$ norm takes the name \emph{mean squared error}, defined as
\begin{equation}
    MSE \left( y_i, \hat{y}_i \right) = \frac{1}{N} \sum_{i=1}^{N} { \left(y_i - \hat{y}_i \right) }^2
    \; ,
    \label{eq:mse} 
\end{equation}
where $N$ is the total number of examples used to compute the MSE.

Similarly, the $L^1$ norm can be generalized to the \emph{mean absolute error}, defined as
\begin{equation}
    MAE \left( y_i, \hat{y}_i \right) = \frac{1}{N} \sum_{i=1}^{N} \left\lvert y_i - \hat{y}_i \right\rvert
    \; .
    \label{eq:mae}
\end{equation}
These generalizations might seem trivial, but they allow for more general formulations, as 
they are in fact \emph{estimators} that measure the average of the outputs. When measuring 
the performance of regression models, the goal is to \emph{minimize} these errors, with 
zero being the optimal value.

\subsection{Approximation Theory}
Before closing this brief overview on ML theory, a note on approximation theory is in 
order. The link to ML has to do with the striking resemblance between both classification 
and regression tasks. Both problems seem to be, in a sense, the same exact problem which is to learn a \emph{map} or \emph{function} that relates an input with an output. However, this problem is not 
new, and in fact, it has been extensively studied before, so much so that is has created a 
branch within mathematics, called \emph{approximation theory}.

The primary goal of approximation theory is simple: to understand how functions can be \emph{best} approximated with even simpler functions~\cite{trefethenApproximationTheoryApproximation2013}. This sounds analogous to the learning tasks, however, there was no mention of simpler functions in the previous discussion. This is because to see this more clearly, there has to be a detailed explanation of each of the models used. For instance, \emph{Gaussian Processes}~\cite{rasmussenGaussianProcessesMachine2006} can approximate an arbitrary function with Gaussian functions, or more precisely, with normal probability distributions. In any case, there is not enough space in this work to talk about all the possible methods and how these are formulated, even if they are formulated based on simpler functions or not.

However, the discussion can be simplified by noting the following. If looked at closely,
\autoref{eq:l2norm} and \autoref{eq:l1norm} look quite similar. Indeed, there is a generalization of this norm, called the $L^p$ norm,
\begin{equation}
    { \left\lVert y - \hat{y} \right\rVert }_{p} = { \left( \sum_{i=1}^{N} { \left(y_i - \hat{y}_i \right) }^{p} \right) }^{1/p}
    \; ,
    \label{eq:lp-norm}
\end{equation}
which generalizes the Euclidean norm to a more general norm, defined now in function spaces 
called $L^p$ spaces~\cite{rudinPrinciplesMathematicalAnalysis2013}. These spaces are, 
roughly speaking, a generalization of vector spaces where the basis that span the linear 
spaces is comprised of functions instead of vectors. There is, however, a particular form of the $L^p$ norm called the \emph{supreme norm}, or equivalently, the \emph{infinity norm} defined as
\begin{equation}
    { \left\lVert y - \hat{y} \right\rVert }_{\infty} = 
    \underset{i}{\max}{\left\lvert y_i - \hat{y}_{i} \right\rvert}
    \; ,
    \label{eq:linf-norm}
\end{equation}
also known as \(L^{\infty}\), and it is a function space that contains all the essentially bounded measurable functions~\cite{taoIntroductionMeasureTheory2011}.

Why is $L^{\infty}$ so important in this context? The answer is one simple, but outstandingly powerful theorem: the Stone-Weierstrass theorem~\cite{stoneApplicationsTheoryBoolean1937, stoneGeneralizedWeierstrassApproximation1948}. In 
1885, Karl Weierstrass proved that any function can be approximated by a polynomial of a 
given degree. This result is so powerful that it created the field of approximation theory. 
And it is so compelling, that it might arguably be the fundamental theorem of ML theory. 
This is the reason for the \(L^{\infty}\), and its relation to the learning tasks should 
become clear once the main theorem is stated.
\begin{theorem}[\textbf{Weierstrass approximation theorem}]
    If \(f\) is a continuous real-valued function on \([a, b]\), and if any \(\epsilon > 0\) is given, then there exists a polynomial \(P\) on \([a, b]\) such that
    \[
        { \left\lVert f(x) - P(x) \right\rVert }_{\infty} < \epsilon
        \]
        for all \(x \in [a, b]\) .
    \label{approx}
\end{theorem}
The proof is left for a specialized text on the subject, for instance the book by Richard 
Beals~\cite{bealsAnalysisIntroduction2004}. Nonetheless, it is important to see the striking resemblance between the kinds of learning discussed so far, and \autoref{approx}.
On one hand, classification and regression both deal with finding the best mapping between 
inputs and outputs. On the other hand, \autoref{approx} states that every function can be 
approximated by another, simpler function. So, in a sense, these problems are related. One 
seeks to find the best mapping, while the other guarantees that there is such a mapping. 
\autoref{approx} helps visualize that the main problem of ML theory is to approximate the 
underlying function between inputs and outputs in a data set. However, the question of how to find such polynomial \(P\) is still open.

With the presentation of the Weierstrass approximation theorem it is now time to move on to 
the case of neural networks, which are in fact a way to build a polynomial \(P\) with the 
help of data. If looked at closely, the theorem only states that there is such polynomial, 
but does not specify how it might be constructed or formulated. There are roughly two 
ways: with pure mathematics or with the help of data. To use mathematics is to turn to the help of approximation theory; in this case 
orthogonal polynomials are used, such as Tchebyshev polynomials, to build such a polynomial 
\(P\). Of course, many other approximation methods can be used~\cite{trefethenApproximationTheoryApproximation2013}.
On the other hand, turn to 
statistics and data, and seek the help of ML theory, such that the data is used to construct
such a polynomial. Before moving on, it should be pointed out that this is not a unique or 
novel way of seeing the problem of learning. Indeed, a more general formulation is based on 
probability theory, which uses much more rigorous arguments to link the relation between ML 
theory and approximation theory. 
The book by Murphy~\cite{murphyMachineLearningProbabilistic2012} is a great reference for 
that.

\section{Neural Networks}
In this section, neural networks (NN), their architectures, training and mathematical formulations are discussed. Although, this is just a short overview of the full theory of NN. In recent years, NN have been the most used, researched and applied method in modern ML. Up to this day, there are unmeasurable number of research articles and books devoted to NN. So this space is obviously quite small to fit all the information about them. It is for this reason that this section will focus primarily on how NN are trained and how they learn a mapping given a data set.
For a more thorough understanding of NN, these references~\cite{mehligMachineLearningNeural2021,goodfellowDeepLearning2016,hastieElementsStatisticalLearning2009,bernerModernMathematicsDeep2021} should suffice.

\subsection{Motivation}
\begin{figure}
    \centering
    \includegraphics[scale=0.4]{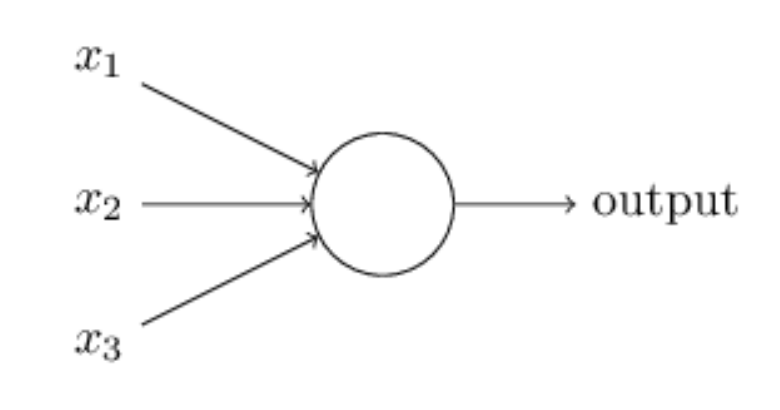}
    \caption{A representation of the original McCulloch-Pitts perceptron. From~\cite{nielsenNeuralNetworksDeep2015}.}
    \label{fig:perceptron}
\end{figure}

\begin{figure}
    \centering
    \includegraphics[scale=0.4]{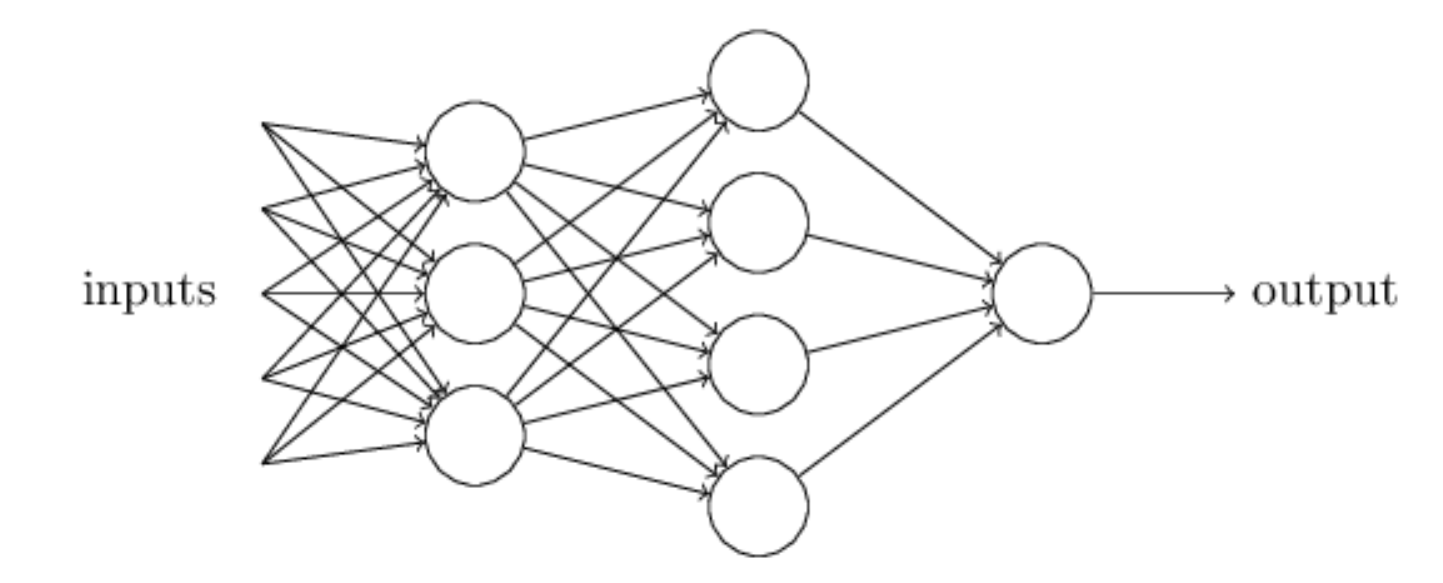}
    \caption{A representation of several perceptrons connected with each other. From~\cite{nielsenNeuralNetworksDeep2015}.}
    \label{fig:multi-perceptron}
\end{figure}

The basic idea of NN is to mimic the way the brain works, and more specifically, the way 
neurons interact with each other. In the modern form of NN, this interaction is highly 
idealized in the sense that the actual human brain does not follow the same form of 
interaction, but it is somewhat a rough approximation. In 1947, Pitts and McCulloch~\cite{pittsHowWeKnow1947} devised the \emph{perceptron}, an idealized model of a neuron. This perceptron can be observed in \autoref{fig:perceptron}. Here, the middle circle is the perceptron or \emph{unit}, and it accepts three inputs \(x_1, x_2, x_3\), or equivalently \(\bm{x}=(x_1, x_2, x_3)\) with \(\bm{x}\) a \(3\)-dimensional input vector. The output of the McCulloch-Pitts perceptron is a single binary output, it can either return a \(0\) or a \(1\). Then, in 1958, Rosenblatt~\cite{rosenblattPerceptronProbabilisticModel1958} extended this idea and introduced a way to compute the output of the perceptron. He introduced what he called \emph{weights}, and gave a weight to each of the three inputs, \(w_1, w_2, w_3\). Later, he proposed to compute the output as a weighted sum of each of the three inputs, such that the output would have the following form,
\begin{equation}
    \text{output} = \begin{cases}
        0 \quad \text{if} \quad \sum_{i} w_i x_i \leq \varepsilon \\
        1 \quad \text{if} \quad \sum_{i} w_i x_i > \varepsilon
    \end{cases}
    \; ,
    \label{eq:output-perceptron}
\end{equation}
where \(\varepsilon\) is a \emph{threshold} value. This model is simple, yet effective. If 
one input would be more important than the other, the weight could be adjusted to reflect 
this. As simple as the model is, it is a good approximation of how decision-making models 
can be constructed efficiently. Furthermore, a more complex decision-making system can be built if several perceptrons are connected with each other. After all, the inputs and outputs of a single perceptron do not have limitations, and they can well be defined to be the inputs and outputs of other perceptrons as well. Instead of using a simple perceptron, a more complex structure can be built upon using several of them. If \emph{layers} of perceptrons are stacked between each other, a more complete model of decision-making is created. This is the case of the \emph{multilayer perceptron}, or MLP, shown in \autoref{fig:multi-perceptron}.

Despite this, the model has serious drawbacks. For one, 
there is no clear understanding of the threshold value and what it is. On the other hand, 
having the perceptron output just two values might make it inaccessible in the most general 
case. After all, the idea of NN is to learn a mapping, a function that in general is 
real-valued. Finally, how are the weights adjusted? Are these values picked randomly? Does 
the user choose them? If in fact NN are learning models, then there is no real reason why 
the user would likely pick the weights for each problem they need to solve. All of this 
questions will be addressed in a later section, when training is discussed in more detail.
For now, architectures and activations functions will be presented next. These topics will
provide answers as to how the output of perceptrons can be modified to return other values,
rather than just a \(0\) or \(1\).

\subsection{Architectures and Activation Functions}
Now that a complete network of perceptrons has been defined, it is time to discuss one 
important issue with them. First, it might be easier to deal with a different notation 
than just using threshold values. It is simpler to use the language of Linear Algebra 
and define the output expression shown in \autoref{eq:output-perceptron} using the dot product as follows,
\begin{equation}
    \text{output} = \begin{cases}
        0 \quad \text{if} \quad \bm{w} \cdot \bm{x} + b \leq 0 \\
        1 \quad \text{if} \quad \bm{w} \cdot \bm{x} + b > 0
    \end{cases}
    \, ,
    \label{eq:output-dot}
\end{equation}
where \(\bm{c} \cdot \bm{d} = \sum_{i}^{N} c_i \, d_i\) is the dot product between 
the $N$-dimensional vectors \(\bm{c}\) and \(\bm{d}\). Now, there is a new value, 
called the \emph{bias} term, defined as \(b\) in \autoref{eq:output-dot}. This new term is 
the value that is added to each perceptron and it can be interpreted as a measure of how 
easy it is to get the perceptron to output \(1\).

\begin{figure}
    \centering
    \includegraphics[scale=0.4]{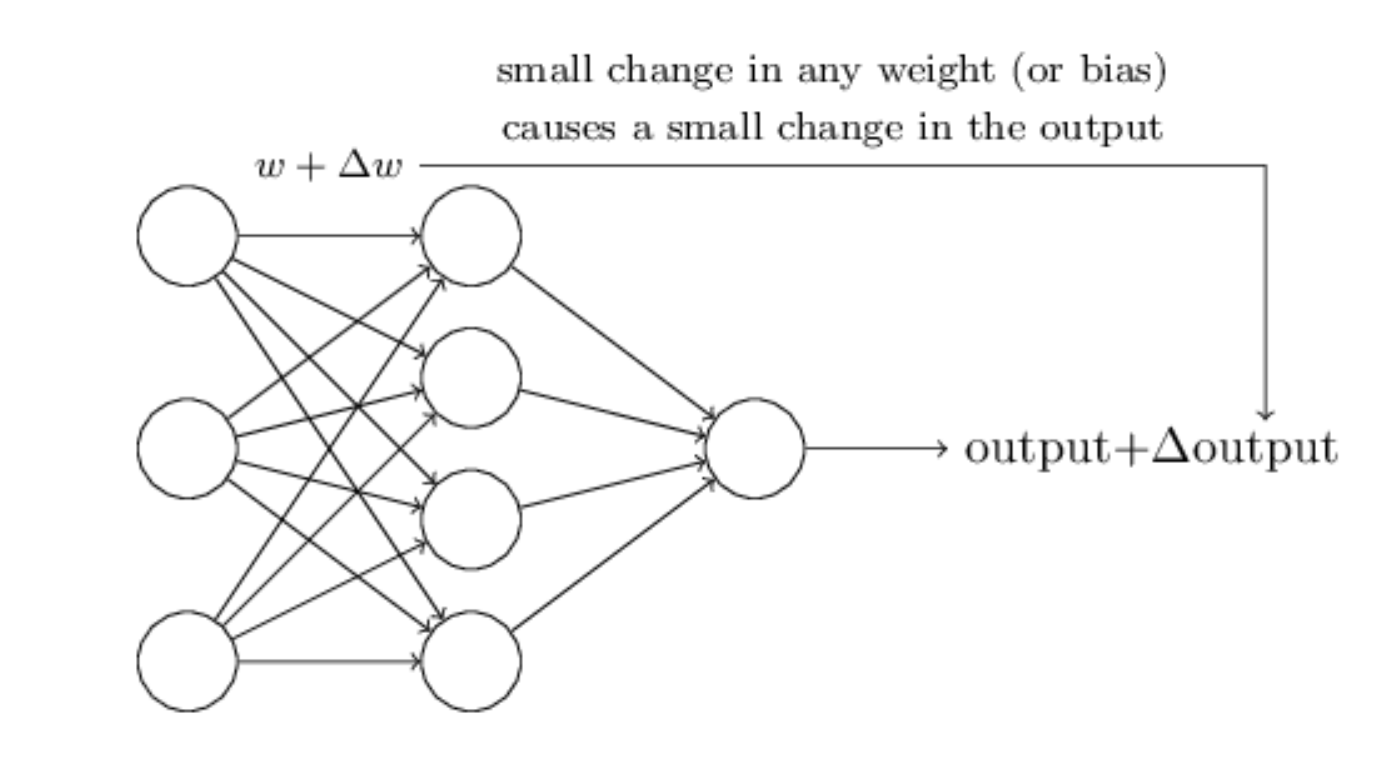}
    \caption{Visualization of the basic training mechanism in a neural network. By inducing a small change to the weights, a small change in the output is expected. From~\cite{nielsenNeuralNetworksDeep2015}.}
    \label{fig:learning-weights}
\end{figure}

There is now the following issue to address, and that is that the output is only \(0\) or 
\(1\). The issue here is that NN are expected to learn complex mappings between inputs and 
outputs, and if the weights are adjusted a minimal amount, for instance a value of 
\(\Delta w\), then the output should change just a small amount, proportional to the change 
in the weights. However, if the output is either a \(0\) or \(1\), then a small change will 
trigger a drastic change, and learning will not be entirely possible. A representation 
of this idea can be seen clearly in \autoref{fig:learning-weights}. The pertaining question 
here is, is there a way to change this behavior? The answer is yes, but to do so, a new 
type of perceptron, as well as a NN \emph{architecture} must be introduced.

The \emph{sigmoid} perceptron is a modification to the McCulloch-Pitts original perceptron which can provide a continuous value to the output by using a continuous function. This function, the \emph{sigmoid} function, is defined as
\begin{equation}
    \sigma(z) \coloneqq \frac{1}{1 + \exp{\left( -z \right)}}
    \; .
    \label{eq:sigmoid}
\end{equation}
Instead of the output being just \(0\) or \(1\), \autoref{eq:output-dot} is now replaced by the following equation, using now \autoref{eq:sigmoid},
\begin{equation}
    \text{output} = \frac{1}{1 + \exp{\left(- \sum_{i} w_i x_i + b\right)}}
    \; .
    \label{eq:output-sigmoid}
\end{equation}
This function might seem to drastically change the behavior and original intention of the 
McCulloch-Pitts perceptron. However, in reality, the sigmoid function is just a smoothed 
version of the original perceptron. With this modification, when a small change is done in 
the weights of the network, a small change will be expected in the output. The amount of 
change, and how this change happens, will be discussed in the next section when training is presented.

If different functions are used instead of the sigmoid function, the network is expected to 
behave differently with each one. The choice of the function is so important, that not only 
have these function taken in a particular name\textemdash \emph{activation functions}\textemdash, but there has been extensive research in this area~\cite{chenUniversalApproximationNonlinear1995,nwankpaActivationFunctionsComparison2018,agostinelliLearningActivationFunctions2015,ramachandranSearchingActivationFunctions2017}. 
In practice, the choice of activation function can make a NN perform better in 
certain classification tasks, e.g. object recognition, segmentation, and others; as well as 
in regression tasks. Activation functions can also help the NN perform faster, which is 
an important aspect of the training mechanisms and practicalities of
NN~\cite{fengPerformanceAnalysisVarious2019,biswasTanhSoftDynamicTrainable2021}.

Finally, it is important to mention that the \emph{topology}, or \emph{architecture}, of 
the network has been subject to significant research in recent times. For instance, if more 
layers are added in between layers, the NN takes in a different name, a \emph{deep neural network}. Deep neural networks have been the quintessential model of the revolutionary and 
extremely popular field of \emph{deep learning}~\cite{goodfellowDeepLearning2016}. In deep 
learning, NN architectures are extended to be larger, deeper, and more complex than simple 
multi-layer perceptrons. These new models can be thought of being several multi-layer 
perceptrons, each stacked upon each other.
Furthermore, deep learning attempts to generalize the simple perceptron model and introduce 
better models that can perform with greater accuracy in certain tasks. The explosion of 
deep learning came when \emph{convolutional neural networks} were applied successfully in 
the AlexNet architecture~\cite{krizhevskyImageNetClassificationDeep2012}. There other kinds 
of important variations in the topology of the network, such as the U-shaped networks use 
for medical image segmentation~\cite{ronnebergerUNetConvolutionalNetworks2015}; V-shaped 
networks used for remote sensing~\cite{abdollahiVNetEndtoEndFully2020}; and very deep 
topologies that can track images in real time~\cite{redmonYOLOv3IncrementalImprovement2018}.

\subsection{Training}
Choosing the correct activation function and the appropriate topology is hard, and a lot of 
experimentation is needed. But even if these could be chosen automatically, there is still 
a latent issue here: how weights can be adjusted. Up until this section, this issue has 
been left aside and it is now time to address it. To do so, it is imperative to remember 
the sole purpose of NN and ML methods for classification and regression tasks: 
\emph{to learn a mapping or function}. 
Assuming it is possible to have a training set \(\mathcal{D}\) with enough examples, 
containing both inputs and outputs, the function that determines the relationship between 
them can be learned by a NN. In general, this is referred to as the 
\emph{universal approximation theorem}, which shall be the topic of the next section.

The topic for this section is to answer the question of how this learning is carried out in 
the first place. And to do so, it is useful to recall the information from 
\autoref{fig:learning-weights} in that, if the weights are modified then the output is 
expected to change. Now, there are several questions that need an answer in that figure. 
For instance, what is the proportion between the change in the weights to the change in the 
output? If the weights are changed by half, is the output expected to change in the same 
amount? For the case of classification, there are only categorical outputs, so how are 
changes to the weights passed on the output? It seems that there needs to be a quantity 
that must inform the \emph{learning} mechanism how much should the weights change in order 
to get the correct output from the network. 

In order to address those questions, there needs to be a relationship between the output, the weights and the bias. Fortunately enough, calculus provides such information in the 
following way,
\begin{equation}
    \Delta \ \text{output} \ \approx \sum_{j} \left[ \frac{\partial \ \text{output}}{\partial w_j} \Delta w_j \right] + \frac{\partial \ \text{output}}{\partial b} \Delta b
    \; ,
    \label{eq:output-combination}
\end{equation}
where the sum is over all the weights in the network, \(w_j\), and \(\partial \ \text{output} \, / \, \partial w_j\) and \(\partial \ \text{output} \, / \, \partial b\) denote 
the partial derivatives of the output with respect to the weights and bias, respectively. 
In other words, these derivatives represent \emph{by how much} the output changes if the 
weights and bias are modified. Yet, there is still even more information from \autoref{eq:output-combination} 
that is extremely useful. If looked at closely, \autoref{eq:output-combination} actually 
represents a \emph{linear combination} of terms that, together, form the change to the 
output. This means that both, the weights and the bias, must contribute to the change in 
the output, and the amount is provided by the partial derivatives. So with just this much, 
the answer to the question of \emph{how much} is now answered. But now there is the burning 
question of how to compute the partial derivatives. Well, this question is easy to answer, 
because assuming that the NN is using sigmoid neurons, then the derivative can be reduced 
to the following expression,
\begin{equation}
    \frac{d}{dz} \sigma (z) = \sigma^{\prime} (z) = \frac{1}{1 + e^{-z}}  \cdot \left(1 - \frac{1}{1 + e^{-z}}\right) = \sigma (z) \, \left(1 - \sigma (z)\right)
    \; ,
    \label{eq:dsigmoid}
\end{equation}
where some steps were skipped to avoid taking too much space. Recalling that the output from 
the sigmoid perceptron is given by \autoref{eq:output-sigmoid}, together with \autoref{eq:dsigmoid}, 
the partial derivatives from \autoref{eq:output-combination} can be obtained without much
effort. In more general cases when the perceptron uses a different activation function than 
the sigmoid, the computation of derivatives is carried out using modern tooling known as 
\emph{automatic differentiation}~\cite{baydinAutomaticDifferentiationMachine2018}.

Now, the issue of addressing how the learning mechanism is informed can be solved by 
introducing the \emph{cost function}. The cost function, also referred to sometimes as 
\emph{loss} or \emph{objective}, is a function that measures how the NN is approximating 
the underlying function. This is, in fact, the same concept as the performance measures 
encountered in \autoref{sec:performance}. In fact, cost functions are metrics that address 
how good the approximation of the NN is. And so this issue has also been addressed. 
However, how is this cost function, and \autoref{eq:output-combination} used to make the NN 
learn the mapping? For this, we turn to optimization and the powerful method of \emph{gradient descent}. To help illustrate the learning mechanism, the mean squared error in 
\autoref{eq:mse} will be used. However, this is not the only metric that can be used, and 
in fact several research advancements have been carried out in this direction~\cite{fontenla-romeroNewConvexObjective2010,liDiversityPromotingObjectiveFunction2016}.

Numerical optimization is a rigorous mathematical subject, and the use of gradient descent techniques can not be summarized in this section. For that, the great book by Nocedal and Wright is an excellent reference for this topic~\cite{nocedalNumericalOptimization2006}.
Here, only the main results will be mentioned and used. In optimization, specifically \emph{unconstrained optimization}, the task is to find the value of an input vector \(\bm{x}\) that \emph{minimizes the function} \(f\), such that \(f(\bm{x}) = 0\). The fact that it is unconstrained is because there is no restriction in the bounds of \(\bm{x}\), or in the form of \(f\), which in general is a non-linear function. In the context of optimization, \(f\) is referred to as a cost function, or objective function.
Now, a relation between optimization and NN can be made by noting the following fact. The 
purpose of NN is to learn the function that relates inputs and outputs in a given data set. 
Or in the context of optimization: \emph{to minimize the error between the true values and 
the approximation values}. The \emph{true values} correspond to the outputs in the data 
set, and the \emph{approximation values} to the outputs from the NN. So, in essence, the 
task is to find the best weights \(w_j\), and the best bias \(b\), that minimize the 
following cost function,
\begin{equation}
    C(\bm{w}, b) = \frac{1}{2 N} \sum_{i=1}^{N} { \left[y_i - \hat{y}_{i}(\bm{w}, b) \right] }^2
    \; .
    \label{eq:cost-nn}
\end{equation}
Here the sum goes over all $N$ training samples in the training data set, and the output 
from the NN \(\hat{y}_{i}(\bm{w}, b)\) depends on the values of the weights
\(\bm{w}\) and bias \(b\).

There is now one step left for the training procedure to be completely characterized, and 
that is to define a way to \emph{update} the weights and bias accordingly so that 
\autoref{eq:cost-nn} is minimized. For this purpose, \emph{gradient descent} is used. 
Gradient descent is an iterative optimization method based on first-order derivative 
information. The idea is to use the information from the gradient of the objective function 
to drive the minimization to minimum, which in general corresponds to a \emph{local minimum}. The derivation of the method is left for a specialized reference on the 
matter~\cite{nocedalNumericalOptimization2006}, and the main results will be used here. In 
the context of NN, the idea to update the weights and bias so as to minimize the cost 
function. Thus, the gradient descent method is introduced, which follows the expression,
\begin{equation}
    \begin{aligned}
        \bm{w}_{j+1} &= \bm{w}_{j} - \eta \frac{\partial C}{\partial \bm{w}_{j}} \\
        b_{j+1} &= b_{j} - \eta \frac{\partial C}{\partial b}
    \end{aligned}
    \label{eq:gradient-descent-nn}
\end{equation}
where the index \(j\) represents each of the steps taken to update each value. These steps 
are also sometimes referred to as iterations. Hopefully, by doing several updates of 
\autoref{eq:gradient-descent-nn}, then \autoref{eq:cost-nn} can be minimized and the 
appropriate value of the weights and bias would have been found.

In practice, this simple rule is not actually followed for one simple reason. If looked at 
closely, \autoref{eq:gradient-descent-nn} actually takes in all the information gathered so 
far, and this includes the training examples, which are encoded in the cost function. But 
if the data set contains thousands of examples, then the optimization can not be carried 
out due to current limitations in modern hardware. So a simple modification is used, where 
instead of choosing all the training examples at once, a small sub-set is chosen randomly 
and then the gradient descent is used on this sub-set. 
This procedure is repeated until all the 
training examples have been selected. This is referred to as a 
\emph{pass}. At the end of this pass, all the obtained values are averaged, and this 
average is used to update the weights and bias. This can be repeated for as many 
\emph{passes} as needed, or wanted. This simple modification is the actual 
method use, and goes by very popular name of \emph{stochastic gradient descent}.
Extensive research has been conducted in this area as well, from introducing simple
accelerating terms to the gradient descent rule. For instance, the Nesterov momentum 
gradient descent method~\cite{ruderOverviewGradientDescent2017}, or the Adam 
method~\cite{kingmaAdamMethodStochastic2017}, which is in fact the method used in this work 
and will be presented at a later part.

To summarize, the training mechanism of NN is as follows. Assuming a training data set 
\(\mathcal{D}\) is available, then a cost function is chosen. In practice, the most common 
has the form of \autoref{eq:cost-nn} but it can be any other. Then, the weights and biases 
are initialized at random. After that, the main loop of the algorithm is to repeat the 
steps shown in \autoref{eq:gradient-descent-nn} while keeping track of the minimum. When a 
certain minimum has been achieved, a performance measure can be used to check whether the NN 
has a good performance for the task given. This concludes, roughly speaking, the common 
practice of modern ML.

\subsection{Universal Approximation Theorem} \label{sec:approximation-thm}
To close this section on ML and NN, an overly important fact of NN must be stated. Before, 
it was mentioned that the goal of tasks such as classification and regression is for the 
model to learn or approximate the function that maps the input to the output for a given 
data set. For this reason, NN were presented, along with their learning mechanism. However, 
what \emph{guarantee} is there that NN \emph{will} approximate the function? Further, 
\emph{can NN, in fact, approximate a function}? In a previous section, it was briefly 
mentioned that NN can in fact \emph{learn} any function, and that will be the topic of this 
section.

Indeed, just as mentioned before in the section for Approximation Theory, there exists a 
similar result for NN. This result, which are actually a series of results, come from the 
mathematical research of NN. The results are known altogether as the \emph{universal 
approximation theorem}. To really understand these results, advanced mathematical results 
should be employed. However, there is no need to extend the current section to such 
lengths. Instead, following the research from Hornik and Cybenko~\cite{hornikMultilayerFeedforwardNetworks1989, hornikApproximationCapabilitiesMultilayer1991, cybenkoApproximationSuperpositionsSigmoidal1989}, 
the main results will be stated here to be interpreted.

\begin{figure}
    \centering
    \includegraphics[scale=0.6]{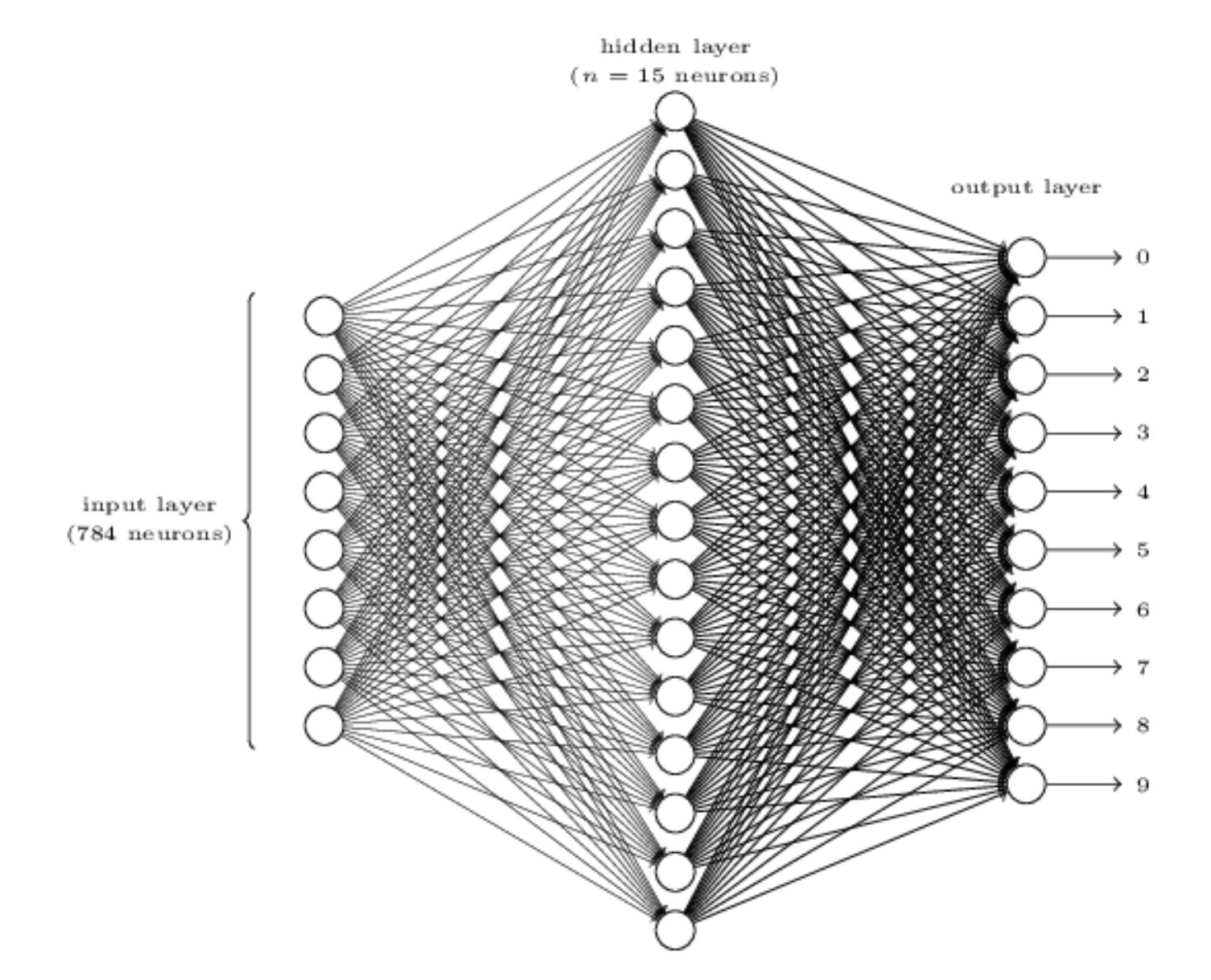}
    \caption{A multi-layer perceptron with three layers: the first layer is the \emph{input layer}; the middle layer is known as the \emph{hidden layer}; and the last layer is known as the \emph{output layer}. From~\cite{nielsenNeuralNetworksDeep2015}.}
    \label{fig:hidden-layer}
\end{figure}

In short, Hornik and Cybenko showed that, given the sigmoid function and a given topology, 
there exists an arbitrary number of nodes that are needed to \emph{approximate} any 
\emph{continuous} function. The topology is a three layer topology, similar to the one in 
\autoref{fig:hidden-layer}, where there is an input, a hidden and an output layers. This 
results is also similar to \autoref{approx}; however, in this case there is no polynomial 
but a NN instead. The arguments that follow this result is beyond the scope of this work, 
but the result remains valid. Then, as the research for the mathematical framework of NN 
advanced, newer results have seen the light that extend these results even more. Now, there 
is a clear understanding that not only the sigmoid function can be used, but any non-affine 
continuous function that it continuously differentiable~\cite{parkMinimumWidthUniversal2020}. Furthermore, there is also no need to restrict the case to just width, but also depth, 
from which the generalization to deep learning 
follows~\cite{zhouUniversalityDeepConvolutional2020,bernerModernMathematicsDeep2021}.
In this work, these guidelines are used to create simple topologies for NN that can 
actually satisfy the conditions of the theorem. A topology similar to 
\autoref{fig:hidden-layer} with a large amount of nodes for each layer, which should 
suffice to satisfy the conditions of the theorem. However, in practice, it has been shown 
that \emph{overparametrization}\textemdash when there are way more weights and bias to adjust than needed \textemdash is the key to generalization in 
NN~\cite{neyshaburRoleOverparametrizationGeneralization2018,cohenLearningCurvesOverparametrized2021}.
This means that most of the time, the common practice is to just use deep topologies, which 
mean a large number of layers and a large number of nodes for each layer, and use this 
overparametrization to leverage the generalization that this provides. So, in a sense, one 
might say that the universal approximation theorem is always satisfied in modern ML practice.

\section{Evolutionary Computation}
In this section, Evolutionary Computation (EC) is presented and discussed. Much like the 
driving force of perceptron is to mimic the brain and its intelligence, the inspiration 
behind EC is the way nature adapts itself through 
evolution~\cite{fogelWhatEvolutionaryComputation2000}. Although, in reality, evolution 
takes millions of years but it eventually creates variations that adapt and \emph{converge} 
to a particular goal. This particular goal, as described by Darwin, is to be fit to its 
surroundings and survive. The idea behind EC is to use this powerful mechanism of evolution 
and apply it to complex problems, with the use of modern computational and mathematical 
resources to facilitate and empower a simplified version of evolution. In the most general 
case, evolution is roughly a two-step process: first comes \emph{variation} and then comes 
\emph{selection}. Algorithms based of EC use these properties and, depending on how these 
variations and selections occur, the algorithms are named and used differently. In 
particular, for this work the focus will be \emph{evolution strategies}, which will be 
address in a later section. However, there is also \emph{genetic algorithms}, 
\emph{evolutionary algorithms}, and other~\cite{kacprzykSpringerHandbookComputational2015}.

\begin{figure}
    \centering
    \includegraphics[scale=0.67]{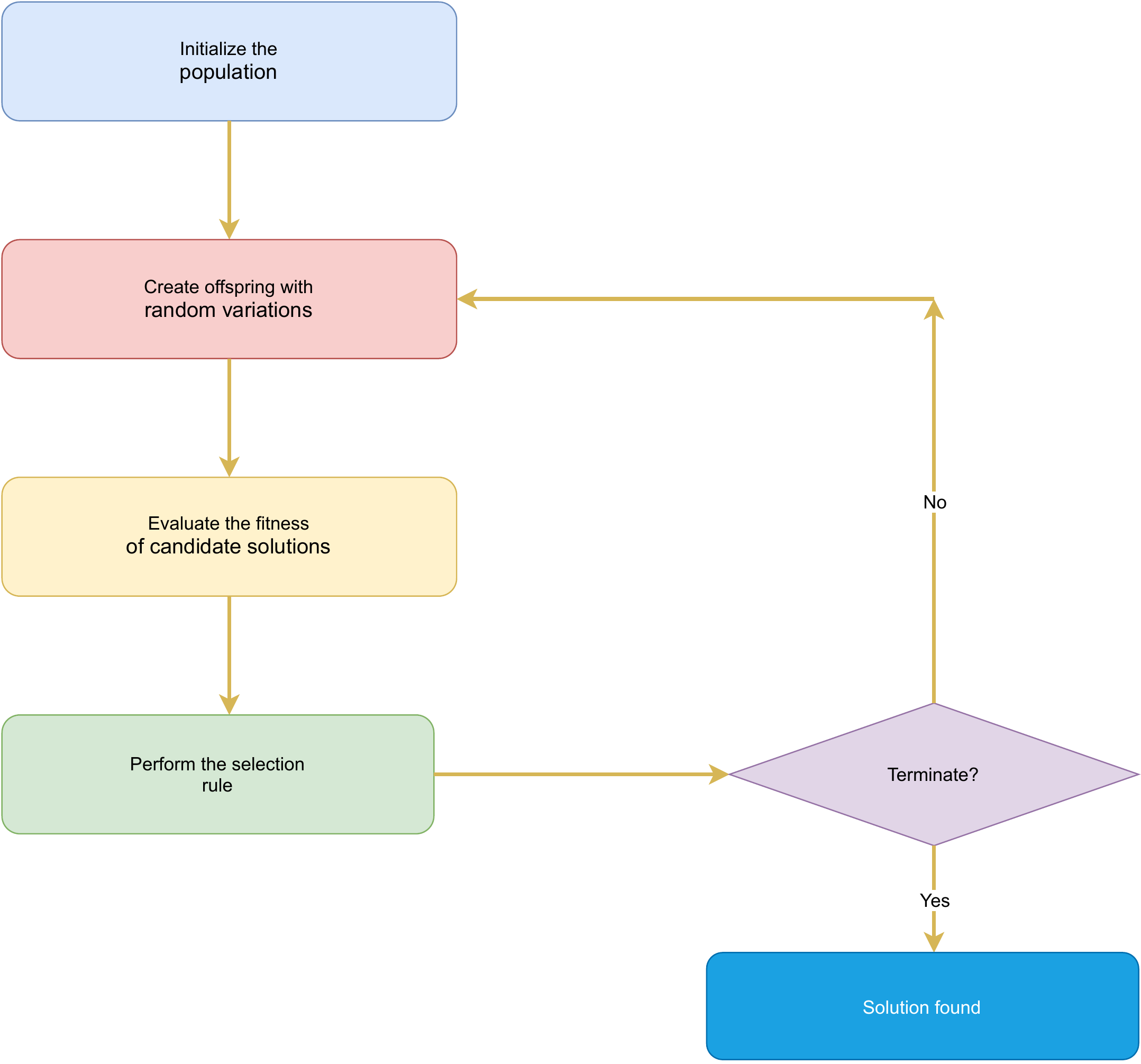}
    \caption{Evolutionary algorithms always start by creating a population of solutions. New solutions are then created by varying those from the original population. Then, the solutions are measured with respect to how well they address the task. Finally, a selection criterion is applied extract the best solutions so far. The process is iterated using the selected set of solutions until a specific criterion is met.}
    \label{fig:evolutionary-optimization}
\end{figure}

As mentioned before, there is a common procedure in evolution, and this has been translated 
to computer science and its application in EC. The basic procedure is as follows: for a 
given problem, an initial population of possible \emph{solution} is created; then, those 
solutions that are \emph{fit} or \emph{approximate} the solution are kept, those that not 
are discarded; finally, the successful candidate are mixed, creating offsprings with 
\emph{variations} of all the other possible solutions. With these steps, a common EC 
algorithm can be constructed, and in general this is case. The advantage of using EC is 
that of \emph{adaptability}. Many of these algorithms are suited for hard problems, some of 
which are impossible to solve with traditional methods, or at least take a long time to 
find a solution, such as the famous traveling salesman 
problem~\cite{dorigoAntColonySystem1997}. Flexibility, adaptability and generalization are 
the advantages of EC algorithms. When used in the popular StarCraft computer videogame, the 
AlphaStar system~\cite{arulkumaranAlphaStarEvolutionaryComputation2019} performed 
surprisingly better than the top players in the world.

However, the lack the general ease of use in some cases, as well as the difficulty to 
implement them might be troublesome in some other cases. Also, EC algorithms tend to have 
many parameters to adjust. For instance the total number of solutions in the population, 
the number of offsprings, and so on. In the case of EC algorithms, there is no \emph{one 
size fits all}, so this can make them hard to deal with. In the present work, EC algorithms 
will be used for non-linear unconstrained optimization, and in particular, in the sub-field 
of \emph{derivative-free optimization}. This will be discussed in the next section.

EC algorithms have been used for optimization in cases where the objective function is not 
\emph{smooth}, does not posses a \emph{continuous derivative}, or when it is too 
\emph{costly} to search for a solution in finite 
time~\cite{kacprzykSpringerHandbookComputational2015}. The basic functionality of EC 
algorithms for optimization follow the diagram from 
\autoref{fig:evolutionary-optimization}. With EC algorithms, the idea is to form an initial 
population of solutions, then to vary those solutions while keeping track of the fitness of 
the objective function. By varying the possible solutions, it is expected that these will 
\emph{follow an evolutionary process}, and eventually reach a minimum. In particular, EC 
algorithms are best suited for \emph{global optimization}, which is the search for minima 
in a large search space, without reaching any local minima along the way. The purpose 
behind this is to \emph{explore} the search space and look for the values that best 
minimize the function. After all, EC algorithms are \emph{approximation algorithms}, i.e. 
this methods do not attempt to find the best solutions, but to 
\emph{find a possible solution} to the problem.

\subsection{Derivative-free and Black-box Optimization}
In optimization tasks, the best algorithms are those that include some kind of information 
about \emph{derivatives}. For instance, the Newton method, and Newton-like methods such as 
BFGS~\cite{nocedalNumericalOptimization2006}, are some of the best for their speed, 
accuracy and convergence properties. Mathematically rigorous frameworks can be used to 
understand these algorithms, and can also be used to understand their convergence 
properties. Another similar optimization method was encountered before, the gradient 
descent method, which by leveraging the gradient information of the objective function a 
local minimum could be found by iterative steps.

However, the derivative information is not always available. When such cases are 
encountered, the only methods capable of providing solutions are known as 
\emph{derivative-free} methods. These methods try to search the best solution, using 
different kinds of strategies, such as the 
trust-region~\cite{byrdTrustRegionAlgorithm1987}, quadratic polynomial 
approximation~\cite{powellUOBYQAUnconstrainedOptimization2002}, and the most recent and 
interesting approach is that of surrogate model 
optimization~\cite{forresterRecentAdvancesSurrogatebased2009}. In each case, the objective 
function is approximated using some form of interpolation, or surrogate model, and the 
original search space is projected onto a new search space where the approximation can be 
optimized efficiently. With this approach, derivatives are no longer needed and 
optimization can be performed, but not without its drawbacks. First and foremost, the 
method relies entirely on function evaluations. If the function is \emph{costly} to 
evaluate, then this methods will take a long time to find a suitable solution. Further, the 
search space will be explored, and this can considerably slow down the optimization 
procedure. After all, these methods try to explore and map the whole search space, but if 
the search space is quite large, the optimization procedure will suffer greatly in terms of 
performance. Finally, the \emph{no free lunch} theorem states that, averaged over all 
optimization problems, all optimization methods perform equally 
well~\cite{adamNoFreeLunch2019}. This means that if a particular derivative-free method is 
not providing sufficiently good results, then there must exist another optimization method 
that should give sufficiently good solutions. In other words, several algorithms must be 
tested against each other, given that there are enough computational resources to do so.

As mentioned before, sometimes there is not derivative information about the objective 
function, but more than that, objective function evaluations are \emph{costly}. This means 
that, for a given input \(\bm{x}\) and an objective function \(f \colon \mathbb{R}^n \mapsto \mathbb{R}\), the evaluation \(f(\bm{x})\) will take a long time. Further, in 
most cases when this is true, the \emph{analytical or functional form} of the objective 
function \emph{is not known}. When a particular objective function satisfies these 
conditions, it is referred to as a \emph{black-box function}. When optimizing costly 
black-box function, the main goal is to \emph{minimize} the number of evaluations of the 
function. This is done by sampling the search space \emph{intelligently} so as to not 
evaluate the objective function where there is no sign of a minimum.

Black-box functions can, in general, be any type of function, either mathematical or 
computational. When the function is computational, this can represent code that performs 
several tasks, each of which might take a considerable amount of time and computing 
resources. In such cases, black-box optimization algorithms are the key to solving the 
problem. Or maybe the black-box function is not too costly, but instead the true form of 
the function is not known. Again, black-box optimization algorithms are the best choice
in that scenario. In the next section, a particular black-box optimization algorithm 
derived from EC methods will be reviewed and presented, which will be extensively used in 
chapter 5 of this thesis.

\subsection{Natural Evolution Strategies}
One algorithm that stands out for its robustness and flexibility in dealing with black-box 
optimization tasks is the \emph{natural evolution strategy} algorithm (NES), presented in 
2014 by Wierstra \emph{et al}~\cite{wierstraNaturalEvolutionStrategies2014a}. This 
algorithm is a derivative of the Evolution Strategies (ES) family of algorithms, which 
follow closely the principles of EC. In ES algorithms, the principal characteristics are 
the ease of dealing with high-dimensional optimization problems, and low number of function 
evaluations. Also, unlike other EC algorithms, ES methods evaluate the fitness of the 
functions in batch instead of individually, allowing for more efficient computation and 
solution space search in general. The most prominent algorithm in the ES family of methods 
is the CMA-ES, or Covariance Matrix Adaptation-Evolution 
Strategy~\cite{hansenCMAEvolutionStrategy2006}.
In general, ES methods do two main things. To perform \emph{variation}, multivariate normal 
random vectors are used. To perform \emph{mutation}, these algorithms modify different 
aspects of such multivariate normal distribution. The case of CMA-ES is to modify and adapt 
the covariance matrix in order to perform better at each step.

However, the method of NES is different. Instead of modifying the covariance directly, the 
information on the \emph{natural gradient} is exploited. To understand this method more 
clearly, let \(f \colon M \subseteq \mathbb{R}^n \mapsto \mathbb{R}\) be a black-box 
function, and the problem of optimization is to search for a vector \(f(\hat{x}) \in M\) 
such that a \emph{local minimum} is found, where the local minimum is defined as,
\begin{equation}
    \exists \; \epsilon > 0 \quad \forall x \in M \; \colon \; 
    \lVert x - \hat{x} \rVert < \epsilon \Rightarrow f(\hat{x}) \leq f(x)
    \: .
    \label{eq:local-minimizer}
\end{equation}
Now, the goal is not to minimize the objective function directly, but to minimize an \emph{expected fitness} under a particular search distribution. This means that the parameters searched will define a continuous probability distribution. This cost function is now expressed as,
\begin{equation}
    J(\theta) = \mathcal{E}_{\theta} \left[ f(\bm{x}) \right] = 
    \int f(\bm{x}) \, \pi (\bm{x} \, \vert \, \theta) \, d \bm{x}
    \; ,
    \label{eq:expected-fitness}
\end{equation}
and using this cost function, together with a gradient descent-like rule, the purpose of NES is to find the best 
\emph{probability distribution} that will eventually be sampled to obtain a minimizer for 
the original objective function. In other words, the NES method is not looking for the 
minimizer itself in a search space, but it is looking for the probability distribution 
that, when sampled, will give in average the minimizer for a particular local minimum of 
the objective function.

Still, the gradient information is unknown, as well as the true form of the search distribution, \(\pi (\bm{x} \, \vert \, \theta)\). But it is also impossible to obtain a gradient estimation without explicit information about the probability distribution. Here is where the NES provides a novel way of dealing with this problem. First, the search distribution is fixed to be a multivariate normal distribution,
\begin{equation}
    \pi (\bm{x} \, \vert \, \theta) = 
    \frac{1}{{(2 \pi)}^{d/2} \text{det} \; \bm{\Sigma} } 
    \cdot
    \exp{\left(- \frac{1}{2} {\left(\bm{x} - \bm{\mu}\right)}^{\top}
    \bm{\Sigma}
    \left(\bm{x} - \bm{\mu}\right) \right)}
    \; ,
    \label{eq:multivariate-gaussian}
\end{equation}
with \(\bm{\Sigma} \in \mathbb{R}^{d \times d}\) the covariance matrix and 
\(\bm{\mu} \in \mathbb{R}^{d}\) 
the mean vector of a multivariate normal distribution. The parametrization 
\(\theta \coloneqq \left\langle \bm{\mu}, \bm{\Sigma} \right\rangle\) is the link between 
these probability distributions. With this information, the gradient is obtained with some important mathematical manipulations, which is approximately,
\begin{equation}
    \nabla_{\theta} J(\theta) \approx \frac{1}{\lambda} \sum_{i=1}^{\lambda}
    f(x_k) \, \nabla_{\theta} \log{\pi(x_k \, \vert \, \theta)}
    \; ,
    \label{eq:search-gradient}
\end{equation}
and it is also known as the \emph{search gradient}, and the gradient descent rule is just
\begin{equation*}
    \theta_{j + 1} = \theta_{j} + \eta \nabla_{\theta} J(\theta)
    \; .
\end{equation*}
The parameter \(\lambda\) in \autoref{eq:search-gradient} is the population size of the 
method, which in general is a small number, \(\lambda \approx 20 - 35\).~\cite{wierstraNaturalEvolutionStrategies2014a}

However, this is just the search gradient, and it has some serious disadvantages. One of 
which is the fact that the gradient descent update rule is not independent of the 
distribution, so the parameters can take different proportions. Another important aspect of 
the gradient descent rule is that it might be slow in terms of convergence for black-box 
objective functions. For this reason, the \emph{natural gradient}~\cite{amariNaturalGradientWorks1998,amariWhyNaturalGradient1998} is introduced, which 
modifies the gradient descent rule to be,
\begin{equation}
    \theta_{j + 1} = \theta_{j} + \eta \cdot \bm{F}^{-1} \nabla_{\theta} J(\theta)
    \; .
    \label{eq:natural-gradient}
\end{equation}
Here, \(\bm{F}\) is the \emph{Fisher information matrix} defined as,
\begin{equation}
    \bm{F} = \int \pi(\bm{z} \, \vert \, \theta) \; 
    \nabla_{\theta} \log{\pi(\bm{z} \, \vert \, \theta)} \;
    {\nabla_{\theta} \log{\pi(\bm{z} \, \vert \, \theta)}}^{\top} \;
    d \bm{z}
    \: ,
    \label{eq:fisher-matrix}
\end{equation}
but in practice it is estimated as,
\begin{equation}
    \bm{F} \approx \frac{1}{\lambda} \sum_{k=1}^{\lambda} 
    \nabla_{\theta} \log{\pi(x_k \, \vert \, \theta)} \;
    {\nabla_{\theta} \log{\pi(x_k \, \vert \, \theta)}}^{\top}
    \; .
    \label{eq:fisher-estimation}
\end{equation}

In summary, NES attempts to estimate the search distribution using the natural gradient 
such that, when sampled, the search distribution will in average output the minimizer to 
the objective function. Here, most of the details have been omitted for the sake of 
brevity, but the original work~\cite{wierstraNaturalEvolutionStrategies2014a} provides all 
the information needed to implement the method. In this work, though, a variant of the NES 
method is used, the \emph{distance-weighted exponential} (DXNES)
NES~\cite{fukushimaProposalDistanceweightedExponential2011,nomuraDistanceweightedExponentialNatural2021}. 
This method is a variation on the way the natural gradient and Fisher information matrix 
are obtained, but the core of the method remains the same. %% Machine Learning
\chapter{Neural networks as an approximation for the bridge function}
\label{Cap4}

%----------------------------------------------------------------------------------------
%	SECTION 1
%----------------------------------------------------------------------------------------

Neural networks can be used as \emph{universal approximators},
i.e., they can take the form of any continuous function if and only if the conditions of the
universal approximation theorem hold (see \autoref{sec:approximation-thm}).
The aim of this chapter is to explore the hypothesis that a neural network might be 
useful as a bridge function parametrization in the closure expression for the 
Ornstein-Zernike equation~\cite{hansenTheorySimpleLiquids2013}.
If this is true, then choosing a particular approximation can be avoided for a given 
interaction potential, and leave the choice of the bridge function to the neural network 
itself, while simultaneously solving for the Ornstein-Zernike equation. It is intended to
explore the implications of two inquiries:

\begin{enumerate}[(a)]
    \item Is it possible to solve the Ornstein-Zernike equation using a neural network?
    \item If it is indeed possible, how good is the quality of the approximation for the pseudo hard sphere potential~\cite{baezUsingSecondVirial2018}?
\end{enumerate}

In this chapter, we show in detail the methodology created to answer these questions, and
the mathematical structure with which a neural network can be used to solve the
Ornstein-Zernike equation.
The results obtained are compared to those from Monte Carlo computer simulations to assess 
the quality of the solution.
In the appendices (\autoref{AppendixA}, \autoref{AppendixB}), the numerical algorithm used 
to solve the Ornstein-Zernike equation
is presented, along with a detailed computation of the gradients used for the
training scheme. Here, we shall focus only on the main results and the algorithm structure
in general.

\section{Parametrization of the bridge function}

The Ornstein-Zernike formalism is given by the following coupled equations~\cite{hansenTheorySimpleLiquids2013}
\begin{equation}
    \begin{aligned}
         & h(\vecr) = c(\vecr) +
        n \int_{V}
        c(\vecr^{\prime})
        h(\lvert \vecr - \vecr^{\prime} \rvert)
        d\vecr^{\prime} \\
         & c(\vecr)
        = \exp{\left[
                -  \beta u(\vecr)
                +  \gamma(\vecr)
                + B(\vecr)
                \right]} -
        \gamma(\vecr)
        - 1 ,
    \end{aligned}
    \label{eq:oz1}
\end{equation}
with the already known notation for each quantity (see~\autoref{sec:ornstein-zernike}).

Let $\nnet$ be a neural network with weights $\theta$. The main hypothesis
of this chapter is that $\nnet$ can replace the bridge function $B(\vecr)$
in the previous equation, which will yield the following expression for
the closure relation

\begin{equation}
    c(\vecr) = \exp{\left[
            -  \beta u(\vecr)
            +  \gamma(\vecr)
            + \nnet
            \right]} -
    \gamma(\vecr)
    - 1 .
    \label{eq:parametrizacion}
\end{equation}

With this new expression, the main problem to solve is to find the weights
of $\nnet$ that can successfully solve the Ornstein-Zernike equation
for a given interaction potential, $\beta u(\vecr)$.

%----------------------------------------------------------------------------------------
%	SECTION 2
%----------------------------------------------------------------------------------------

\section{Training scheme}
Now that a parametrization is defined, a way to fit the weights of the neural network must
be devised. This new numerical scheme must also be able to solve the OZ equation, while
simultaneously finding the appropriate weights for $\nnet$.

\subsection{Cost function}
It was mentioned previously that the main problem is to find the weights of
$\nnet$ that can successfully solve the Ornstein-Zernike equation
for a given interaction potential.
To solve such a problem, a \textbf{cost function} must be defined, and be used as part of
a \emph{minimization} problem.

To define such a function, we consider the successive approximations obtained from the
iterative Piccard scheme to solve the OZ equation, $\left\{\gamma_1(\vecr), \gamma_2(\vecr), \dots, \gamma_n(\vecr)\right\}$.
From this, we expect to have found a solution when each approximation
is \emph{close enough} to the previous one. This can be translated into the following
cost function
\begin{equation}
    J(\theta) = \left[\gamma_{n}(\vecr, \theta) - \gamma_{n-1}(\vecr, \theta) \right]^2 ,
    \label{eq:costo}
\end{equation}
where $\gamma_{n}(\vecr, \theta)$ is the $n$-th approximation of the indirect
correlation function, $\gamma(\vecr)$.
The notation $\gamma(\vecr, \theta)$ indicates that the function now depends
on the weights of the neural network, as seen in \autoref{eq:parametrizacion}.
This means that, if the weights of $\nnet$ change, we should expect a change in the output
from the $\gamma$ function.

Another way of looking at \autoref{eq:costo} is that we require that the last 
two approximations of the $\gamma$ function in each iteration from the numerical scheme to 
be equal within a tolerance value, or upper bound. This will enforce a change on the 
weights every time both approximations deviate between them.

\subsection{Optimization problem}
With a cost function at hand, an optimization problem can be defined such that the
weights of $\nnet$ will be adjusted properly.

This optimization problem is in fact an \emph{unconstrained optimization problem},
and it is defined simply as

\begin{equation}
    \begin{aligned}
         & \underset{\theta \in \mathbb{R}^n}{\text{min}}
         & & J(\theta) .
    \end{aligned}
    \label{eq:optimizacion}
\end{equation}

This formulation is just a search for the best values of the weights that minimize
the squared difference between successive approximations. We denote these optimal values
as the \emph{minimizer}, or $\theta^{*}$, such that $J(\theta^{*})$ is a minimum.
This optimization problem can be solved iteratively, along with the solution of the
OZ equation, whose procedure to get a solution is also through an iterative process.

\subsection{Weight updates}
The iterative method employed to adjust the weights of $\nnet$ is based on the
\emph{gradient descent} method~\cite{nocedalNumericalOptimization2006}.
The most general update rule for a method based on gradient descent reads~\cite{goodfellowDeepLearning2016}
\begin{equation}
    \theta_{n+1} = \theta_n - \eta \nabla_{\theta} J(\theta) ,
    \label{eq:gradiente}
\end{equation}
where $\eta$ is known as the \emph{learning rate}, and it is a hyperparameter
that controls the step size at each iteration while moving toward the minimum
of a cost function. This value needs to be \emph{tuned} accordingly, so
that the method converges properly. By tuning the parameter we mean that this value
should change until the numerical scheme is stable enough, or provides the best
possible answer to the optimization problem.

Regardless the particular expression for the weight updates, every method
based on the gradient descent method \emph{requires} the gradient information from
the cost function with respect to the weights, $\nabla_{\theta} J(\theta)$.
In this particular case, the detailed computation of the gradient is described in
the \autoref{AppendixA}.
Once this information is obtained, all that is left is to build an algorithm that
can correctly use this training scheme and solve the OZ equation.

\subsection{Solving the Ornstein-Zernike equation with neural networks}
\label{subsec:oz-solution}

Having described all the necessary elements needed, a general layout for the solution
of the Ornstein-Zernike using neural networks is now presented.

Thus, we propose the following steps to solve the OZ equation using the parametrization
described by \autoref{eq:parametrizacion}:

\begin{enumerate}
    \item Given a particular interaction potential $\beta u(\vecr)$, \autoref{eq:parametrizacion} is used to obtain the value of the direct correlation function $c(\vecr)$. In this step, an initial value for $\gamma_{n}(\vecr)$ is needed, which is initialized based on the five-point Ng methodology shown in \autoref{AppendixB}. The weights of the neural network $\nnet$ are initialized randomly. The initialization method is discussed in detail in the next section.
    \item The newly found function $c(\vecr)$ is transformed to a reciprocal space by means of the Fourier transform yielding the new function $\hat{c}(\veck)$.
    \item Then, the full OZ equation, \autoref{eq:ornstein-zernike}, is Fourier transformed. Using the information from the previous step, a new estimation of the indirect correlation function is obtained, $\hat{\gamma}_{n+1}(\veck, \theta)$.
    \item The Fourier transform is applied once again to return all the functions to real space. With this operation, a new estimation $\gamma_{n+1}(\vecr, \theta)$ is computed from the transformed function, $\hat{\gamma}_{n+1}(\veck, \theta)$.
    \item Both estimations, $\gamma_{n}$ and $\gamma_{n+1}$, are used to evaluate \autoref{eq:costo}. In this step, the gradient $\nabla_{\theta} J(\theta)$ is computed as well.
    \item The weights $\theta$ are updated using a gradient descent rule, similar to \autoref{eq:gradiente}, and the process is repeated from step 1. In the next iteration, the initial value for the indirect correlation function will be $\gamma_{n+1}$, and a new estimation $\gamma_{n+2}$ will be obtained. This process is repeated until convergence.
\end{enumerate}

\subsection{Convergence criterion}
The procedure described in the previous section is repeated indefinitely until convergence
is achieved. This convergence criterion is defined as follows

\begin{equation}
    \sum_{i=1}^{N} {\left( \gamma^{n+1}_{i} - \gamma^{n}_{i} \right)}^2 \leq \epsilon .
    \label{eq:tolerancia}
\end{equation}

This expression is also known as the \emph{mean squared error}~\cite{goodfellowDeepLearning2016}.
Here, we sum all the $N$ elements of the squared difference between estimates $\gamma_{n+1}$
and $\gamma_{n}$. The parameter $\epsilon$ is a tolerance value that indicates an 
upper bound for the error between estimations. When the computed error is below this 
tolerance value, we consider the algorithm to \emph{have converged to a particular minimum}.
This means that the weights are adjusted until the successive estimations of the $\gamma$
functions are equal between them, up to the defined tolerance $\epsilon$.
Specifically, the numerical tolerance in all the experiments was fixed to be
$\epsilon = \num{1e-5}$ to allow for a robust exploration of the search space, without being
too restrictive. When using a lower value of $\epsilon,$ it was observed that the results 
were not improving at all (data not shown).

%----------------------------------------------------------------------------------------
%	SECTION 3
%----------------------------------------------------------------------------------------
\section{Implementation}
In this section we detail the most important aspects about the implementation of the
method described in the previous section. This includes the topology of the neural network,
the optimization method, and the choice of activation function. The physical parameters
as well as the computer simulations methods used to solve the OZ equation are also outlined.

\subsection{Choice of optimization algorithm}
The general rule for the weight update based on \autoref{eq:gradiente} was
implemented to solve the optimization problem, but numerical inconsistencies rendered this 
method unstable and convergence was almost never achieved.

To solve this issue, the \emph{Adam}~\cite{kingmaAdamMethodStochastic2017} optimization 
method was chosen. This optimization method is an excellent choice for the training
of neural networks, even more when the gradient is expected to be \emph{sparse}, i.e.
most of the elements of the gradient itself are zeros.
The \emph{Adam} method uses several rules to adjust the descent direction of the gradient,
as well as the hyperparameters related to the acceleration mechanism of the method.
Notably, there are two important hyperparameters used in the method; $\beta_1$,
which controls the moving average of the computed gradient; and $\beta_2$, which controls
the value of the gradient squared~\cite{kingmaAdamMethodStochastic2017}. Both parameters 
are necessary for the optimal convergence of the algorithm.

The equations that define the optimization method are the following
\begin{equation}
    \begin{aligned}
        m_{t} &= \beta_1 m_{t-1} - (1 - \beta_1) \nabla_{\theta_{t-1}} J(\theta_{t-1}) \\
        s_{t} &= \beta_2 s_{t-1} + (1 - \beta_2) \nabla_{\theta_{t-1}} J(\theta_{t-1}) \odot \nabla_{\theta_{t-1}} J(\theta_{t-1}) \\
        \hat{m}_{t} &= \frac{m_{t}}{1 - \beta_1^t} \\
        \hat{s}_{t} &= \frac{s_{t}}{1 - \beta_2^t} \\
        \theta_{t} &= \theta_{t-1} + \eta \hat{m}_{t} \oslash \sqrt{\hat{s}_{t} + \varepsilon}
    \end{aligned}
    \label{eq:adam}
\end{equation}
where $\odot$ is the elementwise multiplication, or Hadamard product;
$\oslash$
is the elementwise division, or Hadamard division;
and $\varepsilon$ is a smoothing value to prevent division by zero~\cite{hornMatrixAnalysis2012}.
The index $t$ represents each of the updates, or iterations, done by the algorithm.

In the results presented in this chapter, the parameters were fixed to the ones reported
as optimal in the original work of the \emph{Adam}
method~\cite{kingmaAdamMethodStochastic2017}, which are
$\beta_1=\num{0.9}$ and $\beta_2=\num{0.999}$. It is important to note that this method
has its own mechanisms to control and modify the gradients, as well as the hyperparameters.
This makes it a \emph{hands-off} method, without the need to tune the hyperparameters.
The \emph{learning rate}, $\eta$ in \autoref{eq:gradiente}, was fixed to
$\eta=\num{1e-4}$ for all the experiments because when a larger value was used the 
optimization method did not converge properly, and the results were not useful to be 
presented here. In a more practical situation, the best
way to choose the value of $\eta$ is to employ \emph{grid search} and look for the value
that minimizes the error the most~\cite{hastieElementsStatisticalLearning2009}.

\subsection{Neural network topology}

\begin{figure}[t]
    \includegraphics[width=\textwidth]{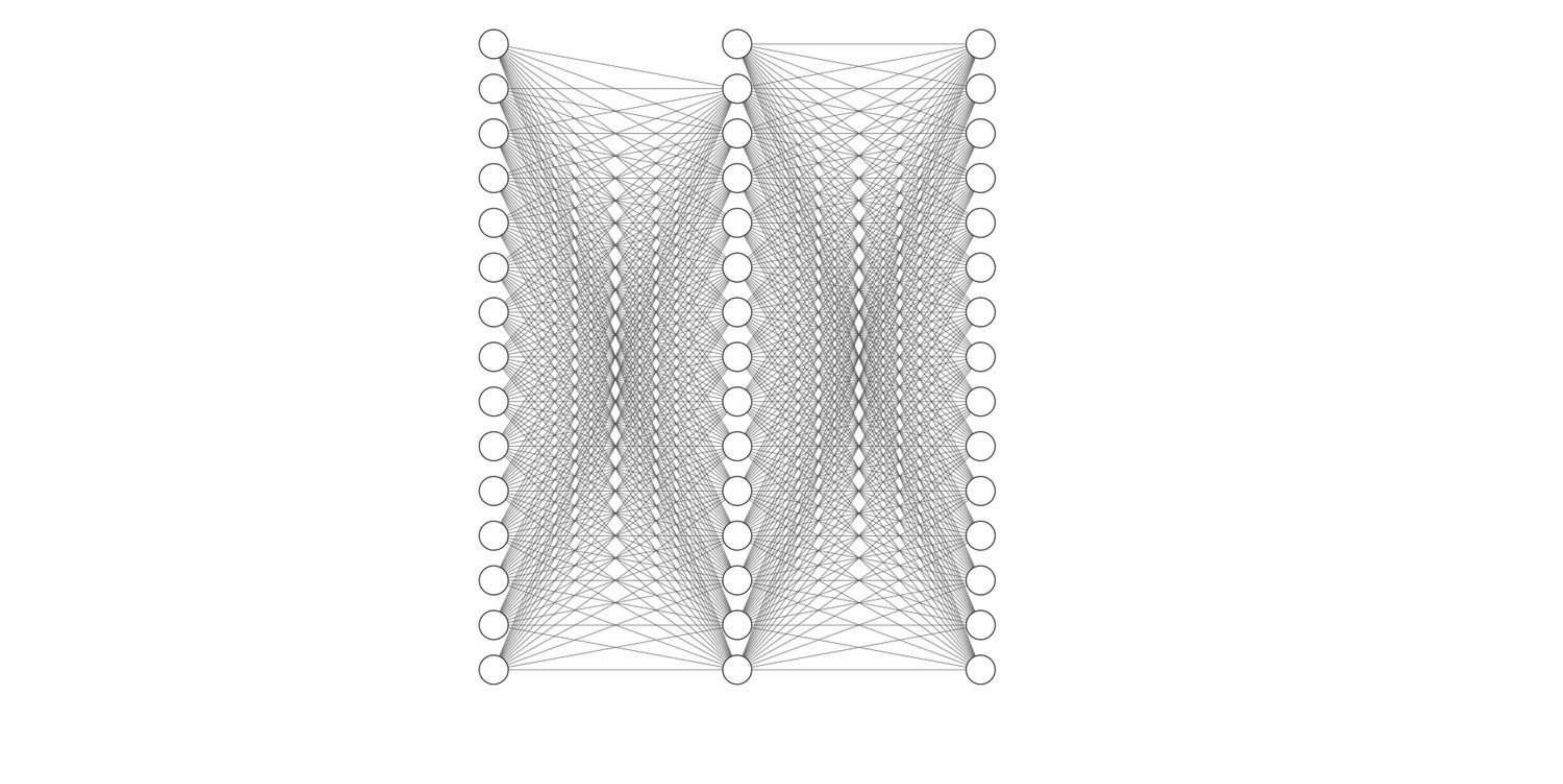}
    \vspace{-1.5cm}
    \caption[General schematics of a neural network.]{Cartoon of a fully connected multilayer neural network. Note that there is one \emph{hidden layer}. The circles represent the \emph{nodes} or \emph{units} used to compute the final output. These nodes are being evaluated by an activation function to account for nonlinearities. The top-most nodes that seem different from the main nodes are known as the \emph{bias} nodes. The real topology used in this chapter is larger, with many more nodes and connections, but the topology is the same.}
    \label{fig:nn-esquema}
\end{figure}

The neural network topology used in all the experiments is identical to the one
shown in \autoref{fig:nn-esquema}, with the exception of the number of nodes in each layer.
In particular, the neural network is made of \emph{three layers} fully connected between 
them.
There is an \emph{input} layer, one \emph{hidden} layer, and a final \emph{output} layer.
All layers have the same number of nodes, which is 4096. This size is set to be the same 
size as the number of points in the discretized distribution functions computed from the OZ 
equation. Additional nodes are added to the final two layers that serve as the \emph{bias} 
terms.

All the weights must be initialized appropriately, and in this case the Glorot uniform
distribution was used~\cite{glorotUnderstandingDifficultyTraining2010}, which has proven
to be an excellent way to help the convergence of neural networks.
When using the Glorot uniform distribution, the weights are initialized as
$
\theta_{ij} \sim \mathcal{U} \left[ -\frac{6}{\sqrt{(in + out)}},
\frac{6}{\sqrt{(in + out)}} \right]
$,
where $\mathcal{U}$ is the uniform probability distribution;
$in$ represents the number of units in the input layer; and $out$ the number of
units in the output layer. All bias nodes were initialized to be zero.

The activation function used was the \emph{ReLU}~\cite{glorotDeepSparseRectifier2011}
function, which has the form
\begin{equation*}
    \text{ReLU}(x) = \max{(0, x)} .
\end{equation*}

This activation function is applied to all the nodes in the layers, with the exception
of the input layer. This function was chosen due to the fact that the other most common
functions ($\tanh, \text{softmax}$, etc.) were numerically unstable in the training process 
of the neural network (data not shown).

\subsection{Physical parameters and simulations}

To solve the OZ equation a cutoff radius of $r_c=7\sigma$ was used, where $\sigma$ is the
particle diameter and it was fixed to be $\sigma=1$.
The interaction potential used was the pseudo hard sphere potential as defined 
in~\autoref{eq:cont-hs}, both for the solution of the OZ equation, as well as the results 
obtained from computer simulations.

Seven different densities were explored in the range $\phi \in [\num{0.15}, \num{0.45}]$, 
with $\Delta \phi = \num{0.05}$.
For each density value, a grid of 70 points was used to ensure convergence of the iterative
algorithm when solving for the OZ equation. This was not the case for the computer 
simulations, where such partition is not needed.

Computer simulations results were obtained using the traditional Monte Carlo simulation
method within the Metropolis scheme for the $NVT$ ensemble. In every experiment, the total 
number of particles was 2197,
the system was equilibrated for a total of 10 million Monte Carlo steps, and the radial
distribution functions were obtained from the sampling of 7 million Monte Carlo steps, after
the system was equilibrated. To reduce the number of computations, a cutoff radius of
half the size of the simulation box was used for the evaluation of the interaction 
potential. Periodic boundary conditions in all spatial dimensions were used accordingly. 
The same pseudo hard sphere potential, \autoref{eq:cont-hs}, was used instead
of the true hard sphere potential, for a fair comparison with the results obtained from the 
OZ equation.

%----------------------------------------------------------------------------------------
%	SECTION 4
%----------------------------------------------------------------------------------------
\section{Results}
It is now time to investigate the results obtained from the proposed methodology, using all the elements previously described.
The main point of discussion will be the radial distribution
function \textemdash $g(r^*)$ \textemdash for different values of densities, both in the
low and high density regimes.

\subsection{Low densities}
\begin{figure}[t]
    \centering
    \includegraphics[width=\textwidth]{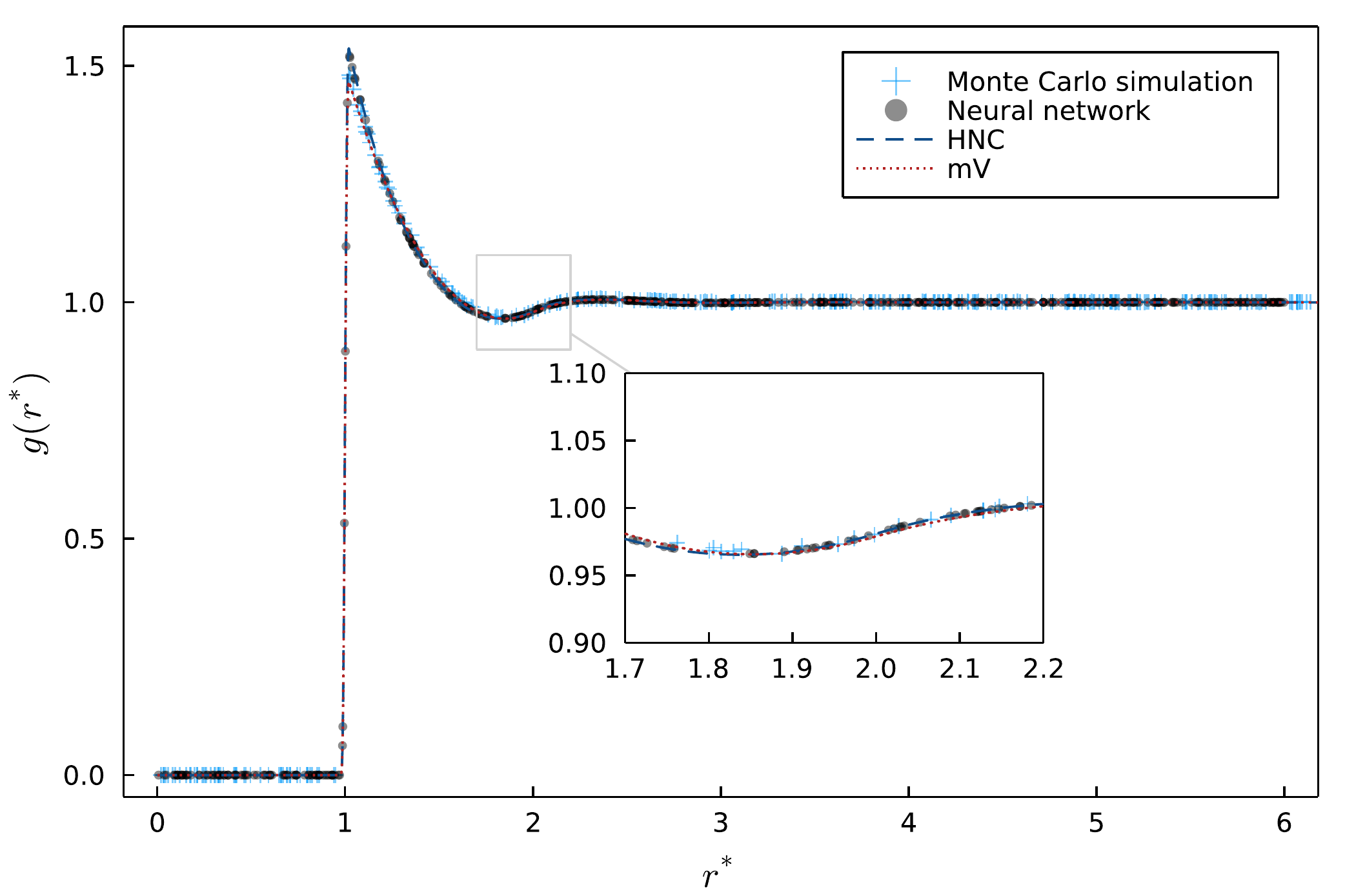}
    \caption[Radial distribution function, $\phi=0.15$.]{Radial distribution function for $\phi=0.15$ obtained from Monte Carlo simulations, and three different approximations:
    \begin{enumerate*}[label=(\alph*),itemjoin={,\enspace}]
        \item \emph{mV}, (modified Verlet)
        \item \emph{HNC}, (Hypernetted Chain)
        \item \emph{NN}, (neural network approximation).
    \end{enumerate*}
    Inset shows the region close to the peak about $r^{*}=2$.
    }
    \label{fig:rdf15}
\end{figure}

\begin{figure}[t]
    \centering
    \includegraphics[width=\textwidth]{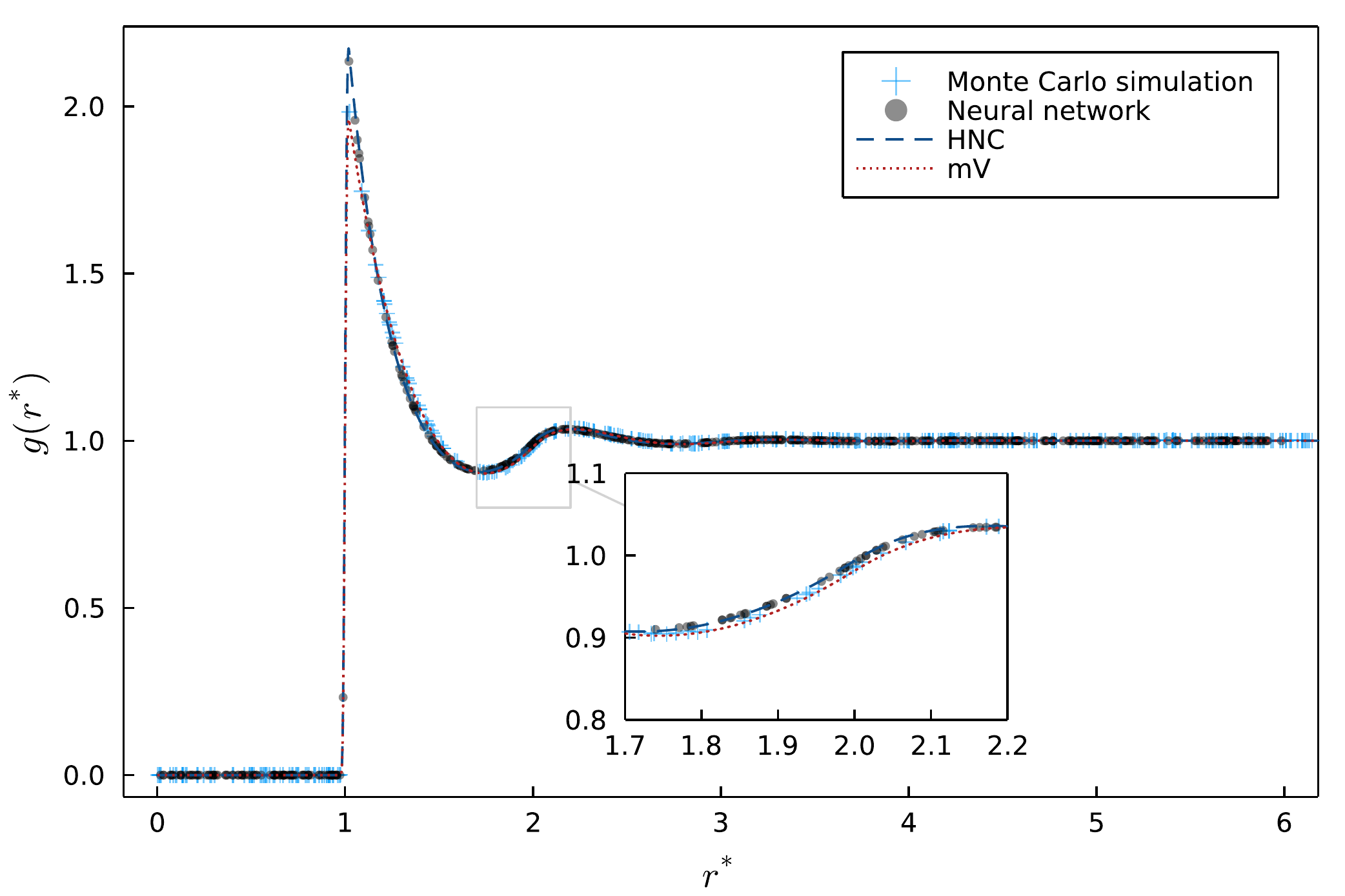}
    \caption[Radial distribution function, $\phi=0.25$.]{Radial distribution function for $\phi=0.25$ obtained from Monte Carlo simulations, and three different approximations:
    \begin{enumerate*}[label=(\alph*),itemjoin={,\enspace}]
        \item \emph{mV}, (modified Verlet)
        \item \emph{HNC}, (Hypernetted Chain)
        \item \emph{NN}, (neural network approximation).
    \end{enumerate*}
    Inset shows the region close to the peak about $r^{*}=2$.
    }
    \label{fig:rdf25}
\end{figure}

In this section we will deal with the low 
density values ranging from $\phi=\numlist[list-pair-separator={\enspace\text{to}\enspace}]{0.15;0.25}$, which are shown in \autoref{fig:rdf15} and \autoref{fig:rdf25}, respectively.
The results show that, at low densities, the HNC and neural network approximations are
more precise than the modified Verlet approximation. Although at first glance, all
approximations seem to fall short compared to computer simulations. This is particularly 
noticeable in the neighborhood around the second peak, which is shown in the insets 
of \autoref{fig:rdf15} and \autoref{fig:rdf25}.
Additionally, it is important to note that the neural network approximation is slightly
more precise than the HNC approximation, which can be qualitatively appraised by observing 
the estimation of the main peak in the radial distribution function. This peak can be found 
in the vicinity of $r^* = 1$. Nevertheless, it is still overestimated, which is the 
same case for the HNC approximation. However, this is not the case for the modified Verlet 
approximation, which underestimates the main peak.

In like manner, the functional form of $g(r^*)$ is important to study closely.
For the HNC and neural network approximations, it appears to have the same form between 
both approximations, and it might as well be the same.
This would imply that, somehow, the weights of the neural network
were updated enough such that a minimum was found, and this minimum was very close to the
HNC approximation. In other words, the results suggest that the weights are very close to
zero, such that when the neural network is evaluated, the output is close to the
result obtained from the HNC approximation.
Another important aspect to observe is that this functional form is marginally different
to the one seen from computer simulations, and that the modified Verlet approximation is 
closer to the form found in the computer simulations results.

\subsection{High densities}
\begin{figure}[t]
    \centering
    \includegraphics[width=\textwidth]{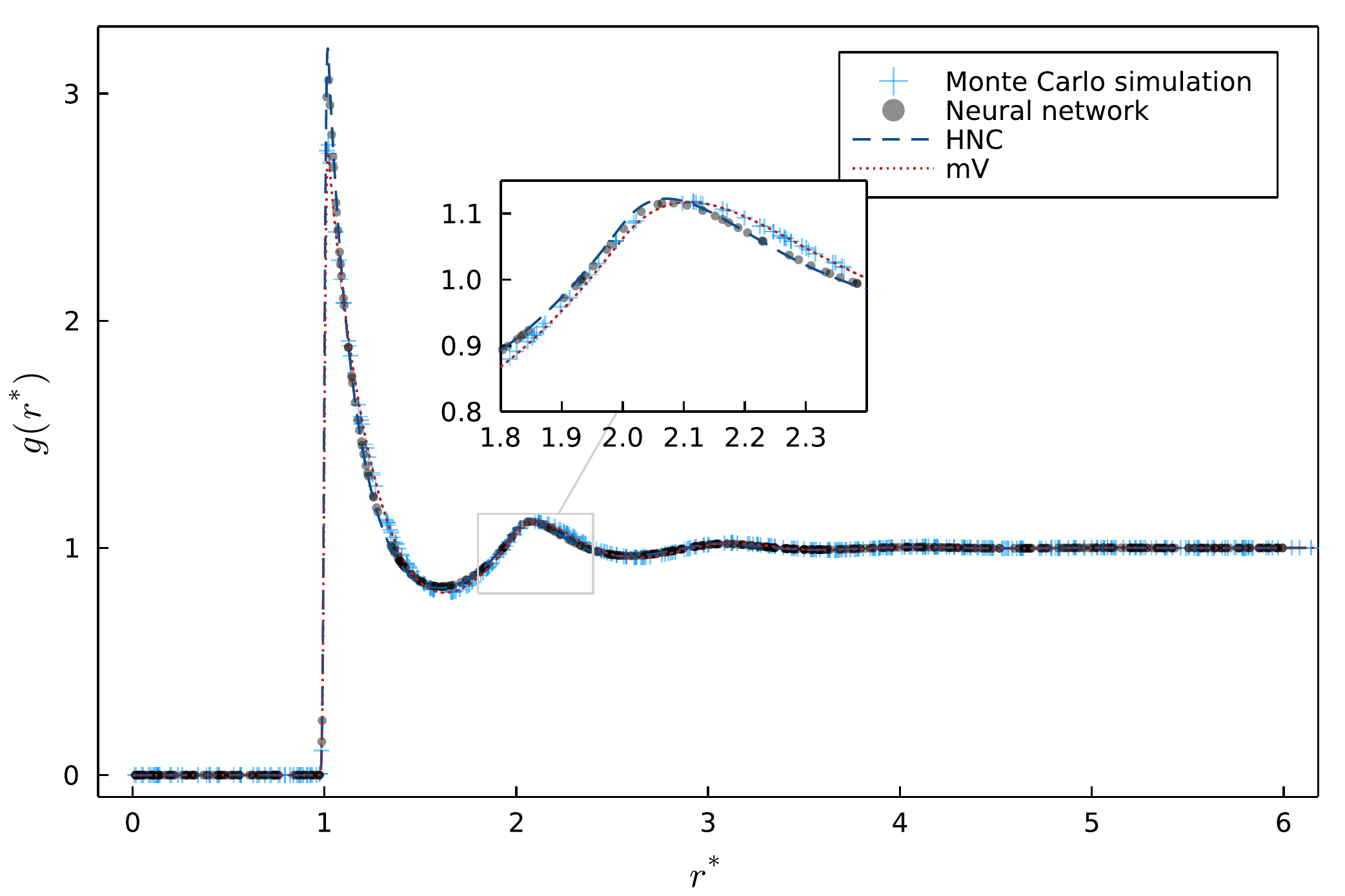}
    \caption[Radial distribution function, $\phi=0.35$.]{Radial distribution function for $\phi=0.35$ obtained from Monte Carlo simulations, and three different approximations:
    \begin{enumerate*}[label=(\alph*),itemjoin={,\enspace}]
        \item \emph{mV}, (modified Verlet)
        \item \emph{HNC}, (Hypernetted Chain)
        \item \emph{NN}, (neural network approximation).
    \end{enumerate*}
    Inset shows the region close to the peak about $r^{*}=2$.
    }
    \label{fig:rdf35}
\end{figure}

\begin{figure}[t]
    \centering
    \includegraphics[width=\textwidth]{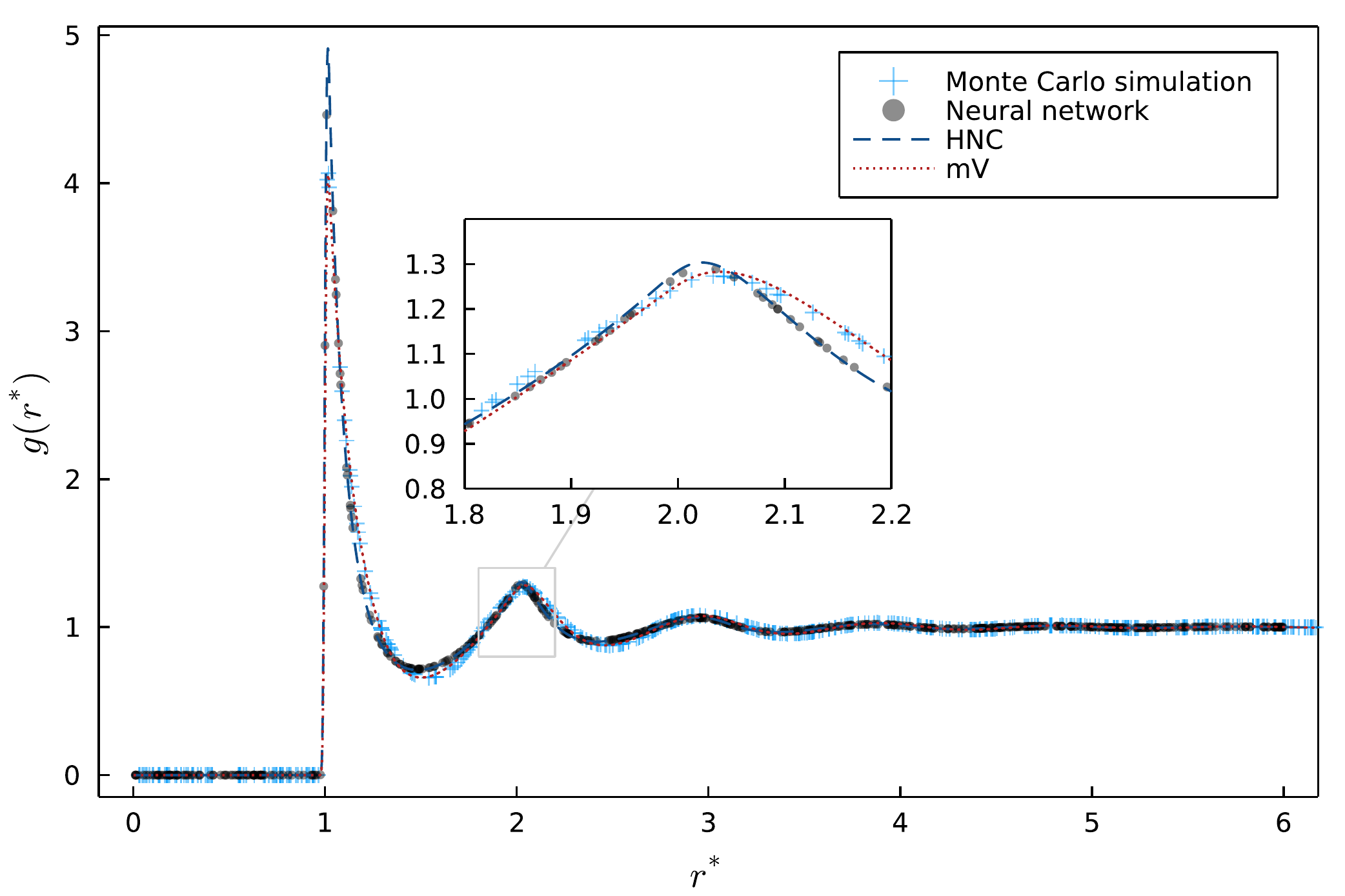}
    \caption[Radial distribution function, $\phi=0.45$.]{Radial distribution function for $\phi=0.45$ obtained from Monte Carlo simulations, and three different approximations:
    \begin{enumerate*}[label=(\alph*),itemjoin={,\enspace}]
        \item \emph{mV}, (modified Verlet)
        \item \emph{HNC}, (Hypernetted Chain)
        \item \emph{NN}, (neural network approximation).
    \end{enumerate*}
    Inset shows the region close to the peak about $r^{*}=2$.
    }
    \label{fig:rdf45}
\end{figure}

We now turn our attention to the high density values, namely,
$\phi=\numlist[list-pair-separator={\enspace\text{and}\enspace}]{0.35; 0.45}$,
represented in \autoref{fig:rdf35} and \ref{fig:rdf45}.
In the same spirit as before with the low densities, the HNC and neural network 
approximations are not precise when compared to computer simulations. In this case,
the modified Verlet bridge function approximation is even more precise, which was expected.
This is because the HNC approximation is a very good approximation for long range
interaction potentials~\cite{hansenTheorySimpleLiquids2013}, whereas the modified Verlet is 
better suited for short range potentials, such as the one studied here.
In this case, modified Verlet is the most precise of the approximations used, which
can be inspected in \autoref{fig:rdf35} and \autoref{fig:rdf45}, where the 
main peak is accurately estimated by the approximation when compared to Monte Carlo 
computer simulation results. However, both HNC and neural network approximations 
overestimate this quantity.

Further, the functional form of $g(r^*)$ computed with the neural network approximation 
is substantially different to the one obtained with computer simulations. Indeed, the result
obtained is similar to the one obtained with the HNC approximation, and both are imprecise 
approximations to the expected functional form; this was also the case for low densities.
This result is important, backing the hypothesis that the
neural network might reduce to the HNC approximation.
This would imply that the neural network is in fact approximating the bridge function
$B(\vecr) \approx 0$. If we now pay attention to the modified Verlet
approximation, again in \autoref{fig:rdf35} and \autoref{fig:rdf45}, 
we can see that the modified Verlet bridge function is the most precise
out of all the set of bridge functions used. In other words, we observe that this
estimation provides a precise prediction of the main peak, as can be seen when compared to 
the results obtained from computer simulations, which are almost identical.

\section{Discussion}
It would seem as though the neural network approximation reduces to the HNC
approximation, as seen in the results from the previous section. In this section
we shall investigate this matter in detail.
We will also continue the discussion of the results presented and try to make sense
of the training dynamics of the neural network. This is an important topic to
address due to the clear results that the neural network provides almost the same
result as the HNC approximation.

\subsection{Weight evolution of the neural network}

\begin{figure}[t]
    \includegraphics[width=\textwidth]{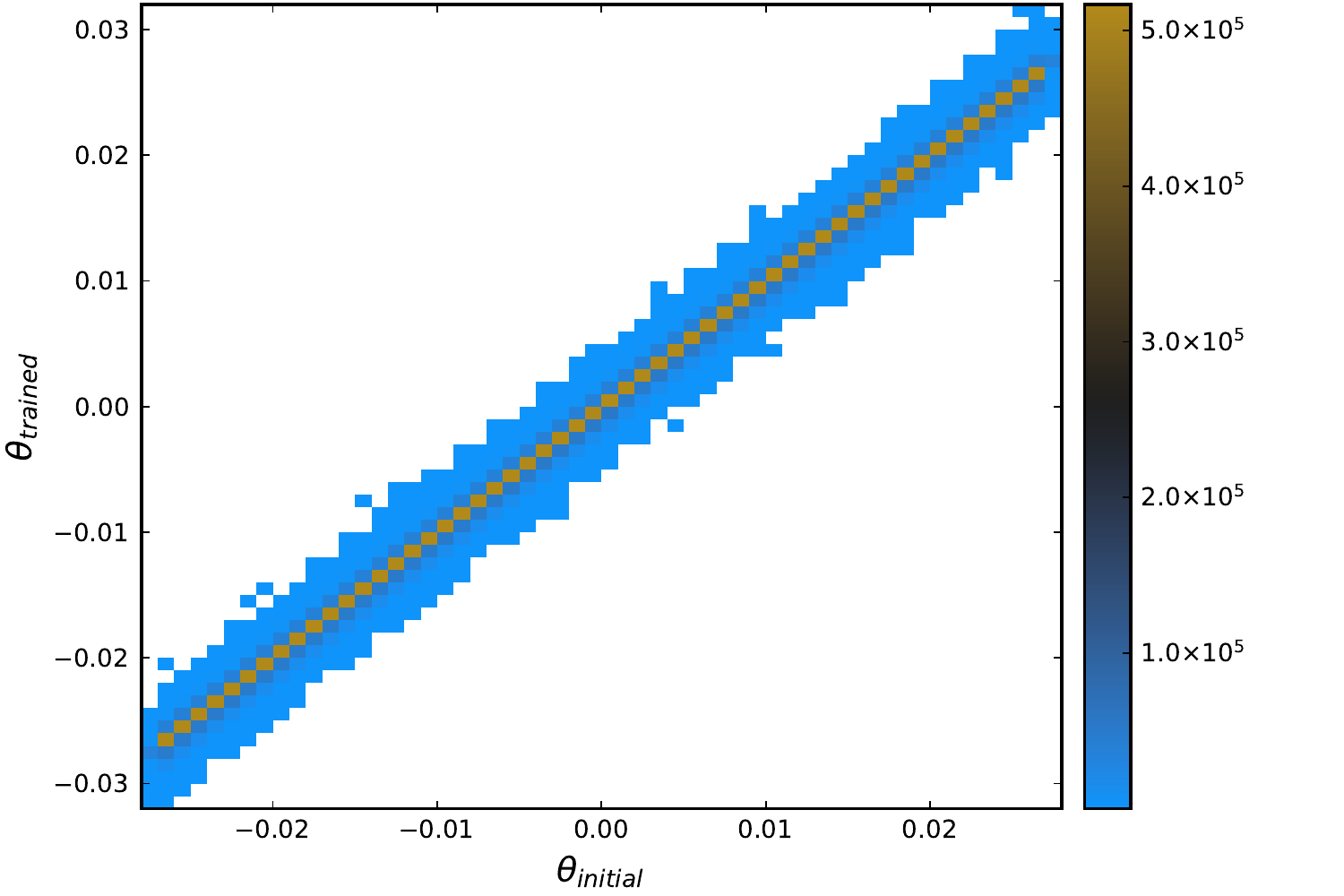}
    \caption[Comparison between weights, $\phi=0.15$.]{Relation between the trained weights and the initial weights of $\nnet$ for $\phi=0.15$. The scale on the right-hand side represents the total number of instances for the trained-initial pair of weights.} 
    \label{fig:pesos15}
\end{figure}

\begin{figure}[t]
    \includegraphics[width=\textwidth]{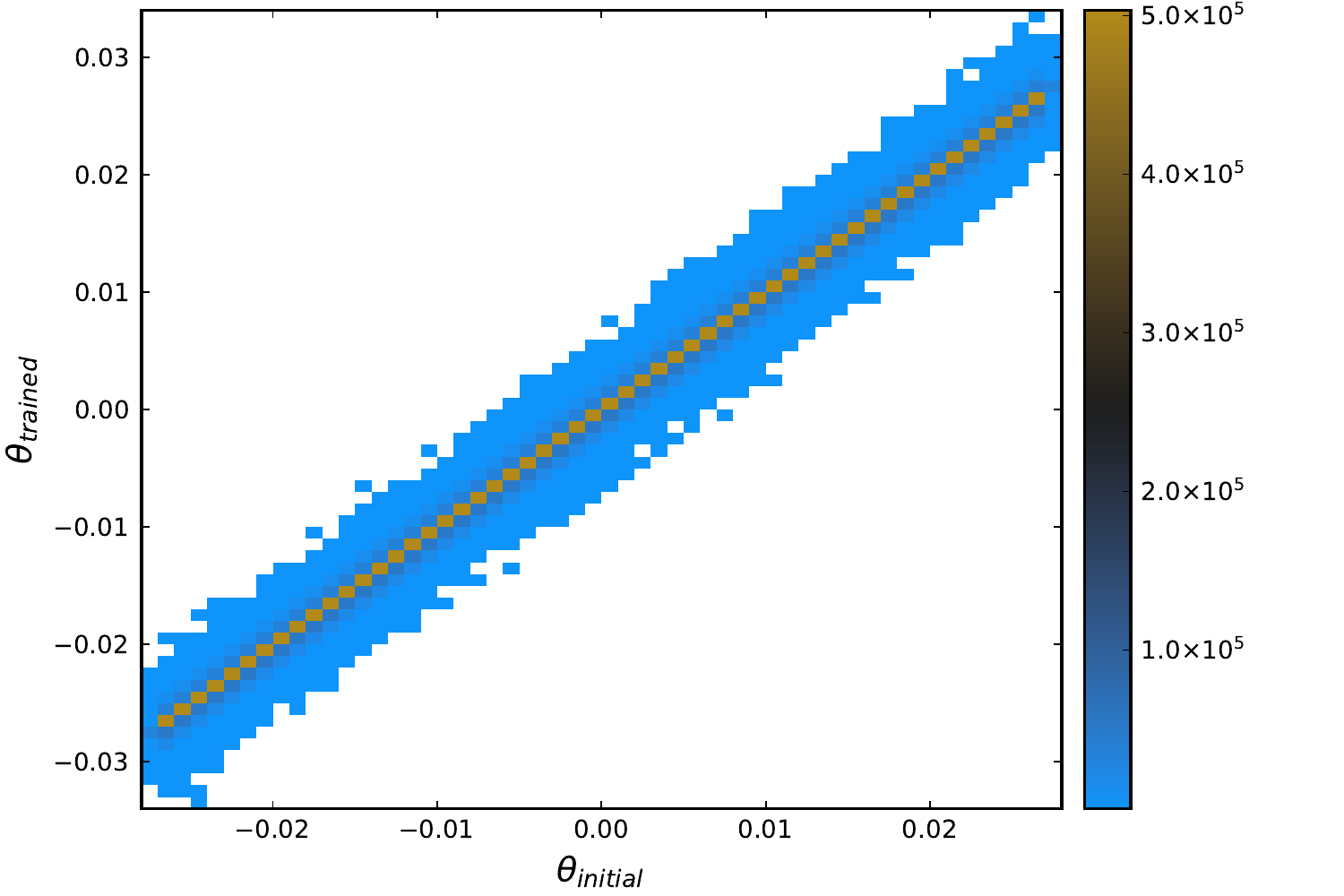}
    \caption[Comparison between weights, $\phi=0.25$.]{Relation between the trained weights and the initial weights of $\nnet$ for $\phi=0.25$. The scale on the right-hand side represents the total number of instances for the trained-initial pair of weights.}
    \label{fig:pesos25}
\end{figure}

\begin{figure}[t]
    \includegraphics[width=\textwidth]{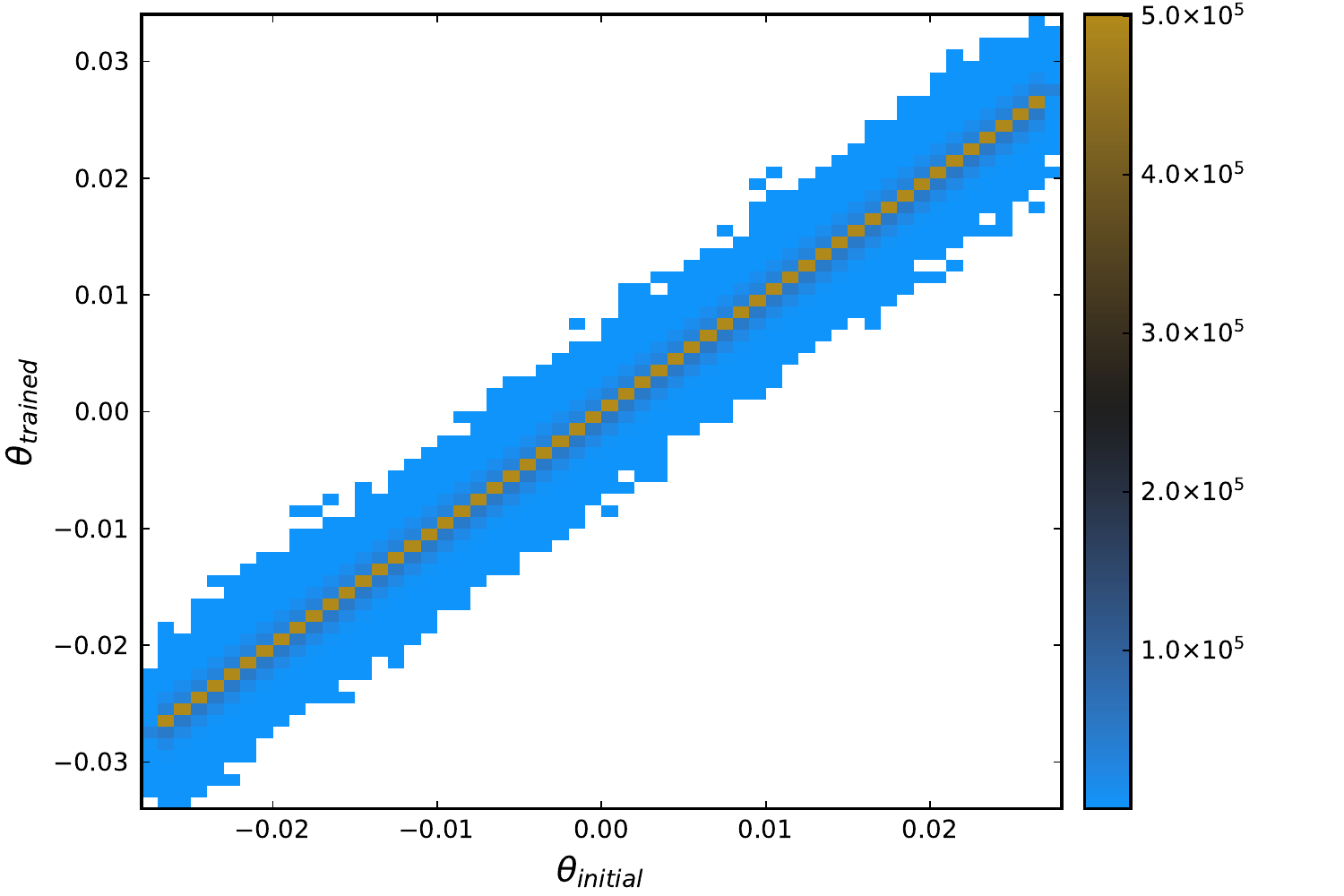}
    \caption[Comparison between weights, $\phi=0.35$.]{Relation between the trained weights and the initial weights of $\nnet$ for $\phi=0.35$. The scale on the right-hand side represents the total number of instances for the trained-initial pair of weights.}
    \label{fig:pesos35}
\end{figure}

\begin{figure}[t]
    \includegraphics[width=\textwidth]{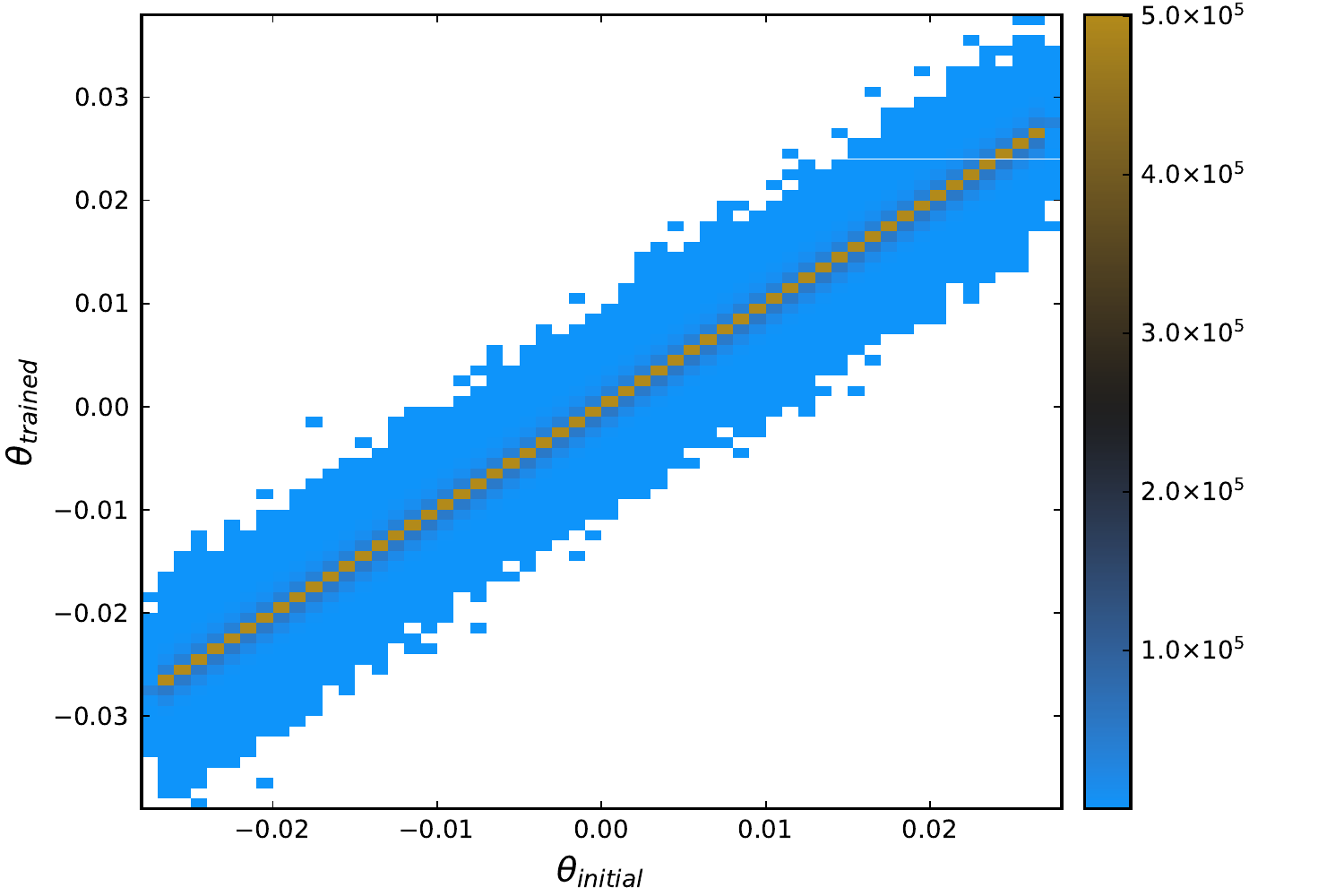}
    \caption[Comparison between weights, $\phi=0.45$.]{Relation between the trained weights and the initial weights of $\nnet$ for $\phi=0.45$. The scale on the right-hand side represents the total number of instances for the trained-initial pair of weights.}
    \label{fig:pesos45}
\end{figure}

We shall now examine the evolution of the weights $\theta$ from
$\nnet$, from the moment it was initialized to the moment its training finalized.
A histogram of this for the density values 
$\phi=\numlist[list-final-separator={\; \text{and} \;}]{0.15;0.25;0.35;0.45}$ 
can be seen in \autoref{fig:pesos15}, \autoref{fig:pesos25}, \autoref{fig:pesos35}, 
and \autoref{fig:pesos45}, respectively.
We can observe that the way the weights show a diagonal represent a linear relationship 
between the initial weights, $\theta_{i}$, and the trained weights, $\theta_{t}$. In other 
words, the weights follow the linear expression
$\theta_{t} = \alpha \theta_{i} + \beta + \epsilon$, with
$\epsilon \sim \mathcal{N}(\mu, \sigma^{2})$ a normal random variable with mean
$\mu$ and variance $\sigma^2$. The noise term can be any other continuous probability 
distribution, but without loss of generality the normal distribution was chosen for
our purposes. For now, we are not interested in the values of $\alpha$ or $\beta$,
but merely on the linear relationship between them.

One thing to notice is the fact that the higher the density value is, the larger
the variance turns out to be. If we observe the variance for the density
$\phi=0.15$ in \autoref{fig:pesos15}
we see that the variance is small due to the fact that the blue shaded region around
the diagonal is close to it. If we now see the same \autoref{fig:pesos45} for the 
density value of $\phi=0.45$ we observe that this shaded region is significantly larger.
This would mean that, at higher densities, the weights of $\nnet$ are more spread out
from the mean, and the neural network might have adjusted its weights to account for
different computations of the bridge function.

The most interesting part of this is the fact that the weights from initialization
do not change much throughout the training scheme, which would imply that a local minimum
has already been found. This might be the case, because HNC is actually a solution of
the OZ equation, and solutions around this particular approximation might as well be
solutions themselves.
This, however, does not answer the question of why the spread is larger when
higher densities are inspected.

\subsection{The Hypernetted Chain approximation as a stable minimum}
It would seem that the way the weights are updated, albeit with minimal change from its
initial values, is due to the fact of already being near a minimum when the training starts.
We must recall that the weight update and neural network training is essentially an
optimization problem, and the main goal is to find a minimum
of the cost function in \autoref{eq:costo}. With the results presented so far, it might be
possible to postulate that the \emph{HNC approximation is a stable minimum} for the
neural network $\nnet$.
This would answer the question of why the weights of the neural network during training
explored in the previous section did not change very much throughout the numerical scheme.
Because if we have already found a minimum, the optimization algorithm might end up
oscillating in the proximity of this value.

On the other hand, this idea could also give answer to the question of why the spread
is larger for higher density values. If we pay close attention to the neural network bridge
approximation results for the \emph{low density} values in \autoref{fig:rdf15}, we can see 
that although all the bridge functions give a low accuracy estimation of the second peak as shown in the inset within the figure. However, for the main peak the neural network 
approximation is accurate.
If we now observe \autoref{fig:rdf45}, which refers to the \emph{high density} value,
we can see that the estimation is a poor one.

Let us now relate this to the weight evolution. For the \emph{low density} regime, the 
weight evolution has a \emph{lower variance}; for the \emph{high density} regime, a \emph{higher variance} is observed in the weight evolution.
This suggests that, for \emph{lower density} values, there was no need to adjust the
weights more than shown in \autoref{fig:pesos15} because the approximation is accurate
enough. However, for the \emph{higher density} values, the approximation is not good enough
and the optimization method was trying to adjust the weights accordingly, even if
unsuccessfully.
Thus, by oscillating near the value of zero, which represents the HNC bridge function,
the neural network does not need additional information and generates a bridge function 
approximation that reproduces the results from the HNC closure relation.

Having a stable minimum when training starts would mean that the neural network does not
learn enough, and will alway keep its weights tightly centered about the mean of this
minimum. Still, this implies that other minima are available for the neural network as long
as the weights are correctly initialized, or a probability distribution centered about
a particular minima is used.

\subsection{Does the neural network reduce to HNC?}
For the low density regimes, HNC is an accurate approximation for the interaction potential.
Hence, the neural network is an accurate approximation. On the contrary, for high density
regimes, both approximations fail to provide an accurate solution.

If the neural network is in indeed oscillating about zero (the HNC approximation), then
it makes sense that both estimations give the results observed. Yet, we cannot guarantee
by any means possible that the neural network reduces to the HNC approximation.
We only possess \emph{statistical evidence} from the training dynamics that the neural
network weights do not change much throughout its training.

This observation might shed light into possibilities of changing the way the neural
network propagates its values and return an output. For example, a modification to the
neural network topology might be in order, such that introducing important 
nonlinearities that are consistent with the physical properties of the system can yield
better results. For the case of hard spheres, the work by Malijevský and Labík~\cite{malijevskyBridgeFunctionHard1987}
shows that the bridge function has particular functional properties, such that the
bridge function is non-negative and oscillating, among others. These properties can then
be consolidated within the neural network structure and investigate if the neural network 
weights change considerably.
From this, one might expect two outcomes. Firstly, the case where the \emph{weights change},
which would indicate that adding nonlinearities according to some aspect of the interaction 
potential prove beneficial for the weight update and training dynamics.
Secondly, the case where the \emph{weights do not change}, in which case we might
be able the have stronger evidence that, regardless of the neural network structure,
this kind of approximation will have a high chance of reproducing the HNC results.
In any case, with either outcome we do not have the information to assess if the neural
network might actually be \emph{more precise} than it is in its current form.

Similar results were found by Goodall and Lee~\cite{a.goodallDatadrivenApproximationsBridge2021}
while using data from simulations. In this approach, a data set was built with several
correlation functions that came from physical properties of the liquid. If the data set was built
just using the indirect correlation function $\gamma(\vecr)$, the neural network trained from this data
set would yield results similar or worse to those obtained with the HNC approximation.
So, in some sense, the proposed methodology here might actually be better than a fully
data-driven methodology. However, this just makes a stronger case for the argument that,
indeed, the neural network might reduce to the HNC approximation and not have enough
information about the system to adjust the weights properly.

\subsection{Neural networks as random polynomial approximations}
An interesting result from this investigation is the fact that a random approximation
provides a solution to the OZ equation. In what way is it random? Take the weights of the
neural network to be the coefficients of some \emph{random polynomial}, i.e. a polynomial 
with coefficients that come from some probability distribution $P$. This is something that 
is not trivial on why it could work as a bridge function approximation, even if the result
is very close to the HNC approximation. This would imply that, for the probability 
distribution $P$ with some finite variance $\sigma^2$ and mean $\mu=0$, the resulting
polynomial would yield a bridge function estimation similar to the HNC closure relation.
Yet, the reason why a random approximation might be a solution to the OZ equation is not
clear.

To understand the significance of this, we must recall that the bridge function can, in 
fact, be understood as a
power series in density~\cite{hansenTheorySimpleLiquids2013},
\begin{equation}
    b(\vecr) = b^{(2)}(\vecr) \rho^2 + b^{(3)}(\vecr) \rho^3 + \cdots ,
    \label{eq:expansion-densidad}
\end{equation}
where the notation $b^{(n)}(\vecr)$ indicates the $n$-particle bridge function, i.e. the
estimation of the bridge function for $n$ interacting particles. It is specially
important to note that the coefficients $b^{(n)}(\vecr)$ are, in general, high-dimensional
integrals that are mostly \emph{intractable}. Almost always, for a given bridge function
approximation, numerical methods are in order if these coefficients are to be determined.
For instance, the work by Kwak and Kofke~\cite{kwakEvaluationBridgefunctionDiagrams2005}
uses a Monte Carlo sampling numerical method to evaluate up to $b^{(4)}(\vecr)$ 
for the hard-sphere fluid. They report that even if the numerical computation is
possible, convergence is slow and computationally costly.

Only for special cases, these coefficients can be determined in closed form, e.g.
for the Percus-Yevick approximation the value of
\[
b^{(2)}(\vecr) = -\frac{1}{2}{\left[ g_{1}(\vecr) \right]}^2
\]
is known from diagrammatic methods~\cite{hansenTheorySimpleLiquids2013}. In this
expression, $g_{1}(\vecr)$ represents the single-particle radial distribution function.

Let us now understand the role of \emph{random polynomials} and their relation to the neural
network bridge function approximation used in this investigation.
Let $p_n$ be an algebraic polynomial of the form
\begin{equation}
    p_n(z) = a_0 + a_1 z + a_2 z^2 + a_3 z^3 + \cdots + a_n z^n, \quad
    z \in \mathbb{C}
    \label{eq:random-poly}
\end{equation}
where the coefficients $a_0, a_1, \dots , a_n$ are independent real-valued random variables
with finite mean and finite variance. This kind of polynomials have found successful
applications in some areas of physics~\cite{houghZerosGaussianAnalytic2009}.
These polynomials have also been the interest of mathematical research~\cite{edelmanHowManyZeros1995}.

Now, by means of the universal approximation theorem, we can regard the neural network
as a power series similar to \autoref{eq:random-poly}. To see this more clearly, we
can think of the weights of the neural network as some coefficients of a power series,
obtained under some transformation that takes in the weights and return the coefficients
needed to build a power series.
Further, if we compare directly \autoref{eq:random-poly}
and \autoref{eq:expansion-densidad}, we can see that these expressions
are related through their coefficients, with the exception of the first two terms, implying 
that random variables might be able to give an answer to the $n$ particle bridge functions 
in the power series.

Indeed, this is a convenient way of estimating the coefficients of the bridge function,
when expressed in a power series of the density or any other correlation functions.
Although, the practical way to estimate these coefficients is to enforce thermodynamic
self-consistency. Such is the case of the work by Vompe and Martynov~\cite{vompeBridgeFunctionExpansion1994},
and the work by Tsednee and Luchko~\cite{tsedneeClosureOrnsteinZernikeEquation2019}, where
the coefficients are found through a minimization problem when the thermodynamic consistency
among different routes has been achieved.

After all, the insight of random polynomials might pave the way for a novel way of 
computing coefficients by using neural networks, or related probabilistic methods.
Even though there is no way to enforce thermodynamic self-consistency when using such
methods, it is interesting to see the striking resemblance with these approaches.
And yet, one of the downsides of this is the fact that the probability
distribution for the coefficients $a_0, a_1, \dots , a_n$ cannot be known even through the
training dynamics of the neural network. A naive way of acquiring the underlying 
probability distribution is to suppose there is
a unique probability distribution and estimate its mean and variance by maximum likelihood
estimation techniques~\cite{hastieElementsStatisticalLearning2009}.
Still, even if there could be a relation between random polynomials and the bridge 
function, there can be no guarantee that the resulting approximation is good enough, or 
that it could reproduce the physics of the problem properly.
This is just mere speculation that a deeper relationship between the neural network 
approximation and the bridge function might be exploitable through the lens of random
polynomials and probability theory.

\section{Concluding remarks}
Even though neural networks can approximate any continuous function,
it is up to the implementation that uses them
to benefit from the underlying domain knowledge of the problem to be solved. In this case,
even if the methodology created is \emph{theoretically possible}, it raises the question of
whether the neural networks are the most suitable solution for the OZ equation.
Such might be the case for the methodology presented here. After all, the results
seem to \emph{strongly suggest} that a neural network reduces to one of the classic 
approximations,
the Hypernetted Chain approximation, which is not the best approximation for all kinds
of interaction potentials, and most certainly is not the case of the pseudo hard sphere
and hard sphere potentials.
At any rate, these results show a promising application of such models, in the form of
understanding how or why certain bridge function approximations provide a solution to the 
OZ equation. Unlike the use of fully data-driven methodologies, like the one from
Goodall and Lee~\cite{a.goodallDatadrivenApproximationsBridge2021},
the proposed methodology might somehow be able to provide intuition into what is happening
with the bridge function itself without the need of other theoretical methods.

One of the main drawbacks of the current proposal is the fact that the number of weights
is too large to effectively train well. This might be a reason of why the spread was
small in the weight evolution of $\nnet$. A way to alleviate this problem is to use
dimensionality reduction techniques such as
Principal Component Analysis~\cite{hastieElementsStatisticalLearning2009}. 
These methods can adequately reduce the total number of weights to be trained, without 
losing too much information from the learning dynamics.

In order to use the physics of the problem to our advantage, a different optimization
problem might be formulated, in which case a thermodynamic consistency cost function
might be able to drive the learning dynamics. In this framework, it is expected to minimize 
the difference between two different routes that compute the pressure, or the isothermal
compressibility, similar to the approaches by Zerah and Hansen~\cite{zerahSelfConsistentIntegral1986}, 
as well as Rogers and Young~\cite{rogersNewThermodynamicallyConsistent1984b}.
A deficiency of such scheme is that the cost function will be
a highly nonlinear function, and possibly a \emph{black-box function}, in which case
the gradient of the function might be hard or even impossible to compute in a timely manner.
This affects not only the computation time but the ability to use gradient descent methods,
or any other optimization method that uses gradients to find an extremum.
Nevertheless, such approach could be a dramatic improvement over the methodology presented
here, even more so if coupled with a dimensionality reduction technique.

In closing, neural networks are capable models for the approximation of any continuous function, 
but domain knowledge of the problem is needed for these models to succeed when no data set 
is available. In the proposal presented in this chapter, we wanted to investigate two things, 
whether neural networks could solve the OZ equation and their quality of approximation. 
It was shown that, indeed, neural networks can solve the OZ equation, but without more 
information on the system itself, neural networks do not generalize well and their training 
dynamics suffer greatly. The bridge function approximation that these models provide is
implied to be as accurate as the Hypernetted Chain bridge function.
The information that these models gather throughout the training 
is minimal, but their training dynamics shed some light into the possibilities of using 
probability methods to approximate bridge functions in liquids. %% Neural networks
\chapter{Evolutionary optimization for the Kinoshita closure}
\label{Cap5}

In the previous chapter, a neural network was used to approximate the bridge function and 
use it in the solution of the Ornstein-Zernike (OZ) equation. From the corresponding 
results, it was implied that, without any more physical information, the neural network 
reduced its approximation to the well-known Hypernetted Chain closure relation. In this 
chapter, the modified Verlet bridge function is introduced, along with its variation, the 
Kinoshita closure. This bridge function contains more information in terms of the indirect 
correlation function, \(\gamma(r)\), which provides a more accurate solution when dealing 
with the hard-sphere fluid. Still, the Kinoshita closure is not thermodynamically 
consistent. Thus, in this chapter, the proposal of parametrizing the Kinoshita is 
introduced. These parameters are then fitted using Evolutionary Computation (EC) method for 
derivative-free optimization tasks. This is in the same spirit as with the 
Rogers-Young~\cite{rogersNewThermodynamicallyConsistent1984b} and the 
Zerah-Hansen~\cite{zerahSelfConsistentIntegral1986} closure relations. The goal of this 
proposal is to alleviate the problems that using a neural network has, in terms of 
incorporating more Physics into the modeling, instead of relying on the neural network to 
learn something by its own.

\section{Kinoshita closure and its parametrization}
The main goal in this chapter is to deal with the modified Verlet closure and a variation 
of it, the Kinoshita modification. Both of these closure relations have been formulated to 
deal with the hard-sphere fluid, for low and large density values. The purpose of this 
section is to look at these closure relations more closely, as well as to define the 
parametrization to be used in later computations and results.
The pseudo hard-sphere model from~\autoref{eq:cont-hs} is used for all the computations 
done in this chapter.

\subsection{Modified Verlet bridge function approximation}
Between 1980 and 1981, Loup Verlet published two papers where he introduced what is now 
called the \emph{modified Verlet} bridge 
function~\cite{verletIntegralEquationsClassical1980,verletIntegralEquationsClassical1981}. 
The idea behind his bridge function approximation was to derive a good approximation based 
on reproducing the exact first five virial coefficients of the hard-sphere equation of 
state. The original proposition for the bridge function by Verlet was,
\begin{equation}
    B(r) = - \frac{A \, \gamma^{2}(r) \, / 2}{1 + B \, \gamma(r) \, / 2}
    \; .
    \label{eq:verlet-params}
\end{equation}
By fitting the exact values of the virial coefficients, Verlet reached the conclusion that 
the values for \(A\) and \(B\) 
that could reproduce the results for the hard-sphere fluid up to the fluid-solid transition 
were the values,
\begin{equation}
    A = 1 \, , \quad B = \frac{4}{5}
    \; ,
    \label{eq:ab-verlet}
\end{equation}
which turns the original closure relation in \autoref{eq:verlet-params} into,
\begin{equation}
    B(r) = - \frac{0.5 \, \gamma^{2}(r)}{1 + 0.8 \, \gamma(r)}
    \; .
    \label{eq:mVerlet}
\end{equation}
This turned out to be an accurate bridge function approximation for the case of the 
hard-sphere fluid, even when hard sphere mixtures are 
considered~\cite{lopez-sanchezDemixingTransitionStructure2013a}.
Some of the advantages of this approximation are the fact that it is simple, efficient and 
there are no free parameters to be fixed, as is the case in the 
Rogers-Young~\cite{rogersNewThermodynamicallyConsistent1984b} and the 
Zerah-Hansen~\cite{zerahSelfConsistentIntegral1986} closure relations. The problem lies, as 
with other closure relations, with the fact that the bridge function in 
\autoref{eq:mVerlet} is not thermodynamically consistent. There are other closure relations 
which are already thermodynamically consistent, as is the case of the reference Hypernetted 
Chain approximation~\cite{ladoSolutionsReferencehypernettedchainEquation1983}. This 
approximation comes from first principles, and the thermodynamic consistency comes from the 
fact that the reference potential is chosen such that the free energy of the fluid is 
minimized. Still, the problem with this closure relation, although accurate, is that it 
might be hard to implement and use in several different scenarios.

\subsection{The Kinoshita variation}
For this reason, the idea of this chapter is to use the modified Verlet approximation, and instead of using the fixed parameters found from theoretical arguments, to let an evolutionary algorithm find the best ones that can also establish thermodynamic consistency. For this reason, the Kinoshita variation~\cite{kinoshitaInteractionSurfacesSolvophobicity2003} to the modified Verlet closure relation is introduced. This variation reads,
\begin{equation}
    B(r) = - \frac{0.5 \, \gamma^{2}(r)}{1 + 0.8 \, \left\lvert \gamma(r) \right\rvert}
    \; ,
    \label{eq:kinoshita-eq}
\end{equation}
and introduces the absolute value \(\left\lvert \cdot \right\rvert\) in the denominator, 
which increases the numerical stability of the closure relation for the case when there are 
very large and negative values of \(\gamma(r)\). However, in this case, the idea is to leave the original form of the closure relation, that is with two free parameters,
\begin{equation}
    B(r) = \frac{\alpha \, \gamma^{2}(r)}{1 + \beta \, \left\lvert \gamma(r) \right\rvert}
    \; ,
    \label{eq:kinoshita-params}
\end{equation}
and instead looking for the best values of both \(\alpha, \beta\) that yield an accurate 
result for the case of the isotropic hard sphere fluid in three dimensions.
To find these values, partial thermodynamic consistency will be enforced through the use of 
the pressure equations, which shall be discussed next.

It is important to note that even though this approach seems to be a good approximation for
the particular case of the hard-sphere fluid, in general a virial expansion based on the
evaluation of the virial coefficients does not lead to a proper equation of state. The 
reason for this is that the virial expansion is an infinite series, but computationally 
only truncated series can be dealt with, and so the series has to be truncated somewhere, 
dropping accuracy and physical relevance for higher order terms in the expansion.

\section{Thermodynamic consistency}
An important part of the results from this chapter is the notion of 
\emph{thermodynamic consistency}, the way it works and how it is implemented numerically. 
This is the goal of the present section.

When approximations to the bridge function are used, together with the OZ equation, most of 
the time the solution shows a phenomenon where, if a thermodynamic observable is computed 
through two or more \emph{routes}, the results may vary drastically. This is not a wanted 
behavior from the closure relations. Rather, it is expected that the closure relations 
could, in fact, provide accurate results regardless the route they are computed with.
In section \autoref{sec:thermodynamics}, some of the ways to compute thermodynamic 
quantities using the radial distribution function, \(g(r)\), were presented. These will be 
presented here once more for the sake of clarity. Particularly, 
it is important to mention the 
two main routes that will be used later, the \emph{virial route} and the 
\emph{compressibility route}.

From \autoref{eq:pressure-equation}, the pressure equation for a $3$-dimensional fluid is,
\begin{equation}
    \beta P = \rho - \frac{2 \pi \rho}{3} \int_{0}^{\infty} \beta u'(r) \, g(r) \, r^3 \, dr \, .
    \label{eq:pressure-gr}
\end{equation}
This equation is also known as the \emph{virial equation} because it can be derived 
formally from the virial theorem (see \autoref{sec:thermodynamics}). Henceforth, 
this equation shall be known as the \emph{virial pressure equation}.

Another quantity of interest is the \emph{isothermal compressibility}, which was already 
introduced in \autoref{sec:thermodynamics}, and can be computed using the \(c(r)\) function 
as follows~\cite{hansenTheorySimpleLiquids2013},
\begin{equation}
    \frac{\beta}{\left( \chi_{T} \,  \rho \right)} = 1 - \rho \int d \vecr \, c(r)
    \; .
    \label{eq:compress-gr}
\end{equation}
For this reason, this equation shall be known throughout the rest of this thesis as the 
\emph{compressibility equation}. Using Thermodynamics, one can relate the pressure \(P\) of the system to the isothermal compressibility using the expression,
\begin{equation}
    \chi_{T} = - \, \frac{1}{V} { \left( \frac{\partial V}{\partial P} \right) }_{T} =
    \frac{1}{\rho} { \left( \frac{\partial \rho}{\partial P} \right) }_{T}
    \; ,
    \label{eq:chi-thermo}
\end{equation}
which is the same as in \autoref{eq:isothermal-chi}, again from 
\autoref{sec:thermodynamics}.

By enforcing that the pressure computed from \autoref{eq:pressure-gr} and the pressure 
computed from \autoref{eq:compress-gr} and then using \autoref{eq:chi-thermo}, are equal to 
some arbitrary precision, closure relations can be \emph{thermodynamically consistent}. 
This would mean that both routes will yield the same value. In reality, not many closure 
relations can do this, except for some special cases, which were already mentioned in the 
previous section. Thus, the standard way of alleviating the problem of thermodynamic 
inconsistency is to compute both routes and adjust some parameter, or function, 
such that both routes provide the same result. It need not be just the pressure equations, 
also the energy equations can be used, as well as other thermodynamic quantities, such as 
the Helmholtz free energy or the chemical 
potential~\cite{tsedneeClosureOrnsteinZernikeEquation2019}.

\subsection{Numerical implementation}
In order to compute the pressure from both routes, the \emph{virial} route and the \emph{compressibility} route, the following approach was followed.

For the case of the \emph{virial} route, \autoref{eq:pressure-gr} was computed as is, using 
the Romberg method for computing the integral~\cite{hammingNumericalMethodsScientists2012}. 
Due to the fact that the interaction potential is a continuous function (see 
\autoref{eq:cont-hs}), the exact derivative was computed and implemented. Then, to compute 
the isothermal compressibility from \autoref{eq:chi-thermo}, the derivative was 
approximated using a finite central difference 
scheme~\cite{hammingNumericalMethodsScientists2012},
\begin{equation}
    \chi_{T}^{-1} = \rho { \left( \frac{\partial P}{\partial \rho} \right) }_{T}
    \approx \rho \lim_{\Delta \rho \to 0} \frac{P(\rho + \Delta \rho) - P(\rho - \Delta \rho)}{2 \, \Delta \rho}
    \; ,
    \label{eq:central-difference}
\end{equation}
where \(\chi_{T}^{-1}\) is the inverse isothermal compressibility. The notation 
\(P(\rho + \Delta \rho)\) means that, for a particular value of density given by 
\(\rho^{\prime}=\rho + \Delta \rho\), as well as for a fixed temperature \(T\), 
\autoref{eq:pressure-gr} was used to compute the 
pressure through the virial route. Here, \(\Delta \rho\) is a small value, fixed to be
\(\Delta \rho=\num{1e-10}\) in the results presented in this chapter. Other central 
differences were explored, such as the five stencil 
approximation~\cite{hammingNumericalMethodsScientists2012}, however, the results obtained 
were as accurate as those obtained with \autoref{eq:central-difference}, so there was no 
reason to justify the extra computational resources needed to do a five stencil difference 
scheme.

For the case of the \emph{compressibility} route, \autoref{eq:compress-gr} was computed as 
is, using once again the Romberg method for computing the integral.

\section{Black-box optimization problem implementation}
\begin{figure}
    \centering
    \vspace{-2cm}
    \includegraphics[scale=0.3]{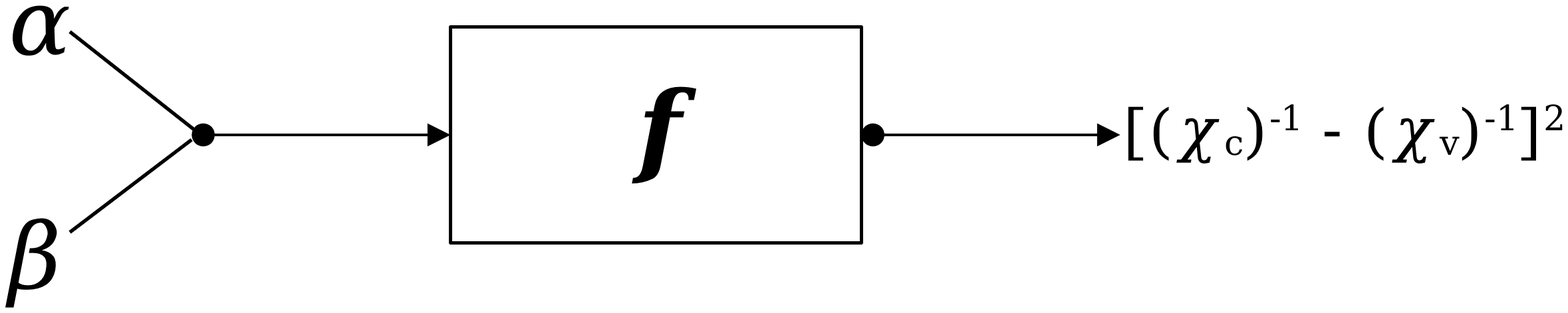}
    \vspace{-3cm}
    \caption{Schematics of the black-box function implementation. The function takes two parameters, \(\alpha, \beta\) as input, and yields the output given by the squared difference of the inverse isothermal compressibility through the virial and compressibility routes.}
    \label{fig:black-box-function}
\end{figure}

In order to solve the problem of finding the best set of values \(\alpha, \beta\) in 
\autoref{eq:kinoshita-params} that would provide thermodynamic consistency, a special type 
of setup must be implemented. It must be recalled that the main objective is for the virial 
and compressibility routes to provide the same value, thus, an \emph{optimization problem} 
can be setup. The problem can be formulated as an \emph{unconstrained optimization} problem,
\begin{equation}
    \underset{\alpha \, , \beta}{\text{minimize}} \: {\left[
        {\left(\chi_{T}^{c} \left(\alpha, \beta\right) \right)}^{-1} - {\left(\chi_{T}^{v} \left(\alpha, \beta\right) \right)}^{-1} \right]}^2
    \; ,
    \label{eq:optimiziation-chi}
\end{equation}
where \({\left(\chi_{T}^{c}\right)}^{-1}\) is the inverse isothermal compressibility 
computed through the \emph{compressibility} route; \({\left(\chi_{T}^{v}\right)}^{-1}\) is 
the inverse isothermal compressibility computed through the \emph{virial} route; and the 
notation \({\left(\chi_{T}^{c, v} \left(\alpha, \beta\right) \right)}^{-1}\) represents the 
dependency of the parameters \(\alpha, \beta\) that would yield such value of the 
compressibility. In other words, for a fixed set of parameters \(\alpha, \beta\), the 
inverse isothermal compressibility is computed via both routes, and its difference squared 
is computed. The goal of the optimization problem is to \emph{minimize} this difference. If 
a good set of parameters is found, then the difference will be, ideally zero or close to 
zero, which in turn would mean that both routes are providing the same result, up to an 
arbitrary numerical tolerance.

In order to perform the optimization procedure, all the necessary computation must be bundled together into a single \emph{black-box function}. This function, which is modeled in \autoref{fig:black-box-function}, will take as input two values and will output the difference between the inverse isothermal compressibility values computed using both routes. Thus, the function will perform the following steps:
\begin{enumerate}
    \item Take as input two fixed values, \(\alpha^{\prime}, \beta^{\prime}\), and use them to define the Kinoshita closure in \autoref{eq:kinoshita-params}.
    \item Solve the OZ equation (see \autoref{AppendixB}) with the closure relation provided in the previous step. This will output two important quantities, the radial distribution function, \(g(r)\), and the direct correlation function, \(c(r)\).
    \item Using \autoref{eq:pressure-gr} and the details presented in the previous section, \({\left(\chi_{T}^{v}\right)}^{-1}\) is computed.
    \item Then, using \autoref{eq:compress-gr} and the details presented in the previous section, \({\left(\chi_{T}^{c}\right)}^{-1}\) is computed.
    \item After both routes have been computed, its squared difference is computed, i.e., \({\left[ {\left(\chi_{T}^{c} \left(\alpha, \beta\right) \right)}^{-1} - {\left(\chi_{T}^{v} \left(\alpha, \beta\right) \right)}^{-1} \right]}^2 \).
\end{enumerate}
Thus, the previous steps comprise the \emph{black-box function} shown in 
\autoref{fig:black-box-function}, which will later be minimized.

\subsection{Optimization procedure}
To solve the optimization problem in \autoref{eq:optimiziation-chi}, Evolutionary 
Computation methods were employed. However, there are two parts to the solution of this 
problem. First, the function is treated as a \emph{black-box function}, which means 
that no derivatives can be computed, and that it is assumed that each function evaluation 
is costly. Although the numerical methods are quite optimized, it is safe to assume that 
each function evaluation is costly, and use this argument to select a good optimization 
method that can effectively reduce the number of function evaluations.

For this reason, Evolution Strategy methods were employed, in particular, Natural Evolution 
Strategies (NES) (see \autoref{Cap3}). These methods are robust, albeit slow to reach a 
minimum. 
Furthermore, these methods are \emph{global optimization methods}, meaning that these 
methods will only look for global values throughout the function landscape, and if any 
minima is found, it will not look deeper into that particular set of parameters. For this 
reason, \emph{local optimization methods} are used, to perform a more exhaustive search 
into the previously found global minima. Thus, the optimization procedure looks like the 
following,
\begin{enumerate}
    \item Select an initial search space of \(\left[-75, 75\right]\) for each parameter, and choose an initial set of parameters randomly from within the search space.
    \item Using the DXNES global optimization method~\cite{nomuraDistanceweightedExponentialNatural2021}, perform a search for all the global minima in the interval \(\rm{x} \in \left[a, b\right]\).
    \item Keep the minima that satisfy the condition \(\rm{x} \leq 0.5\).
    \item Using a local optimization method, the BOBYQA method~\cite{powellUOBYQAUnconstrainedOptimization2002}, take as input the parameters found by the previous two steps, \(\rm{x}^{\prime}\). Use these within a shorter interval of \(\pm 5\) for each parameter, and look for a new set of parameters.
    \item Stop the optimization procedure when the new set of parameters satisfy the condition \(\rm{x}^{*} \leq \num{1e-7}\).
\end{enumerate}

To see this more clearly, an example is worked out now. First, a density value is fixed, 
\(\phi = 0.4\), and all the necessary parameters to solve the OZ equation are fixed as 
well. Then, the search space is fixed to be \(\rm{x} = \left(\alpha, \beta\right) \in \left[-75, 75\right]\). 
An initial set of parameters is chosen randomly from within this search space, for 
instance, \(\rm{x} = \left(-10, 42\right)\). These pair of values are then fixed to be the 
value of \(\alpha, \beta\) in \autoref{eq:kinoshita-params}. With these new values, the 
closure relation is fixed and the OZ equation is solved. If the OZ equation can be solved 
using these values, then the result of the OZ equation is the pair of functions 
\(g(r), c(r)\). Using these values, and the details from the previous section, the squared 
difference 
\[
    \Delta \chi^{-1}_{T} = {\left[ {\left(\chi_{T}^{c} \left(\alpha, \beta\right) \right)}^{-1} - {\left(\chi_{T}^{v} \left(\alpha, \beta\right) \right)}^{-1} \right]}^2 
\; ,
\] 
is computed, and the global optimization method registers this value. If the value 
\(\Delta \chi^{-1}_{T}\) satisfies the condition \(\Delta \chi^{-1}_{T} \leq 0.5\), then 
the set of values \(\rm{x} = \left(-10, 42\right)\) is used for the local optimization 
method. However, in this case, the search space is bounded about these values by a factor 
of \(5\). In this example, that would mean that the new search space for the local 
optimization method would be \(\alpha \in [-15, -5]\) for the first parameter, and 
\(\beta \in [37, 47]\) for the second parameter. Then, the local optimization method would 
start looking for the minimum in this search space, and when a set of values satisfy the 
condition \(\rm{x}^{*} \leq \num{1e-8}\), then the optimization procedure has converged, 
and the set of values are regarded as the best values that provide thermodynamic 
consistency.

\section{Results}
In this section, the results obtained are presented; these results are also summarized in \autoref{tab:global-opt} and \autoref{tab:local-opt}. The results obtained are the two parameters computed through the optimization procedure described in the previous section, as well as the values of the isothermal compressibility computed using both routes.

\subsection{Kinoshita parameters}
\begin{table}
    \centering
    \begin{tabular}{|c|c|c|}
    \hline
    \(\eta\)  & Parameters \(\left(\alpha, \beta\right)\)  & Fitness (minimum)    \\ \hline
    0.10 & {[}5.05, 7.89{]}    & 0.11          \\ \hline
    0.15 & {[}-1.74, 17.87{]}   & 0.02          \\ \hline
    0.20 & {[}-1.15, 5.16{]}    & 0.02   \\ \hline
    0.25 & {[}-6.43, 39.74{]}   & 0.45  \\ \hline
    0.30 & {[}-12.65, 39.23{]} & 0.1  \\ \hline
    0.35 & {[}-16.91, 47.15{]}    & 0.21  \\ \hline
    0.40 & {[}-27.44, 74.81{]}  & 0.08  \\ \hline
    0.45 & {[}-28.99, 71.32{]}  & 0.03 \\ \hline
    \end{tabular}%
    \caption{Results obtained directly from the optimization procedure for different values of volume fractions, \(\eta\). Here, the fitness or minimum represents the best value of \(\Delta \chi^{-1}_{T}\) that satisfies the condition \(\Delta \chi^{-1}_{T} \leq 0.5\).}
    \label{tab:global-opt}
\end{table}

\begin{table}
    \centering
    \begin{tabular}{|c|c|c|c|}
    \hline
    \(\eta\)  & Parameters \(\left(\alpha, \beta\right)\) & \(\beta \, {\left(\chi^{v}_{T} \, \rho\right)}^{-1}\) (Virial)  & \(\beta \, {\left(\chi^{c}_{T} \, \rho\right)}^{-1}\) (Compressibility)  \\ \hline
    0.10 & [-1.39, 9.82]   & 2.19777  & 2.19778 \\ \hline
    0.15 & [-3.11, 17.87]  & 3.26186 & 3.26187  \\ \hline
    0.20 & [-1.35, 5.09]   & 4.79978  & 4.79972  \\ \hline
    0.25 & [-10.52, 41.28] & 7.28143  & 7.28142 \\ \hline
    0.30 & [-12.64, 42.16] & 10.96744 & 10.96747 \\ \hline
    0.35 & [-16.78, 49.43] & 16.64003 & 16.64005 \\ \hline
    0.40 & [-27.87, 74.52] & 25.4941 & 25.4943 \\ \hline
    0.45 & [-28.81, 71.35] & 39.0225  & 39.0226  \\ \hline
    \end{tabular}
    \caption{Results obtained from the local optimization procedure. The last two columns represent the inverse isothermal compressibility, from the virial and compressibility routes respectively.}
    \label{tab:local-opt}
\end{table}

For the parameters \(\alpha, \beta\), the optimization procedure of the previous section 
was followed, first using a global optimizer, and then a local optimizer, until the 
precision for the minimum was at least \(\leq \num{1e-8} .\) 
In \autoref{tab:global-opt}, the values for the global optimization procedure are shown. 
There, the minimum is at least less or equal than \(0.5\). 
Up to this point, the results between the virial and compressibility routes were not distant 
from each other, numerically speaking. However, when the parameters found were used to 
search for more accurate minima, the results were promising. This can be seen in 
\autoref{tab:local-opt}, where the results using the local optimizer are shown. It is now 
clear that both routes give the same results, which shows that the Kinoshita closure is now 
thermodynamically consistent, at least partially through the pressure equations.

\subsection{Isothermal compressibility of the hard sphere fluid}
\begin{figure}
    \centering
    \includegraphics[scale=0.32]{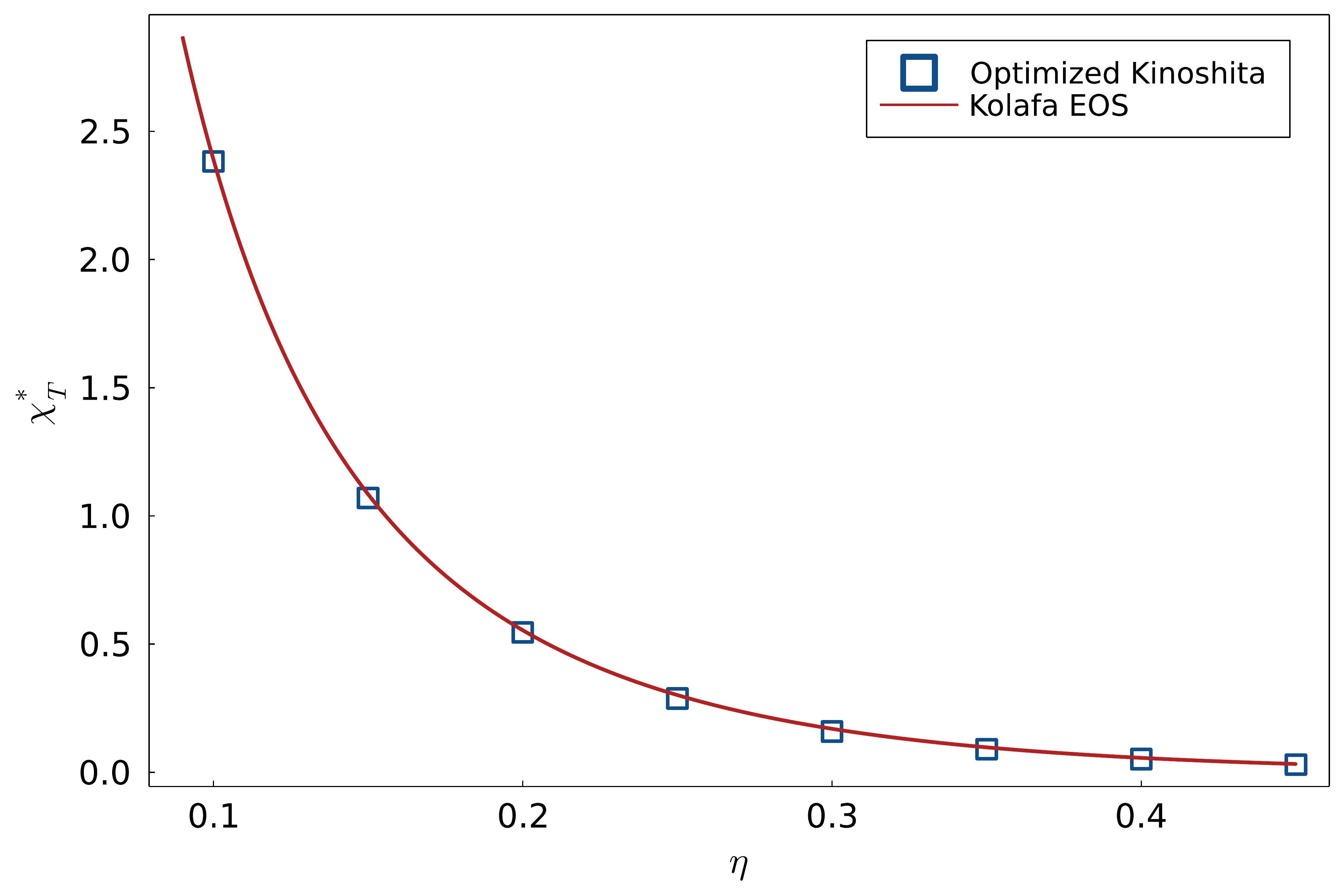}
    \caption{Isothermal compressibility of the hard sphere fluid for different density values, expressed in terms of the volume fraction, \(\eta .\) The solid line is the isothermal compressibility obtained from the Kolafa equation of state (Kolafa EOS)~\cite{kolafaAccurateEquationState2004}. The squares are obtained from the optimization procedure, reported in \autoref{tab:local-opt}.}
    \label{fig:isothermal-results}
\end{figure}

To showcase the results obtained, a direct comparison of the isothermal compressibility 
values are plotted against the Kolafa equation of state (Kolafa EOS) in 
\autoref{fig:isothermal-results}. To obtain this plot, the isothermal compressibility was 
computed in its reduced form,
\begin{equation}
    \chi_{T}^{*} =  \frac{\chi_{T}}{\beta \, \sigma^3}
    \; ,
    \label{eq:reduced-chi}
\end{equation}
and also the results from \autoref{tab:local-opt} were multiplied by the reduced particle 
number density, \(\rho^{*} = \rho \, \sigma^3 .\)

In the case of the Kolafa EOS, the equation reads,
\begin{equation}
    Z = \frac{P V \beta}{\rho} = \frac{1 + \eta + \eta^2 - \frac{2}{3} (\eta^3 + \eta^4)}{{\left(1 - \eta\right)}^{3}}
    \label{eq:kolafa}
\end{equation}
with \(\eta = \frac{\pi}{6} \rho^{*} .\) This is an important fact about hard sphere 
fluids; in general their equation of state can always be represented in terms of the 
geometry of the system, i.e., its volume fraction.
Then, to compute the isothermal compressibility, the equation of state needs to be 
differentiated, and the expression is~\cite{liuCarnahanStarlingTypeEquations2021},
\begin{equation}
    \chi_{T}^{*} = {\left[\left(\eta \, \frac{\partial Z}{\partial \eta} + Z\right) \cdot \rho^{*}\right]}^{-1}
    \; .
    \label{eq:chi-eos}
\end{equation}
Using \autoref{eq:chi-eos}, the solid line in \autoref{fig:isothermal-results} was 
computed, along with the results from \autoref{tab:local-opt}. It is simple to see that the 
results are in excellent agreement with the results obtained from the Kolafa EOS, which is 
considered to be a very accurate equation of state for the hard sphere fluid.

\subsection{Radial distribution functions}
\begin{figure}
    \centering
    \includegraphics[scale=0.35]{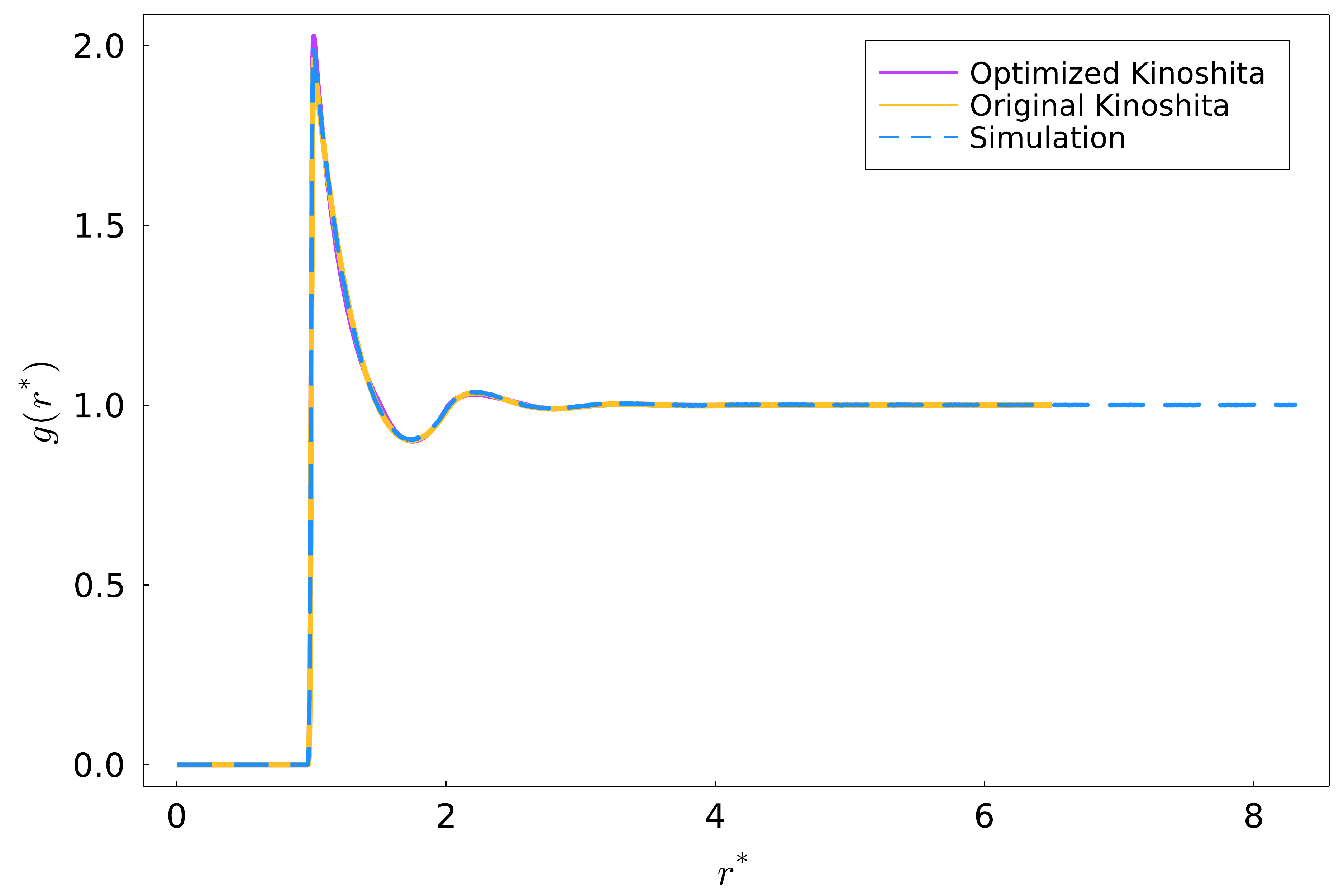}
    \caption{Radial distribution function comparison between the Kinoshita parameters found by optimization, \emph{Optimized Kinoshita}, the original parameters from~\autoref{eq:kinoshita-eq}, \emph{Original Kinoshita}, and Monte Carlo simulations, \emph{Simulation}, for the density value of \(\phi=0.25 .\)}
    \label{fig:rdf-kinoshita-025}
\end{figure}

\begin{figure}
    \centering
    \includegraphics[scale=0.35]{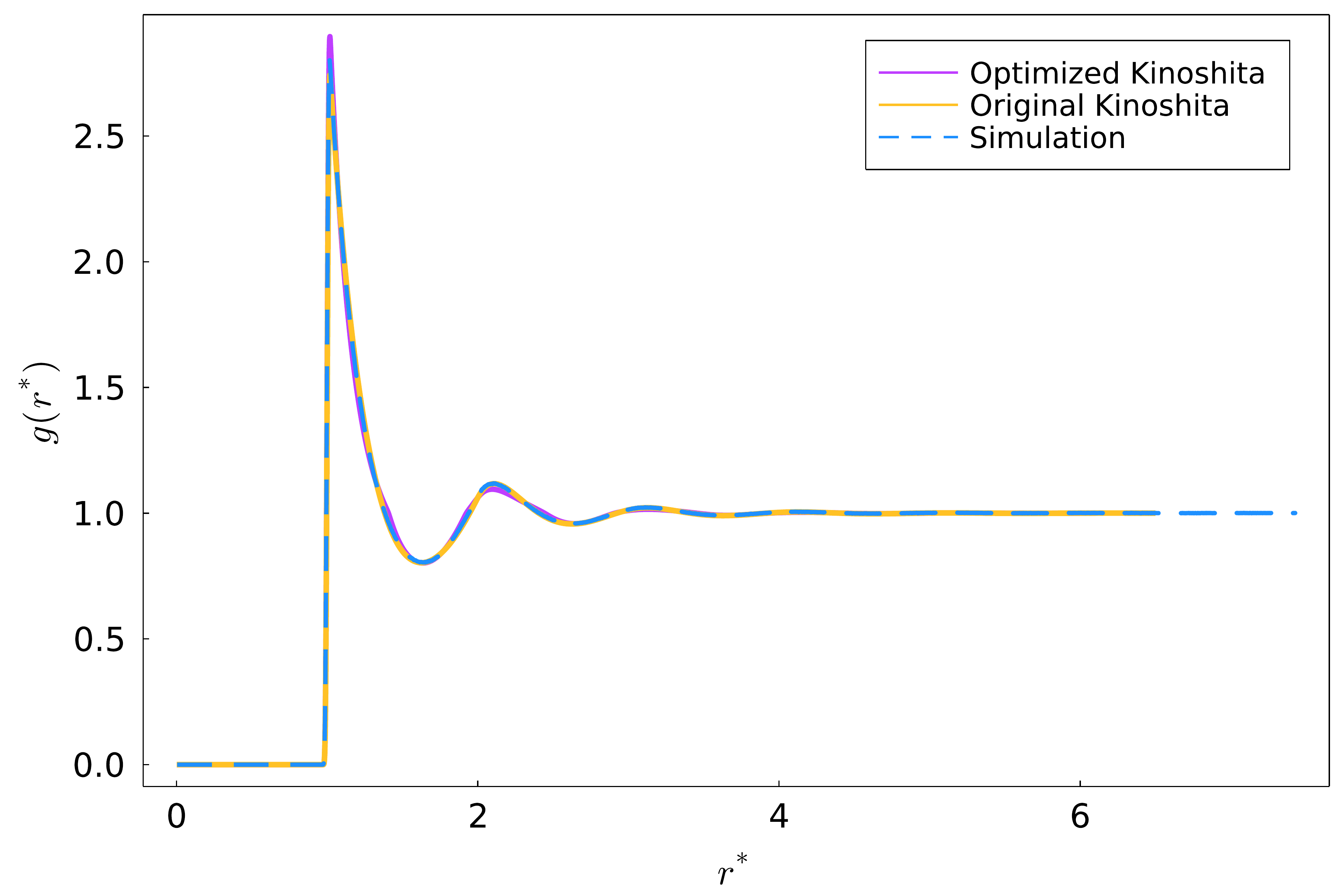}
    \caption{Radial distribution function comparison between the Kinoshita parameters found by optimization, \emph{Optimized Kinoshita}, the original parameters from~\autoref{eq:kinoshita-eq}, \emph{Original Kinoshita}, and Monte Carlo simulations, \emph{Simulation}, for the density value of \(\phi=0.35 .\)}
    \label{fig:rdf-kinoshita-035}
\end{figure}

\begin{figure}
    \centering
    \includegraphics[scale=0.35]{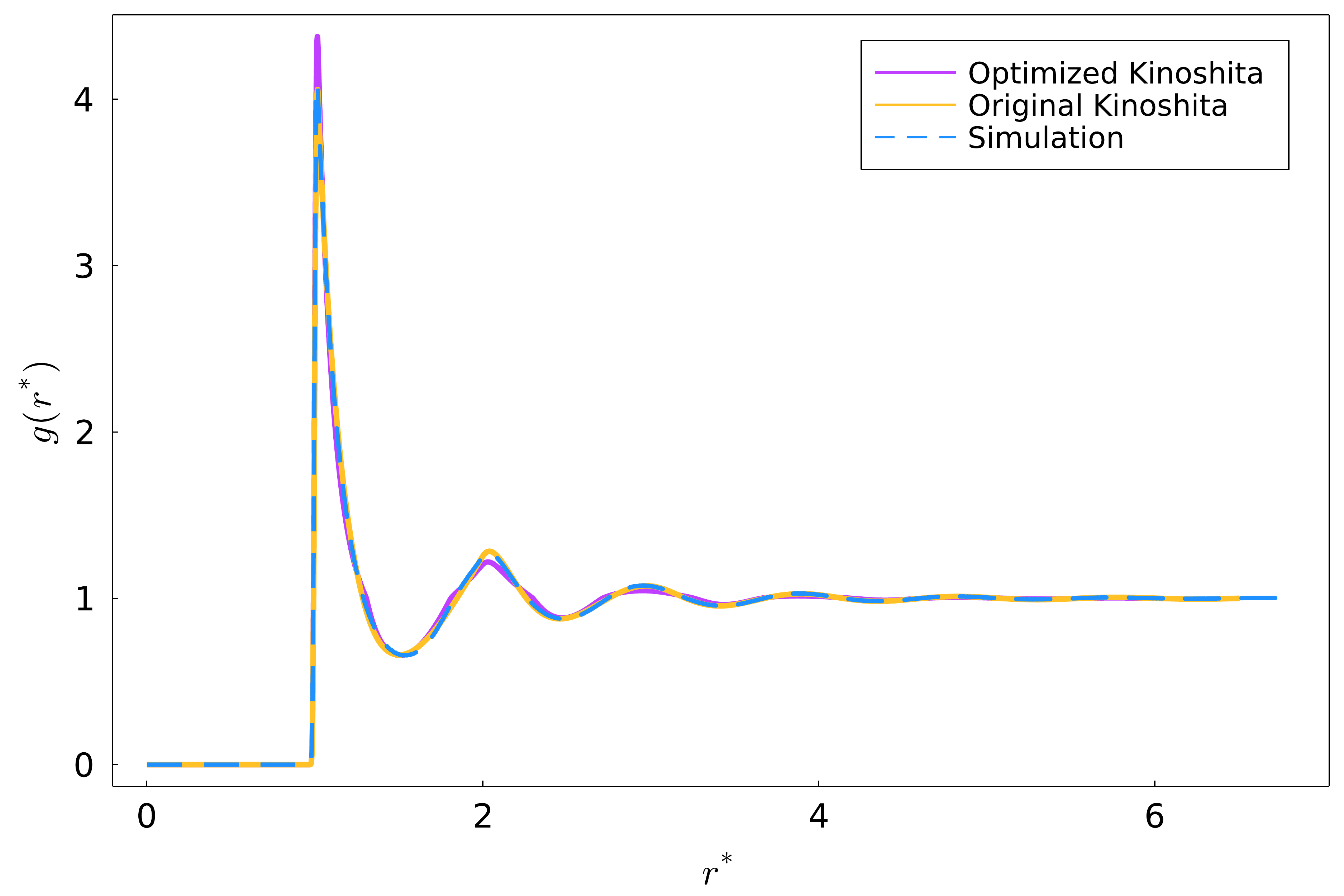}
    \caption{Radial distribution function comparison between the Kinoshita parameters found by optimization, \emph{Optimized Kinoshita}, the original parameters from~\autoref{eq:kinoshita-eq}, \emph{Original Kinoshita}, and Monte Carlo simulations, \emph{Simulation}, for the density value of \(\phi=0.45 .\)}
    \label{fig:rdf-kinoshita-045}
\end{figure}

The radial distribution functions are now presented, but only for a small subset of 
densities, namely 
\(\phi=\numlist[list-final-separator={\enspace\text{and}\enspace}]{0.25; 0.35; 0.45} .\)
In these figures,~\autoref{fig:rdf-kinoshita-025},~\autoref{fig:rdf-kinoshita-035}, 
and~\autoref{fig:rdf-kinoshita-045}, the radial distribution functions found using the 
optimization procedure are compared directly with the ones found by means of Monte Carlo 
computer simulations, and using the original parameters for the Kinoshita closure 
in~\autoref{eq:kinoshita-eq}. One important aspect that is clear is that, at higher 
densities, the radial distribution function does not seem to follow the same form as that 
found via computer simulations.

\section{Discussion}
It seems that, with the results shown in \autoref{fig:isothermal-results}, this approach is 
a great proposal for alleviating the thermodynamic consistency problem that permeates the 
bridge function literature. Indeed, the results show a lot of promise for the
\emph{optimization procedure} itself, but there are some important aspects that need to be 
pointed out.

\subsection{Thermodynamic consistency as an optimization problem}
The proposal of this chapter is to extend the Kinoshita closure in 
\autoref{eq:kinoshita-params}, and find the best values for the closure that can provide 
partial thermodynamic consistency. However, this proposal is not a new one, and in fact, 
Verlet himself established a proposal quite similar using his own bridge function 
approximation~\cite{verletIntegralEquationsClassical1981}. Still, in his proposal there 
are \emph{three free parameters} to adjust, and the optimization procedure is quite 
different. Rogers-Young~\cite{rogersNewThermodynamicallyConsistent1984b} and 
Zerah-Hansen~\cite{zerahSelfConsistentIntegral1986} are two of the fundamental proposals 
using the same idea, with just one free parameter two adjust, and interpolating between two 
well-established bridge functions. Further, the recent contribution by Tsednee and 
Luchko~\cite{tsedneeClosureOrnsteinZernikeEquation2019} of fixing the parameters of a 
series expansion was also posed as an optimization problem. However, in their work, there 
is no clear expression of how or why the optimization problem is handled, and just a hint 
of the local optimizer used is mentioned, which was the \verb|fmin| function in the 
MATLAB programming language.

In this thesis, the idea is simply to free up the fixed parameters in the Kinoshita bridge 
function approximation, which has shown the potential to be extended to mixtures of hard 
sphere fluids and other more complex systems. Further, the idea of an optimization 
problem based on black-box modeling seems to be a great way of thinking about the problem 
itself, given that the consistency enforced here is just partial, there are a lot more 
thermodynamic quantities that \emph{should} be adjusted as well, such as the chemical 
potential or the free energy. This is because more advanced techniques can be used, such as 
surrogate-base optimization~\cite{forresterRecentAdvancesSurrogatebased2009}, which 
attempts to build a surrogate, or substitute, model of the original black-box function, 
which can now be differentiable and more easily optimized. But this leads to another 
important point of discussion.

\subsection{The search space and the solution of the Ornstein-Zernike equation}
Although the results seem promising, one of the drawbacks encountered in this chapter when 
computing the results was the \emph{definition of the search space}, and the solution of 
the OZ equation in such a search space. Evolutionary Computation 
methods are \emph{population-based}, meaning that these methods generate a fixed set of 
possible solutions, and all solutions are tested and tried out by the optimization method. 
Though, it turns out that the OZ equation is \emph{not solvable} for a lot of parameters 
\(\alpha, \beta\) that are searched and sampled by the optimization method. In fact, at 
least 40\% of the parameters searched that were tried as possible solutions for the 
optimization problem in \autoref{eq:optimiziation-chi}, rendered the numerical solution of 
the OZ equation impossible. The data for this argument is not shown here, because there is 
not much to show regarding numerical errors and numerical instabilities.

This finding is quite important, because it shows that the OZ equation is only solvable for 
certain value of the parameters \(\alpha, \beta\), but no clear pattern is shown or easy to 
see. The only clue to this phenomenon is that the actual problem, which are just numerical 
instabilities of the numerical scheme. This would mean that, as in the previous chapter, a 
different numerical scheme could be used, maybe based of Krylov sub-space methods, 
following the original proposal by Gilles Zerah~\cite{zerahEfficientNewtonMethod1985}. Or 
it could also mean that there are several values and search subspaces for which the 
parameters \(\alpha, \beta\) do not provide a good bridge function approximation. But it is 
not clear \emph{how} or \emph{why} the bridge function approximation would not provide at 
least a solution to the OZ equation.

The search space is another important part of the problem. A large search space means that 
the optimization procedure must look into very different places until a particular minimum 
has been found. This could be solved by restricting the bounds to particular bounds in both 
\(\alpha, \beta\), instead of using a single large search space for both parameters. 
However, this is not simple to see, or to define. One way to do this would be rely on the 
same argument used by Verlet~\cite{verletIntegralEquationsClassical1980}, and inform the 
optimization procedure of the virial coefficients for the hard sphere fluid.
However, as it was previously stated, this definition does not lead to a correct bridge 
function, because a virial expansion needs to be truncated due to its infinite nature.

The problem of the search space can be further explored with the radial distribution 
functions found, and the differences between the densities used. For instance, for a low 
density of \(\phi=0.25\) volume fraction, the results agree quite well with the Monte Carlo 
computer simulations and the original parameters from the Kinoshita closure. This result 
can be clearly seen in~\autoref{fig:rdf-kinoshita-025}. Here, the search space is enough for 
the optimization procedure to find a correct value of the parameters, and it is easily seen 
that the result is quite good as well in~\autoref{fig:isothermal-results}. But when the 
density starts to rise, then the local minima found is not the best, and a the search space 
is then implied to change. This can be clearly seen in the radial distribution functions 
for \(\phi=0.35\) in~\autoref{fig:rdf-kinoshita-035}, and for \(\phi=0.45\) 
in~\autoref{fig:rdf-kinoshita-045}. It is now clear that, as the density starts to get 
higher, the search space for the optimization procedure becomes extremely important. One 
aspect of this is that the radial distribution functions do not match the results obtained 
from Monte Carlo computer simulations, nor they match the results from the original 
Kinoshita parameters.

\subsection{On the use of Natural Evolution Strategies}
Another drawback of the proposed method is the fact that Evolutionary Computation methods 
are slow. This has already been said in previous sections, but in this case this is 
important to mention briefly. Natural Evolution Strategies have to sample distribution 
probabilities, and this is costly. Not to mention that the actual black-box function might 
be costly as well. The method itself should just find some minima in a small period of 
time, but in this case, a large portion of the time was spent with failed solutions of the 
OZ equation. Indeed, several restarts of the numerical method were 
needed in order to find values within the search space, and most of the time, numerical 
instabilities were encountered. These methods are good enough for simple problems, such as 
smooth and analytical high-dimensional functions, but other methods might be able to solve 
the problem more quickly and efficiently, such as Differential 
Evolution~\cite{dasDifferentialEvolutionSurvey2011} methods, also part of the 
Evolutionary Computation family of methods.

However, a strong metric to compare different methods in this will be needed, specially 
when dealing with the Physics of the problem. This is because it might be possible that not 
\emph{all} the minima found will have Physical meaning, or it might just be that there are 
some local minima for which the bridge function yields accurate results. However, these 
methods will not discriminate between either of those, so a metric that could drive the 
optimization problem to Physical meaning would be a novel way to solve the problem of 
bridge function approximations. This is needed as well given the results obtained from the 
radial distribution functions in~\autoref{fig:rdf-kinoshita-025},~\autoref{fig:rdf-kinoshita-035}, 
and~\autoref{fig:rdf-kinoshita-045}. For small density values, there seems to be no need to 
drive the optimization method with more Physics information about the system. But at higher 
density values, it seems that the radial distribution function start to distort and does 
not seem to match the results from Monte Carlo computer simulations. A way to deal with 
this problem might be to incorporate more information into the loss function 
in~\autoref{eq:optimiziation-chi} in terms of the radial distribution function, maybe 
regarding the principal peaks, or even more information from the solution of the OZ 
equation itself.

\section{Concluding remarks}
The results shown here are clearly better than those shown in the previous chapter. Indeed, 
incorporating more information about the Physics of the system provides more robustness to 
the Computational Intelligence method. There needs to be a reformulation as well, but in 
general it is safe to say that this method is better in terms of accuracy, as is shown in
\autoref{fig:isothermal-results}, which would not be the case if the radial distribution 
functions were not estimated correctly.
In a sense, it is expected to see these results, because it is known that the Verlet bridge 
function, as well as the Kinoshita variation, are good functions for the hard sphere fluid.
What might not be too obvious, and a novel way of seeing the problem, is the fact that the
Kinoshita closure relation can be, in fact, thermodynamically consistent.

However, the results shown here are just a minimal working example of what can be done. 
An important aspect to study is the fact that the Neural Network approach of the previous 
chapter could be \emph{informed}, or combined, with the Kinoshita bridge function 
approximation, and instead of reducing to the Hypernetted Chain, it could potentially 
reduce to the Kinoshita closure relation. And yet, this seems like more work to do, because 
the Kinoshita closure only has two free parameters, compared to the thousands of parameters 
that appear in a Neural Network. Still, the Neural Network is faster to fit and solve than 
using Evolutionary Computation methods, which could be extended to other systems as well.

The results here show great promise to be extended to other spatial dimensions and systems. 
For instance, in highly asymmetric mixtures~\cite{zhouLocalStructureThermodynamics2019}, 
where several parameters could be found. The method shown here would not change at all, due 
to the fact that Evolutionary Computation methods naturally handle high-dimensional 
optimization problems. The only thing that would need to change is the solution to the OZ 
equation. Another aspect that could be explored is the study of hard disks and their 
mixtures, which is the generalization of the hard sphere fluid to a \(2\)-dimensional 
space. It would be interesting to see if the method itself holds, or if there are a lot 
more numerical instabilities that might show when different dimensions are handled.
An extension to other interaction potentials might be possible with this method as well. 
The study of the Lennard-Jones fluid, the Yukawa fluid, and many others, could be studied 
using this method, just to showcase the true robustness of the method.

Finally, another line of future research could be the comparison of different optimization 
algorithms for this kind of problems. A robust implementation that could be translated into 
a highly efficient code package could serve well the community of Liquid Theory and Soft 
Matter, in general. Different optimization methods could yield better or worse results, or 
it could shed some light into what might be happening to the search space itself. %% Optimization
\chapter{Conclusions}
\label{Cap6}

Liquid State Theory is at the core of most modern Soft Matter research fields. As such, it 
seems important to clearly understand its underlying theoretical frameworks and their main 
results. Also, it is important to have a clear understanding of the methods used to provide 
accurate solutions to specific problems. Without the exact theoretical framework of the 
Ornstein-Zernike equation, most Colloidal Soft Matter research fields would struggle with 
other theories which are more complicated and harder to implement. Rather, the 
Ornstein-Zernike equation is simple, effective, and has the outstanding ability to be able 
to generalize well, beyond its original goal, which was the description of simple liquids. 
It is the purpose of this thesis to showcase the richness of the Ornstein-Zernike 
formalism, and to inject novel ways of thinking about it, using it, and most importantly, 
of solving it.

Throughout this thesis, the Ornstein-Zernike formalism was introduced as an integral 
equation theory that is capable of describing the structure of a simple liquid, in 
particular we focused on the fluid model that results the cornerstone of the Statistical 
Mechanics of fluids,  namely,  the hard-sphere fluid. Using the theory of equilibrium 
Statistical Mechanics, the Ornstein-Zernike equation was described, along with the 
important quantity, the \emph{bridge function}. This function plays a fundamental role in 
trying to solve the Ornstein-Zernike equation, because it provides a closed expression  for 
the indirect correlations between particles the system that is the subject of study. 
However, the reality is that this function is still an approximation, and to correctly 
describe a particular fluid, this approximation must be chosen arbitrarily, based primarily 
on computer simulations and previous research. But this seems cumbersome, 
and although lots of literature reflect this, it is still used based on experience and 
common knowledge. This is because finding an exact bridge function is very hard, it cannot 
be done either numerically or analytically, thus approximations are needed if a solution is 
wanted for the fluid under scrutiny.

For this reason, Computational Intelligence methods are introduced in this thesis. First, 
instead of trying to approximate the bridge function, a more general approximation is 
presented in the form of a Neural Network. Due to the 
\emph{Universal Approximation Theorem}~\cite{hornikMultilayerFeedforwardNetworks1989,hornikApproximationCapabilitiesMultilayer1991,cybenkoApproximationSuperpositionsSigmoidal1989}, 
Neural Networks can approximate any continuous and well-defined, smooth function. Using 
this knowledge, instead of defining a particular approximation for the bridge function, a 
new way of solving the Ornstein-Zernike equation is devised, that attempts to use the 
Neural Network to find the best approximation for the hard sphere fluid. The results show 
that the Neural Network reduces to a particular bridge function approximation, the 
Hypernetted Chain solution, for the case of the hard sphere fluid. This points toward to 
the direction that without any more information from the system, the Neural Network tries 
to stay in a particular minimum, which in this case corresponds to the Hypernetted Chain 
solution. However, this also means that if there is more information about the underlying 
Physics of the systems, more information can be used along with the Neural Network, 
potentially given a better solution to the problem.

Instead of relying on a universal approximator, one can effectively use the underlying 
Physics of the problem, and use different Computational Intelligence methods that can solve 
the same problem. This proposal is shown in this thesis, using the Verlet bridge function 
approximation, together with the Kinoshita variation, to provide more information to the 
Ornstein-Zernike equation. Nevertheless, instead of using these bridge functions as they 
are, the whole problem was turned into a black-box optimization problem that attempts to 
find the best variation of the Kinoshita bridge function that can also provide partial 
thermodynamic consistency to the problem. It seems that thermodynamic consistency is a 
better way of dealing with the problem, and the results show this. The black-box 
optimization problem was solved using Natural Evolution Strategies, a type of optimization 
method that comes directly from the family of Evolutionary Computation optimization 
algorithms. When computing the isothermal compressibility for different density values of 
the hard sphere fluid, the results obtained from these optimization methods provides 
excellent agreement with the theoretical results. This means that, by reformulating the 
problem to an optimization problem with less parameters to find, the solution is better and 
more accurate. One of the drawbacks of this method is that it is slow and computationally 
expensive. Nonetheless, the method can be extended to other systems and interaction 
potentials.

\section{Future work}
To close this thesis, a few ideas will be presented which can serve as future work that can 
carry on from the main proposals presented in this work. Some of these ideas serve to close 
off the points made in the results presented, and others might serve as inspiration for the 
reader interested in the topics developed in this work.

\subsection{A functional approach to the bridge function}
In \autoref{Cap4}, the bridge function was presented as a series expansion on the density 
in \autoref{eq:expansion-densidad}. A different idea, related to this one, is that the 
approximation can be worked out from a series expansion where the terms are themselves 
neural networks. This would look like,
\begin{equation}
    b \left[r, N_{\theta}(r)\right] = \sum_{i=1}^{\infty} a_{i} \, N_{\theta \, , i} (r)
    \; ,
    \label{eq:functional-bridge}
\end{equation}
where the notation \(N_{\theta,i}(r)\) represents a different neural network for each 
\(i\) in the expansion. In other words, instead of having just a polynomial series 
expansion in the bridge function, the terms themselves would be constructed from 
\emph{functions}, thus providing a \emph{functional approach} to the bridge function. This 
would have all the mathematical power and generalization of simple neural networks, and 
provide more information to the bridge function itself. However, this is not a simple 
proposal and the mathematical framework to work this out might be daunting and difficult to 
use.

\subsection{Surrogate-based modeling}
The Ornstein-Zernike equation can be solved efficiently using the Fast Fourier 
Transform~\cite{hammingNumericalMethodsScientists2012}. But sometimes the solution is 
difficult to obtain using the same iterative Piccard scheme. Further, if the black-box 
optimization method is implemented to enforce partial thermodynamic consistency, then the 
solution can become difficult to obtain. A solution to this problem is to propose a 
\emph{surrogate-based modeling} approach. Surrogate 
models~\cite{forresterRecentAdvancesSurrogatebased2009} as substitute models that 
\emph{cleverly} interpolate between the input and output of difficult-to-evaluate 
functions. The meaning of clever here means that most of these substitute models rely on 
modern Machine Learning techniques, such as Neural Networks, Gaussian Processes, and many 
others.

What would happen if, instead of solving the Ornstein-Zernike equation directly, a 
surrogate model could be built from previous known answers? At least for simple state 
points, the surrogate model could be expected to perform well. But when difficult phenomena 
arise, such as phase transitions, spinodal lines and others, it might be unreasonable to 
expect too much from the surrogates. For this, additional information based on the 
underlying Physics might serve helpful in defining better surrogate models. This can also 
help in the optimization process of finding better parameters for the bridge function 
approximation.

\subsection{Closing remarks}
Finally, it is important to note that the use of Computational Intelligence methods in 
Liquid State Theory is novel, and is still in its exploration phase. Modern Soft Matter 
research has been constantly adopting new methods brought from the Machine Learning and 
Deep Learning communities, and this is creating a trend in this research field. This is 
good, because Machine Learning methods, and in general Computational Intelligence methods, 
have proven useful in cases where one can use all the previous Physics together with more 
powerful and robust methods, such that new ways of dealing with the same problems arise. 
These new ways can be better, more efficient, or simply put, easier to understand and 
implement. It is the main purpose of this thesis to showcase this important idea: that new 
Computational Intelligence methods and technologies can help boost modern Soft Matter and 
Liquid State Research, as long as the elemental Physics of the system are used 
appropriately. Hopefully, in years to come, these ideas will become the standard way of 
dealing with the fantastic research in Liquid State Theory and Soft Matter Physics. %% Conclusions

%----------------------------------------------------------------------------------------
%	THESIS CONTENT - APPENDICES
%----------------------------------------------------------------------------------------

\appendix % Cue to tell LaTeX that the following "chapters" are Appendices

% Include the appendices of the thesis as separate files from the Appendices folder
% Uncomment the lines as you write the Appendices

\chapter{Gradient Computation}
\label{AppendixA}

In \autoref{Cap4} a training scheme was developed to adjust the weights of a neural
network while simultaneously solving for the OZ equation. A crucial part of
this algorithm is the \emph{gradient computation}. In this section, we shall work
out the details of this computation.

\section{Mathematical development}
Recall that we want a neural network $\nnet$ with weights $\theta$ to work as a
parametrization in the closure expression of the OZ equation, as defined in
\autoref{eq:parametrizacion}. To find the weights of the neural network, an
unconstrained optimization problem was presented, which was meant to be solved with
an iterative procedure known as \emph{gradient descent}, which has the following
general rule
\[
\theta_{n+1} = \theta_{n} - \eta \nabla_{\theta_{n}} J(\theta_{n})
\]
where the cost function $J(\theta)$ is defined in \autoref{eq:costo}; $\nabla_{\theta}$
is the gradient of $J(\theta)$ with respect to the weights, $\theta$; and $\eta$ is known
as the learning rate, which controls the length of the step taken by the gradient
towards a minimum. The iteration step is accounted for with the index $n$, which runs until
convergence has been achieved.

We are now interested in the closed form of $\nabla_{\theta} J(\theta)$. If we use the
definition of the cost function as defined in \autoref{eq:costo}, we have for the
gradient the following expression
\begin{equation}
    \nabla_{\theta} J(\theta) = \nabla_{\theta} \left[\gamma_{n}(\vecr, \theta) - \gamma_{n-1}(\vecr, \theta) \right]^2
    \label{eq:grad1}
\end{equation}
where $\gamma_{n}(\vecr, \theta)$ is the $n$-th approximation of the indirect
correlation function, $\gamma(\vecr)$.
The notation $\gamma(\vecr, \theta)$ indicates that the function now depends
on the weights of the neural network. When the weights $\theta$ are modified, the output of
$\gamma(\vecr, \theta)$ must change as well.

We now apply the gradient operation to \autoref{eq:grad1} to obtain the following
expression
\begin{equation}
    \nabla_{\theta} J(\theta) = 2 \left[\gamma_{n}(\vecr, \theta) - \gamma_{n-1}(\vecr, \theta) \right]
    \left[ \partial_{\theta} \gamma_{n}(\vecr, \theta) - \partial_{\theta} \gamma_{n-1}(\vecr, \theta) \right]
    \label{eq:grad2}
\end{equation}
which results from direct application of the chain rule. Now, an expression for
$\partial_{\theta} \gamma_{n}(\vecr, \theta)$ needs to
be found by some other route. Notably, this expression should be expressed in terms
of a quantity that \emph{depends on the weights}, $\theta$. In this case, 
this would mean that we seek some expression in terms of the neural
network $\nnet$, which has a direct dependency on the weights.

In order to find this new expression, we invoke \autoref{eq:parametrizacion} and
differentiate it with respect to the weights
\begin{equation}
    \frac{\partial c(\vecr, \theta)}{\partial \theta} = \frac{\partial}{\partial \theta}
    \left[ e^{p(\vecr, \theta)} - \gamma(\vecr, \theta) - 1 \right] =
    e^{p(\vecr, \theta)} \partial_{\theta} p(\vecr, \theta) - \partial_{\theta} \gamma(\vecr, \theta)
    ,
    \label{eq:grad3}
\end{equation}
here we have for $p(\vecr, \theta)$
\begin{equation}
    p(\vecr, \theta) = -\beta u(\vecr) + \gamma(\vecr, \theta) + \nnet
    \label{eq:pr}    
\end{equation}
with derivative with respect to the weights
\begin{equation}
    \partial_{\theta} p(\vecr, \theta) = \partial_{\theta} \gamma(\vecr, \theta) + \partial_{\theta} \nnet
    .
    \label{eq:grad-pr}
\end{equation}

We have now found a closed form for the value of
$\partial_{\theta} \gamma_{n}(\vecr, \theta)$
which is essentially the same as \autoref{eq:grad3}, but in a slightly different form,
which reads
\begin{equation}
    \boxed{
    \frac{\partial \gamma(\vecr, \theta)}{\partial \theta} =
    e^{p(\vecr, \theta)} \partial_{\theta} p(\vecr, \theta) - \partial_{\theta} c(\vecr, \theta)
    .
    }
    \label{eq:grad4}
\end{equation}
Note, however, that this new expression depends on the value of
$\partial_{\theta} c(\vecr, \theta)$
which we do not readily have at this point. It is at this step that we propose
to compute the value of $\partial_{\theta} c(\vecr, \theta)$ using numerical
differentiation with a finite central difference scheme, which should be enough to get a good estimate of the derivative.

\subsection{General solution scheme}
In order to carry out the detailed explanation of the numerical approximation approach,
we must recall the order in which the general solution to the OZ equation is achieved.
We need to do this in order to understand how the numerical approximation of the derivative
for $c(\vecr, \theta)$ can be obtained.
This is already described in detail in \autoref{subsec:oz-solution}, but
we need to recall it here briefly.

In short, the general solution to the OZ equation with a neural network 
parametrization looks like this:

\begin{itemize}
    \item For the first part, we solve the OZ equation in a classical fashion, employing Fourier transforms in order to build a set of approximations for the $\gamma(\vecr, \theta)$ function. At this step we have found also an approximation for $c(\vecr, \theta)$, which is crucial to remember.
    \item When we have found said approximations, we compute the gradient as shown in \autoref{eq:grad2}. Due to the fact that we have found previous approximations to the correlation functions, we can use them as part of our gradient computation.
    \item Finally, with the value of the gradient, we compute the loss function and adjust the weights $\theta$ for the neural network. We then repeat the process until convergence is reached.
\end{itemize}

As we can see, given the fact that the OZ solution scheme already provides numerical
approximations for the correlation functions, we can use the value of $c(\vecr, \theta)$
in other numerical schemes to find the value of \autoref{eq:grad4}.

\section{Numerical differentiation approach}
We will now outline the scheme used in the current work to find the value of 
\autoref{eq:grad4} using a numerical differentiation scheme for
$\partial_{\theta} c(\vecr, \theta)$.

To achieve such goal, we must briefly outline the method of numerical differentiation
for functions of several variables~\cite{hammingNumericalMethodsScientists2012}.
We wish to work with the $c(\vecr, \theta)$ function, which is essentially a function of
two variables. The first variable, $\vecr$ is the \emph{Euclidean distance}
between two particles, or in other words $\vecr={\lvert \vecr_1 - \vecr_2 \rvert}^2$.
For the second variable $\theta$, this represents the weights of $\nnet$, which is
in fact a vector of its own of size $n$. The value of $n$ depends on the number of
nodes in the neural network, but here we will use a more general approach instead of
defining a specific value of $n$.

So, with this in mind, we can define the function to be 
$c(\vecr, \theta) : \mathbb{R} \times \mathbb{R}^{n} \to \mathbb{R}$.
For a multivariate function, its partial derivative is computed numerically,
for a particular value of distance $\bar{\vecr}$, as
\begin{equation}
    \frac{\partial c(\bar{\vecr}, \theta)}{\partial \theta} \approx
    \frac{c(\bar{\vecr}, \theta + h) - c(\bar{\vecr}, \theta)}{h}
    \label{eq:num-diff}
\end{equation}
with $h \in \mathbb{R}$ usually a small number. However, we already know the value
of $c(\bar{\vecr}, \theta)$ from the approximation obtained with the solution of
the OZ equation. We need only the value of $c(\bar{\vecr}, \theta + h)$ which can
be easily obtained by modifying the weights of $\nnet$ by the value $h$ and evaluating
the closure relation
\[
    c(\bar{\vecr}, \theta + h) = \exp{\left[-\beta u(\bar{\vecr}) + \gamma(\bar{\vecr}) + N_{\theta + h}(\bar{\vecr})\right] - \gamma(\bar{\vecr}) - 1} .
\]
It is important to point out that all operations are elementwise.

From here, we simply compute the rest of \autoref{eq:grad4} using the numerical 
approximation of $\partial_{\theta} c(\bar{\vecr}, \theta)$ and we use this information
to compute the value of the gradient in \autoref{eq:grad2}. We can then continue with the
development of the closed form of this gradient.
Putting all this information together we are able to find an expression for
$\partial_{\theta} \gamma_{n}(\vecr, \theta)$.
We take \autoref{eq:grad4} together with \autoref{eq:grad-pr} to obtain the following
expression
\begin{equation}
    \frac{\partial c(\vecr, \theta)}{\partial \theta} = 
    e^{p(\vecr, \theta)} \left[ \partial_{\theta} \gamma(\vecr, \theta) + \partial_{\theta} \nnet \right]
    - \partial_{\theta} \gamma(\vecr, \theta)
    \label{eq:grad5}
\end{equation}
and now to find an expression for $\partial_{\theta} \gamma(\vecr, \theta)$ we perform
the following steps:
\begin{subequations}
    \begin{align*}
        \partial_{\theta} c(\vecr, \theta) &=
        e^{p(\vecr, \theta)} \left[ \partial_{\theta} \gamma(\vecr, \theta) + \partial_{\theta} \nnet \right]
        - \partial_{\theta} \gamma(\vecr, \theta) \\
        \partial_{\theta} c(\vecr, \theta) &= e^{p(\vecr, \theta)} \partial_{\theta} \gamma(\vecr, \theta) + 
        e^{p(\vecr, \theta)} \partial_{\theta} \nnet -
        \partial_{\theta} \gamma(\vecr, \theta) \\
        \partial_{\theta} \gamma(\vecr, \theta) \left[ 1 - e^{p(\vecr, \theta)} \right] &=
        e^{p(\vecr, \theta)} \partial_{\theta} \nnet - \partial_{\theta} c(\vecr, \theta)
    \end{align*}
    \begin{equation}
        \boxed{\partial_{\theta} \gamma(\vecr, \theta) = \frac{e^{p(\vecr, \theta)} \partial_{\theta} \nnet - \partial_{\theta} c(\vecr, \theta)}{1 - e^{p(\vecr, \theta)}}}
        \label{eq:grad6}
    \end{equation}
\end{subequations}
An equation for the derivative $\partial_{\theta} \gamma(\vecr, \theta)$ has now been found 
which does satisfy the conditions of being dependent on the weights explicitly, and with
the numerical approximation of $\partial_{\theta} c(\vecr, \theta)$.

To finish up the gradient expression, take \autoref{eq:grad2}
\begin{equation*}
    \nabla_{\theta} J(\theta) = 2 \left[\gamma_{n}(\vecr, \theta) - \gamma_{n-1}(\vecr, \theta) \right]
    \left[ \partial_{\theta} \gamma_{n}(\vecr, \theta) - \partial_{\theta} \gamma_{n-1}(\vecr, \theta) \right] ,
\end{equation*}
and plug in \autoref{eq:grad6} to obtain a closed form of the gradient we seek
\begin{equation}
    \boxed{
    \nabla_{\theta} J(\theta) = 2 \left[\gamma_{n}(\vecr, \theta) - \gamma_{n-1}(\vecr, \theta) \right]
    \left[ \frac{e^{p_{n}(\vecr, \theta)} \partial_{\theta} \nnet - \partial_{\theta} c(\vecr, \theta)}{1 - e^{p_{n}(\vecr, \theta)}} - \frac{e^{p_{n-1}(\vecr, \theta)} \partial_{\theta} \nnet - \partial_{\theta} c(\vecr, \theta)}{1 - e^{p_{n-1}(\vecr, \theta)}} \right] .
    }
    \label{eq:closed}
\end{equation}
In practice, a smoothing factor of $\varepsilon=\num{1e-7}$ was used in the denominator of 
the gradient expression from \autoref{eq:closed} to avoid division by zero. Again, the 
multiplication and division in the same expression are elementwise operations. %% Gradient computations
\chapter{Numerical solution to the Ornstein-Zernike equation}
\label{AppendixB}

The Ornstein-Zernike (OZ) equation is usually solved using a particular closure for a given 
interaction potential. In this thesis, two closures were explored and discussed. However, 
the solution to the OZ equation remained the same. The solution is based on using the Fast 
Fourier Transform and the convolution theorem to solve algebraic equations instead of an 
integral equation. In this appendix, the complete numerical scheme used in this thesis will 
be described in detail, for the particular case of \(3\)-dimensional space.

The method used is a derivative of the so-called \emph{Piccard iterative} methods, with a 
variation due to Ng~\cite{ngHypernettedChainSolutions1974}, referred to in this work as the 
\emph{five-point method} of Ng. A slight variation was implemented in practice, which is 
just a straightforward extension to the Ng method.

\section{Fourier Transform of the Ornstein-Zernike equation}
The OZ for an isotropic fluid is, recalling from \autoref{eq:ornstein-zernike},
\begin{equation}
    h(r) = c(r) + \rho \int_{V} c(r') \, h(\lvert \vecr - \vecr' \rvert) \, d \vecr'
    \; ,
    \label{eq:oz-appendix}
\end{equation}
with \(\rho\) the particle number density, and \(h(r), c(r)\) are the total and direct correlation functions, respectively.

The \emph{Fourier transform} defined as,
\begin{equation}
    \hat{f}(k) = \int_{- \infty}^{\infty} f(x) \, e^{-2 \pi \, i k x} \, dx
    \; ,
    \label{eq:fourier-transform}
\end{equation}
and the \emph{inverse Fourier transform} is defined as,
\begin{equation}
    f(x) = \int_{- \infty}^{\infty} \hat{f}(k) \, e^{2 \pi \, i k x} \, dk
    \; .
    \label{eq:inv-fourier-transform}
\end{equation}

When using \autoref{eq:fourier-transform} in \autoref{eq:oz-appendix}, and using the 
\emph{convolution theorem}~\cite{kornerFourierAnalysis1989}, the result is the following 
algebraic equation,
\begin{equation}
    \hat{h}(k) = \hat{c}(k) + \rho \, \hat{c}(k) \, \hat{h}(k)
    \; .
    \label{eq:algebraic-oz}
\end{equation}
As is the case when dealing with Fourier transforms, solving \autoref{eq:algebraic-oz} is 
much easier than solving \autoref{eq:oz-appendix} directly. Not only because the equation 
is now an algebraic equation, but because there exist efficient numerical methods that can 
compute the Fourier transform in its discrete form.

\section{Piccard method}
The Piccard method is an iterative numerical method that can provide solutions to a 
\emph{linear system} based on previous iterations. The idea is to build a set of possible 
solutions that, when measured against each other, the error is minimized.

To see this more clearly, the OZ equation can be regarded as the linear system,
\begin{equation}
    A \, f = f
    \; ,
    \label{eq:linear-operator}
\end{equation}
with \(f \colon \mathbb{C} \mapsto \mathbb{C}\) is in general a complex-valued function, 
and \(A\) is a generic \emph{linear operator} defined on some functional space that acts on 
\(f\). In order to find the solutions to \autoref{eq:linear-operator}, the Piccard 
iterative method provides the following iterative algorithm,
\begin{equation}
    A \, f_{n+1} = f_{n}
    \; ,
    \label{eq:piccard}
\end{equation}
with \(n \in \mathbb{Z}\). When an initial function \(f_1\) is provided and plugged into
\autoref{eq:piccard}, several other functions are generated \(f_2, f_3, \dots\). If this 
sequence of functions converge to a particular limiting function \(g\), then that is 
defined as the solution to \autoref{eq:linear-operator}. However, in the case of fluids, 
and particularly liquids, the Piccard method often oscillates and diverges, thus no 
solutions is found. The method of Ng provides stability to the numerical method, and it 
shall be described next.

\section{The method of Ng}
Let \(g_n\) be defined as,
\begin{equation}
    g_n \coloneqq A \, f_n
    \: ,
    \label{eq:gn}
\end{equation}
and let \(d_n\) be defined as,
\begin{equation}
    d_n \coloneqq g_n - f_n = (A - 1) f_n
    \; .
    \label{eq:dn}
\end{equation}
In the numerical analysis jargon, function \(f_n\) is usually referred to as the 
\emph{input}, and function \(g_n\) is referred to as the \emph{output}. In both cases, 
\(n\) is the \(n\)-th iteration of the method. Further, \(d_n\) is usually employed as a 
\emph{metric} to measure the accuracy of the solution when obtaining its \(L^2\) norm in 
functional space, that is,
\begin{equation}
    {\left\lVert d_n \right\rVert}^{2} = \int {\left\lvert d_n (x) \right\rvert}^{2} \, dx
    \; .
    \label{eq:precision}
\end{equation}

Now, for the method of Ng, suppose that the following functions are known, for \(n \geq 6\),
\(f_{n-i}, g_{n-i}\) with \(i = 0,1,2,3,4,5 \, .\) The goal is to use all of these 
functions to generate a good estimation for the \emph{input} function to the iterative 
method, thus the following function is used,
\begin{equation}
    f=(1-c_{1}-c_{2}-c_{3}-c_{4}-c_{5}) f_n + \sum_{i=1}^{5} c_i f_{n-i}
    \; ,
    \label{eq:finit}
\end{equation}
with \(c_i\), \(i = 1,2,3,4,5\), arbitrary constants. Here, the goal is to find the values 
that will make \autoref{eq:finit} a good solution to \autoref{eq:linear-operator}.
Thus, pluggin \autoref{eq:finit} into \autoref{eq:linear-operator},
\begin{equation}
    A \, f = (1-c_{1}-c_{2}-c_{3}-c_{4}-c_{5}) g_n + \sum_{i=1}^{5} c_i g_{n-i}
    \; ,
    \label{eq:ginit}
\end{equation}
hence,
\begin{equation}
    \Delta \coloneqq \left\lVert A \, f - f \right\rVert =
    \left\lVert d_n - \sum_{i=1}^{5} c_i d_{0i} \right\rVert
    \; ,
    \label{eq:deltas}
\end{equation}
where
\begin{equation}
    d_{0i} = d_n - d_{n-i} \quad i=1,2,3,4,5
    \; .
    \label{eq:dzeros}
\end{equation}

Now, to find the best set of coefficients \(c_i\), the error \(\Delta^2\) must be minimized 
with respect to the coefficients \(c_i\), which gives,
\begin{equation}
    D_{ij} \cdot c_j = \left(d_n, d_{0i}\right)
    \; ,
    \label{eq:linear-system}
\end{equation}
where \(D_{ij} = \left(d_{0i}, d_{0j}\right)\) are the elements of the \(5 \times 5\) 
matrix, and the indices take the values \(i=j=1,2,3,4,5\). Here, \(\left(u, v\right)\) 
determines the \emph{inner product} defined to be,
\begin{equation}
    \left(u, v\right) = \int u(x) \, v(x) \, dx
    \: .
    \label{eq:inner-product}
\end{equation}

With this, the Ng method is now complete. The goal is to find the best values of \(c_i\) 
using \autoref{eq:linear-system} such that the \(n+1\)-th iteration of the \emph{input 
function}, \(f_{n+1}\),
\begin{equation}
    f_{n+1} = (1-c_{1}-c_{2}-c_{3}-c_{4}-c_{5}) g_n + \sum_{i=1}^{5} c_i g_{n-i}
    \; ,
    \label{eq:best-input}
\end{equation}
is the best approximation to solve \autoref{eq:linear-operator}.

\section{Numerical algorithm for solving the Ornstein-Zernike equation}
The numerical algorithm to solve the OZ equation will now be described, which uses the Ng 
method. For this, it is assumed that there is a \emph{closure relation} available that 
relates \(c(r)\) and \(\gamma(r) = h(r) - c(r)\) through a bridge function \(B(r)\). With 
this information, the numerical algorithm in \autoref{alg:cap} is used to solve the OZ equation.
\begin{algorithm}
    \caption{Piccard method for the OZ equation}\label{alg:cap}
    \begin{algorithmic}
        \State \(c(r) \gets \exp{\left[\beta u(r) + B(r) + \gamma(r)\right]} - \gamma(r) - 1\) \Comment{Using the bridge function} \\
        \State \(\mathcal{F} \left[c(r)\right] \gets \hat{c}(k)\) \\
        \State \(\hat{\gamma}(k) \gets \frac{\rho \, \hat{c}^{2}(k)}{1 - \rho \, \hat{c}(k)}\) \\
        \State \(\mathcal{F}^{-1} \left[\hat{\gamma}(k)\right] \gets \gamma^{\prime}(r)\) \\
        \State \(c^{\prime}(r) \gets \exp{\left[\beta u(r) + B(r) + \gamma^{\prime}(r)\right]} - \gamma(r) - 1\) \Comment{New \(c^{\prime}(r)\) using previous value of \(\gamma(r)\)}
    \end{algorithmic}
\end{algorithm}

In \autoref{alg:cap}, the symbol \(\mathcal{F} [f(x)]\) represents the Fourier transform of \(f(x)\), while \(\mathcal{F}^{-1} [\hat{f}(k)]\) represents the inverse Fourier transform of \(\hat{f}(k)\).
However, the solution to the OZ equation is quite sensible to the density of the system, 
which is why it makes it hard for the iterative method to converge properly, specially at 
high densities. A useful way of solving this issue is by constructing a grid of density 
values, and to solve the OZ equation at each intermediate value, until the target density 
is reached. This makes the method convergent, but at the cost of high computational 
resources, which stem from performing \autoref{alg:cap} as many times as there are density 
values in the grid. With this information, the complete algorithm will now be described 
next.

\begin{enumerate}
    \item \label{item:1} A grid of density values is built, \(\{\rho_i\} \, , i=1,2,\dots,M\); where \(\rho_{M}\) is the target density of the system, and \(\Delta n = n_{i+1} - n_{i}\).
    \item \label{item:2} Using the first density value \(n_1\) and \(\gamma_0 = 0\), \autoref{alg:cap} is used to obtain the functions \(\gamma_i (r) \, , i=1,2,3,4,5,6\). These functions are then mapped, \(\gamma_6 \mapsto g_n\), \(\gamma_5 \mapsto g_{n-1}\), \(\gamma_4 \mapsto g_{n-2}\), \(\gamma_3 \mapsto g_{n-3}\), \(\gamma_2 \mapsto g_{n-4}\), \(\gamma_1 \mapsto g_{n-5}\).
    \item \label{item:3} Then, using \autoref{eq:linear-system}, all the coefficients \(c_i\) are then computed, such that \autoref{eq:best-input} can then be obtained.
    \item \label{item:4} Using \autoref{eq:dn}, expression \(d_n = f_{n+1} - g_n\) is evaluated. If \(\lVert d_n \rVert > \num{1e-3}\), then the functions are updated as \(f_{n+1} \mapsto \gamma_6 \, , g_5 \mapsto \gamma_4 \, , g_4 \mapsto \gamma_3 \, , g_3 \mapsto \gamma_2 \, , g_2 \mapsto \gamma_1\). Then, the previous two steps are repeated until \(\lVert d_n \rVert \leq \num{1e-3}\).
    \item \label{item:5} Finally, once the condition \(\lVert d_n \rVert \leq \num{1e-3}\) is met, a new density value is taken \(n_2 = n_1 + \Delta n\), and the steps \ref{item:2} to \ref{item:4} are repeated. However, this time around, the initial \emph{input} is updated as \(\gamma_{0} \mapsto \gamma_6\), which resulted from the previous step. From here on, this process is repeated until convergence, which usually means until all the density values in the grid are used. However, the function update \(\gamma_{0} \mapsto \gamma_6\) is changed to be a linear extrapolation between the previous two results.
\end{enumerate}

\section{Extensions to the method}
In the actual method used in this work, the method of Ng is extended to use not only five 
input functions, but an arbitrary \(m\) number of functions. It turns out that 
this extension does not provide more accurate results, at least for the case of the hard 
sphere fluid. It does make it more stable for some high density values, numerically 
speaking. It would be interesting to see if the actual extension makes sense for other 
types of interaction potentials.

To solve the linear system in \autoref{eq:linear-system}, the common way of doing this is 
through the use of the LU decomposition~\cite{trefethenNumericalLinearAlgebra1997}. The 
main issue with the LU decomposition method is that is it slow, i.e., it needs a lot of 
floating point operations, and can also be numerically unstable when dealing with 
ill-conditioned matrices. Instead, in this work a faster method was employed which is based 
on the Krylov sub-space iterative family of methods. This method, called 
DIOM~\cite{saadPracticalUseKrylov1984a}, is an upgrade to the basic LU method, and can be 
used with ill-conditioned, symmetric matrices. %% Numerical scheme, OZ eqn

%----------------------------------------------------------------------------------------
%	BIBLIOGRAPHY
%----------------------------------------------------------------------------------------

\printbibliography[heading=bibintoc]

%----------------------------------------------------------------------------------------

\end{document}